\documentclass[aps,prx,twocolumn,notitlepage,superscriptaddress]{revtex4-2}

\usepackage{amsmath,amssymb, times}
\usepackage{mathtools}
\usepackage{color}
\usepackage{bm}
\usepackage{bbm}
\usepackage{graphicx}
\usepackage[normalem]{ulem}
\usepackage{enumitem}
\usepackage{makecell}
\usepackage{amsthm}
\usepackage{thmtools}

\usepackage{here}
\usepackage{amsthm}
\usepackage{thmtools}
\usepackage{blindtext}
\usepackage{array}
\usepackage{bbding}
\usepackage[colorlinks=true,citecolor=blue,linkcolor=blue]{hyperref}

\usepackage{tikz}
\usepackage{tikz-3dplot}
\usepackage{tikz-cd}
\usetikzlibrary{external}
\usetikzlibrary{matrix,positioning, decorations.pathreplacing,calligraphy,decorations.markings, knots, arrows.meta,decorations.markings}
\definecolor{tensorcolor}{rgb}{1., 1., 1.}
\definecolor{btensorcolor}{rgb}{0.65,0.77,0.95}

\tikzset{
    mid arrow/.style args={#1}{
        decoration={
            markings,
            mark=at position #1 with {\arrow{Stealth}}
        },
        postaction={decorate}
    }
}

\declaretheoremstyle[
shaded={bgcolor=\color{rgb}{0.9,0.9,0.9}}
]{definition}

\begin{document}

\preprint{APS/123-QED}

\title{Fermionic Gaussian Unitary Network}

\author{Wenqing Xie}
\affiliation{Department of Physics, Hong Kong University of Science and Technology, Clear Water Bay, Hong Kong SAR, China}
\affiliation{Center for Theoretical Condensed Matter Physics, Hong Kong University of Science and Technology, Clear Water Bay, Hong Kong SAR, China}

\author{Seishiro Ono}
\affiliation{Institute for Solid State Physics, University of Tokyo, Kashiwa 277-8581, Japan}
\affiliation{RIKEN Center for Interdisciplinary Theoretical and Mathematical Sciences (iTHEMS), RIKEN, Wako 351-0198, Japan}
\affiliation{RIKEN Quantum, Transformative Research Innovation Platform of RIKEN platforms (TRIP) Headquarters, RIKEN, Wako 351-0198, Japan}

\author{Hoi Chun Po}
\email{hcpo@ust.hk}
\affiliation{Department of Physics, Hong Kong University of Science and Technology, Clear Water Bay, Hong Kong SAR, China}
\affiliation{Center for Theoretical Condensed Matter Physics, Hong Kong University of Science and Technology, Clear Water Bay, Hong Kong SAR, China}

\begin{abstract}
We introduce fermionic Gaussian unitary networks as variational
ansatz for fermionic Gaussian unitaries. Unlike circuit-based
constructions, these networks allow directed loops while
preserving global unitarity up to an overall normalization
factor. We derive contraction rules that operate entirely at
the single-particle level, using orthogonal matrices in the
Majorana basis or unitary matrices in the particle-number-conserving
case. For translation-invariant networks, the contraction
decomposes into independent momentum sectors, enabling efficient
variational optimization over the Brillouin zone. We demonstrate the expressive power of the ansatz by
transforming product states into approximations of free-fermion
ground states, including those of the Qi--Wu--Zhang model and
a one-dimensional metal. For the latter, we introduce a
scale-invariant layered architecture that applies the same
unitary network at successive scales to approximate gapless
states at increasing system sizes. We further investigate real-time evolution under quadratic
Hamiltonians and find that unitary networks are particularly
well suited to flat-band dynamics, where entanglement growth
is typically suppressed. By converting optimized
Gaussian tensors into their many-body representations and
combining them with on-site interaction gates, we extend the
construction to interacting dynamics. We demonstrate this
approach for the one-dimensional Hubbard model and benchmark
its accuracy against time-evolving block decimation (TEBD).
\end{abstract}

\maketitle

\section{Introduction}

The Hilbert space of a quantum many-body system grows exponentially with the system size, making a direct representation of generic states or operators computationally intractable. Tensor networks provide an efficient alternative by decomposing many-body wave functions into networks of local tensors. This idea underlies several successful variational ansatzes, including \textit{matrix product states} (MPS) \cite{perezgarcia2007matrixproductstaterepresentations, Schollw_ck_2011}, \textit{projected entangled-pair states} (PEPS)~\cite{verstraete2004renormalizationalgorithmsquantummanybody, Verstraete_2008, Cirac2021Matrix}, and the \textit{multi-scale entanglement renormalization ansatz} (MERA)~\cite{Vidal2007Entanglement, Vidal2008Class}. Their
efficiency is closely tied to the entanglement structure of physical states, especially the area-law behavior of ground states of local gapped Hamiltonians~\cite{Hastings_2007, Eisert2010Colloquium}.

Beyond representing quantum states, tensor networks can also be used to
represent quantum operators and unitary transformations. \textit{Matrix product
operators} (MPOs) and unitary tensor-network circuits provide compact descriptions of
time evolution, variational state preparation, and renormalization
transformations~\cite{Vidal2007Entanglement,Haghshenas2021Optimization,Haghshenas2022Variational}.
Imposing exact unitarity on an MPO is a nontrivial constraint and has been
studied within the framework of \textit{matrix product unitaries}
(MPUs)~\cite{Chen2011Two-dimensional,Ignacio_Cirac_2017,Burak2018Matrix}.
Unitary networks~\cite{xie2025unitarynetworktensornetwork} provide a versatile construction by combining local unitary tensors according to a causal
ordering, which guarantees global unitarity while allowing representations
beyond uniform MPUs. However, when directed loops are present, this causal
ordering is lost, and the global unitarity of the contracted network is no
longer guaranteed. This issue is particularly relevant for translation-
invariant networks with periodic boundary conditions, where directed loops
arise naturally.

Fermionic Gaussian unitaries 
\cite{bravyi2004lagrangian, guaita2024representationtheorygaussianunitary, Hackl_2021}
are many-body unitary operators generated by quadratic fermionic 
Hamiltonians. Although they act on the full fermionic Fock space, they are characterized, up to a projective
sign, by their linear action on fermionic operators \cite{guaita2024representationtheorygaussianunitary, Hackl_2021}. In the Majorana 
representation, this action is described by an orthogonal matrix. When particle number is conserved, the transformation reduces to a unitary matrix acting on the annihilation  operators. This finite-dimensional single-particle representation makes 
fermionic Gaussian unitaries both physically relevant and 
computationally tractable.

Fermionic generalizations of several tensor-network ansatzes 
have been developed, including fermionic MERA 
\cite{Fermionic2009Corboz,interacting2010Corboz}, fermionic MPS 
\cite{Bultinck2017Fermionic}, and fermionic PEPS 
\cite{Kraus2010Fermionic,gu2010grassmanntensornetworkstates,
Simulation2010Corboz}. Gaussian fermionic PEPS, in particular, 
have been used to represent chiral topological states 
\cite{Wahl2013Projected,Dubail2015Tensor,Chiral2024Niu} 
and states with Fermi surfaces \cite{Tensor2022Mortier}.

Here, we focus on tensor-network representations of fermionic 
Gaussian unitaries, which represent operators rather than 
quantum states. In our previous work 
\cite{xie2025unitarynetworktensornetwork}, we applied the framework of unitary networks to construct fermionic Gaussian unitary, and showed that successive 
applications of the cosine-sine decomposition to the 
single-particle unitary transformation yield a compact tensor 
network that captures the locality of the transformation. 
In that construction, the local tensors are still arranged in a causal order (from an underlying directed acyclic graph), closely 
relating these networks to quantum circuits 
\cite{interacting2010Corboz,Disentangling2025Wu}. 
Fermionic Gaussian unitary networks are also related to fermionic Gaussian isometric 
tensor-network states 
\cite{wu2025alternatinggaussianfermionicisometric}, 
but differ in that the latter represent quantum states 
rather than operators.

In this work, we further develop the theory of fermionic Gaussian unitary networks, and deploy them 
as variational ansatz for fermionic Gaussian unitaries. 
In contrast to the adaptation of the general construction of unitary network to the fermionic Gaussian case, as was done in our 
work \cite{xie2025unitarynetworktensornetwork}, here we show that the condition on causal ordering of the local tensors can be lifted. This freedom allows for more general tensor-network 
geometries than those accessible to conventional quantum 
circuits \cite{interacting2010Corboz,Disentangling2025Wu}. Importantly, for the fermionic Gaussian unitary networks considered here,  directed loops in the network do not destroy the global unitarity of the 
contracted operator but contribute only an overall scalar 
normalization factor. 
This property allows us to extend the construction beyond 
directed acyclic networks and consider translation-invariant 
networks with periodic boundary conditions.

For translation-invariant networks built from uniform local tensors, the single-particle contraction further decomposes into independent momentum sectors. The contraction result at each momentum $\mathbf{k}$ is obtained directly from the local tensor data. This momentum-space formulation turns the variational optimization into a data-driven problem over the Brillouin zone: different momentum points $(\mathbf{k}, U_\mathbf{k})$ serve as training samples, making stochastic optimization possible.

We demonstrate the expressive power of the fermionic Gaussian unitary-network ansatz using two representative free-fermion models. For the Qi--Wu--Zhang model~\cite{Qi2006Topological}, we show that a network with finite bond dimension, independent of system size, can accurately approximate an entangling Gaussian unitary that maps a simple product state to the target ground state. For the gapless half-filled one-dimensional nearest-neighbor metal, we introduce a scale-invariant layered architecture based on the zipper entanglement renormalization scheme~\cite{wong2022zipperentanglementrenormalizationfree}, which shares the multiscale structure of MERA~\cite{Vidal2007Entanglement, Vidal2008Class}. By applying the same unitary-network transformation at each layer, the construction generates approximations to the target state at progressively larger system sizes while keeping the local tensor unchanged.

We also apply the framework to real-time evolution under quadratic fermionic Hamiltonians. For generic dispersive bands, the resulting entanglement growth eventually exceeds the expressive capacity of a fixed-bond-dimension network, whereas flat-band Hamiltonians, with vanishing group velocity and periodic dynamics, are particularly well suited to this representation. We then extend the construction beyond purely Gaussian dynamics by converting the optimized Gaussian network into a many-body tensor-network operator and composing it with local interaction gates. As a proof of principle, we apply this scheme to the one-dimensional Hubbard model and benchmark the resulting time evolution against the \textit{time-evolving block decimation} (TEBD) method \cite{Vidal_2004, A_J_Daley_2004}.

\section{Fermionic Gaussian unitary operators}

Fermionic Gaussian unitaries are generated by quadratic fermionic
Hamiltonians and are characterized, up to an overall sign, by
their linear action on fermionic operators
\cite{bravyi2004lagrangian,guaita2024representationtheorygaussianunitary,
surace2022fermionic,Hackl_2021}.
We collectively denote these operators by
$\hat{\boldsymbol{\xi}}$, which can be expressed in the
Majorana basis $\hat{\boldsymbol{\xi}}=(\hat{\mathbf{q}},\hat{\mathbf{p}})$
or the creation-annihilation basis
$\hat{\boldsymbol{\xi}}=(\hat{\mathbf{c}},\hat{\mathbf{c}}^\dagger)$.
A fermionic Gaussian unitary $\hat{U}$ acts by conjugation as
\begin{equation}
    \hat{U}^\dagger \hat{\xi}^a \hat{U}
    = R^a{}_b \hat{\xi}^b.
\end{equation}
For $N$ fermionic modes, $R\in\mathrm{SO}(2N)$ in the Majorana
basis, whereas in the creation-annihilation basis it describes
a Bogoliubov transformation preserving the canonical
anticommutation relations.

For transformations connected to the identity, one may write $R=e^{K}$, where $K$ belongs to the corresponding single-particle Lie algebra. The associated many-body unitary operator is obtained through the projective representation $\hat{\rho}$,
\begin{equation}
\label{eq: projective representation}
\begin{aligned}
    &\hat{\rho}(e^{K}) = \pm e^{\hat{K}}, \\
    &\hat{K} = \frac{1}{2} g_{ac} K^{c}_{\; b}\,\hat{\xi}^{\dagger}_{a}\hat{\xi}_{b},
\end{aligned}
\end{equation}
where $g_{ac}$ is the inverse of the matrix $G^{ab}$ defined by the anticommutation relations
\[
 \{\hat{\xi}^{a}, \hat{\xi}^{b}\} = G^{ab}\hat{I}.
\]
Thus, the single-particle transformation determines the corresponding many-body Gaussian unitary up to an overall sign.

In the particle-number-conserving case, the transformation reduces to an $N\times N$ single-particle unitary matrix $U=e^{-ih}$. The relation between $U$ and the corresponding many-body operator $\hat{U}$ is
\begin{equation}
\begin{aligned}
    \hat{\rho}(e^{-ih}) &= \pm e^{-i\hat{H}}, \\
    \hat{H} &= h_{ab}\hat{c}_a^\dagger \hat{c}_b .
\end{aligned}
\end{equation}
Throughout this paper, hatted symbols, such as $\hat{U}$, denote many-body operators, while unhatted symbols, such as $U$, denote their single-particle representations. Further details of fermionic Gaussian unitary transformations are summarized in Appendix~\ref{Append: Fermionic Gaussian unitary operations}.

\section{Fermionic Gaussian Unitary Network}

Fermionic Gaussian unitaries admit single-particle
representations as orthogonal matrices in the Majorana basis
or unitary matrices in the particle-number-conserving case.
We extend this correspondence to local Gaussian tensors and
introduce compatible contraction rules for their
single-particle matrices. This allows fermionic Gaussian
unitary networks to be constructed and contracted entirely
at the single-particle level, substantially simplifying
tensor-network calculations.

In our previous work~\cite{xie2025unitarynetworktensornetwork},
we restricted the underlying graph $G[U_{\mathrm{Net}}]$ to
a directed acyclic graph (DAG). This structure provides a
causal ordering of the local unitary tensors, allowing the
network to be assembled sequentially while ensuring that
the resulting global tensor is unitary.

In this work, we lift the DAG restriction and allow directed
loops. This generalization is particularly useful for
implementing translation invariance in finite systems with
periodic boundary conditions, where the network naturally
contains loops connecting opposite boundaries. For example,
a translation-invariant unitary-network ansatz can be
constructed as follows:
\begin{align} \label{fig:1D PBC UN}
    \begin{array}{c}
    \tikzsetnextfilename{1D_UN}
    \begin{tikzpicture} 
    \node[draw, rectangle, rounded corners=2pt, minimum width=0.6cm, minimum height=0.6cm, fill=tensorcolor, thick] (a) at (0,0) {};
    \node[draw, rectangle, rounded corners=2pt, minimum width=0.6cm, minimum height=0.6cm, fill=tensorcolor, thick] (b) at (1.2,0) {};
    \node[draw, rectangle, rounded corners=2pt, minimum width=0.6cm, minimum height=0.6cm, fill=tensorcolor, thick] (c) at (2.4,0) {};
    \node[draw, rectangle, rounded corners=2pt, minimum width=0.6cm, minimum height=0.6cm, fill=tensorcolor, thick] (d) at (3.6,0) {};
    \draw[mid arrow=0.65] (a.90) -- ++(0,0.6);
    \draw[mid arrow=0.65] ([yshift=-0.6cm]a.270) -- ++(0,0.6);
    \draw[mid arrow=0.65] (b.90) -- ++(0,0.6);
    \draw[mid arrow=0.65] ([yshift=-0.6cm]b.270) -- ++(0,0.6);
    \draw[mid arrow=0.65] (c.90) -- ++(0,0.6);
    \draw[mid arrow=0.65] ([yshift=-0.6cm]c.270) -- ++(0,0.6);
    \draw[mid arrow=0.65] (d.90) -- ++(0,0.6);
    \draw[mid arrow=0.65] ([yshift=-0.6cm]d.270) -- ++(0,0.6);
    \draw[mid arrow=0.65] ([yshift=0.1cm]a.0) -- ([yshift=0.1cm]b.180);
    \draw[mid arrow=0.65] ([yshift=-0.1cm]b.180) -- ([yshift=-0.1cm]a.0);
    \draw[mid arrow=0.65] ([yshift=0.1cm]b.0) -- ([yshift=0.1cm]c.180);
    \draw[mid arrow=0.65] ([yshift=-0.1cm]c.180) -- ([yshift=-0.1cm]b.0);
    \draw[mid arrow=0.65] ([yshift=0.1cm]c.0) -- ([yshift=0.1cm]d.180);
    \draw[mid arrow=0.65] ([yshift=-0.1cm]d.180) -- ([yshift=-0.1cm]c.0);
    \draw[mid arrow=0.5] ([yshift=0.1cm]d.0) --  ++(0.3,0)-- ++(0,-0.5) -- ++(-4.8,0) -- ++(0,0.5) --([yshift=0.1cm]a.180);
    \draw[mid arrow=0.5] ([yshift=-0.1cm]a.180) --  ++(-0.4,0)-- ++(0,-0.5) -- ++(5,0) -- ++(0,0.5) --([yshift=-0.1cm]d.0);
\end{tikzpicture}
\end{array}.
\end{align}
As illustrated in the diagram above, directed loops are necessary to realize translational invariance. In a generic unitary network, however, such loops generally destroy global unitarity. Fermionic Gaussian unitary networks are exceptional: even in the presence of directed loops, the contracted tensor remains proportional to a unitary operator, with the loops contributing only an overall normalization factor.

\subsection{Fermionic Gaussian Unitary Tensor as a Single-Particle Unitary Matrix}

The tensor-product structure of local fermionic Hilbert spaces is $\mathbb{Z}_2$-graded ~\cite{Bultinck2017Fermionic}. We denote the graded tensor product of two fermionic Hilbert spaces by $\mathcal{H}_1 \otimes_g \mathcal{H}_2$.  A unitary tensor can be obtained from a unitary operator by splitting its incoming and outgoing Hilbert space into multiple legs \cite{ferris2012perfect, xie2025unitarynetworktensornetwork}. For example, as shown in panel (a) below, the many-body unitary operator $\hat{U}:\mathcal{H}_{\mathrm{in}}\to
\mathcal{H}_{\mathrm{out}}$ can be regarded as a rank-4 tensor, with its the output Hilbert space decomposes as
$\mathcal{H}_{\mathrm{out}}=\mathcal{H}_{A}\otimes_{g} \mathcal{H}_{B}$ and the
input Hilbert space as
$\mathcal{H}_{\mathrm{in}}=\mathcal{H}_{C}\otimes_{g} \mathcal{H}_{D}$:
\begin{align}
\begin{array}{c}
\tikzsetnextfilename{mode_unitary}
\begin{tikzpicture}
    \begin{scope}  [local bounding box=graph_a]
        \node[draw, rectangle, rounded corners=2pt, minimum width=1.0cm, minimum height=0.6cm, fill=tensorcolor, thick] (hat_U){$\hat{U}$};
        \draw[mid arrow=0.6] ([xshift=-0.3cm]hat_U.90) -- ++(0,0.4) node[left] {$A$};
        \draw[mid arrow=0.6] ([xshift=0.3cm]hat_U.90) -- ++(0,0.4) node[right] {$B$};
        \draw[mid arrow=0.6] ([xshift=-0.3cm,yshift=-0.4cm]hat_U.270) node[left] {$C$} -- ++(0,0.4) ;
        \draw[mid arrow=0.6] ([xshift=0.3cm,yshift=-0.4cm]hat_U.270) node[right] {$D$} -- ++(0,0.4) ;
    \end{scope}
    \begin{scope}  [shift={(4, 0)}, local bounding box=graph_b]
        \node[draw, rectangle, rounded corners=2pt, minimum width=1.0cm, minimum height=0.6cm, fill=tensorcolor, thick] (U){$U$};
        \draw[mid arrow=0.6] ([xshift=0.3cm]U.90) -- ++(0,0.4) node[right] {$B$};
        \draw[mid arrow=0.6] ([xshift=-0.3cm]U.90) -- ++(0,0.4) node[left] {$A$};
        \draw[mid arrow=0.6] ([xshift=-0.3cm,yshift=-0.4cm]U.270) node[left] {$C$} -- ++(0,0.4);
        \draw[mid arrow=0.6] ([xshift=0.3cm,yshift=-0.4cm]U.270) node[right] {$D$} -- ++(0,0.4);
    \end{scope}
    \node  at ([xshift=-0.2cm, yshift=0.2cm]graph_a.north west) {(a)};
    \node at ([xshift=-0.2cm, yshift=0.2cm]graph_b.north west) {(b)};
\end{tikzpicture}
\end{array},
\end{align}
For fermionic Gaussian unitaries, the same leg structure can
be represented at the single-particle level
\cite{xie2025unitarynetworktensornetwork}. Let $U$ denote
the single-particle  matrix associated with $\hat{U}$,
such that $\hat{\rho}(U)=\pm\hat{U}$.  Partitioning the output
modes into $A,B$ and the input modes into $C,D$, we obtain
\[
U =
\begin{pmatrix}
U_{AC} & U_{AD} \\
U_{BC} & U_{BD}
\end{pmatrix}.
\]
This block matrix form is reflected in the leg arrangement shown
in panel (b). Although
panels (a) and (b) share the same leg arrangement, they
represent different mathematical objects: the many-body
tensor $\hat{U}$ in panel (a) and the single-particle matrix
$U$ in panel (b). In panel (b), outgoing and incoming legs
represent blocks of rows and columns, respectively.

\subsection{Fermionic Gaussian Unitary Networks}

A fermionic Gaussian unitary network $\hat{U}_{\mathrm{Net}}$
admits a corresponding single-particle representation
$U_{\mathrm{Net}}$, obtained by connecting the single-particle
matrices associated with its local Gaussian tensors and
applying the contraction rules introduced below.

This network-level correspondence follows from the compatibility
of $\hat{\rho}$ with elementary tensor-network operations.
First, combining two disconnected tensors corresponds to a
direct sum of their single-particle matrices. Second, contracting
a shared mode leg at the single-particle level corresponds to
contracting the associated many-body tensor legs:
\begin{equation}
\begin{aligned}
   \hat{\rho}(U_{1}\oplus U_{2})
   &= \pm \hat{\rho}(U_{1})\otimes_g\hat{\rho}(U_{2}),\\
   \hat{\rho}(U_{1}\times_B U_{2})
   &= \pm \hat{\rho}(U_{1})\times_B\hat{\rho}(U_{2}),
\end{aligned}
\end{equation}
where $U_{1}\times_B U_{2}$ denotes contraction over the
shared mode leg $B$. The overall signs arise from the
projective nature of the representation. For networks with directed loops, we also require a
self-contraction rule, which relates the many-body
supertrace to its single-particle counterpart:
\begin{equation}
  \operatorname{STr}_{B}[\hat{U}]
  = \mathcal{N}\,\hat{\rho}\bigl(
  \operatorname{str}_{B}[U]\bigr),
\end{equation}
where $\mathcal{N}$ is an overall normalization factor.

We formulate the contraction rules directly in terms of
single-particle unitary matrices, allowing a network
$U_{\mathrm{Net}}$ to be constructed from local matrices
$\{U_1,U_2,\ldots,U_n\}$. We first consider the
particle-number-conserving case. The generalization to
particle-number-nonconserving Gaussian transformations is
discussed in Sec.~\ref{sec: Without fermion number conserving}.

\subsubsection{Parallel composition}
When two fermionic Gaussian unitary operators, $\hat{U}$ and $\hat{V}$, act in parallel on disjoint sets of modes, their combined action is given by the graded tensor product $\hat{U} \otimes_{g} \hat{V}$. At the single-particle level, this operation corresponds to the direct sum $U \oplus V$:
\begin{align}
        \begin{array}{c}
        \tikzsetnextfilename{direct_sum_mode_unitary}
        \begin{tikzpicture}
        \begin{scope}  [local bounding box=graph_a, shift={(0,0)}]
        \node[draw, rectangle, rounded corners=2pt, minimum width=1.2cm, minimum height=0.6cm, fill=tensorcolor, thick] (U){$U$};
        \draw[mid arrow=0.6] ([xshift=-0.4cm]U.90) -- ++(0,0.4);
        \draw[mid arrow=0.6] (U.90) -- ++(0,0.4);
        \draw[mid arrow=0.6] ([xshift=0.4cm]U.90) -- ++(0,0.4);
        \draw[mid arrow=0.6] ([xshift=-0.4cm,yshift=-0.4cm]U.270) -- ++(0,0.4);
        \draw[mid arrow=0.6] ([yshift=-0.4cm]U.270) -- ++(0,0.4);
        \draw[mid arrow=0.6] ([xshift=0.4cm,yshift=-0.4cm]U.270) -- ++(0,0.4);
    \end{scope}
    \begin{scope}  [local bounding box=graph_b, shift={(1.6,0)}]
        \node[draw, rectangle, rounded corners=2pt, minimum width=1.2cm, minimum height=0.6cm, fill=tensorcolor, thick] (U){$V$};
        \draw[mid arrow=0.6] ([xshift=-0.4cm]U.90) -- ++(0,0.4);
        \draw[mid arrow=0.6] (U.90) -- ++(0,0.4);
        \draw[mid arrow=0.6] ([xshift=0.4cm]U.90) -- ++(0,0.4);
        \draw[mid arrow=0.6] ([xshift=-0.4cm,yshift=-0.4cm]U.270) -- ++(0,0.4);
        \draw[mid arrow=0.6] ([yshift=-0.4cm]U.270) -- ++(0,0.4);
        \draw[mid arrow=0.6] ([xshift=0.4cm,yshift=-0.4cm]U.270) -- ++(0,0.4);
    \end{scope}
    \end{tikzpicture}
    \end{array} = U \oplus V =  \left( \begin{array}{cc}
    U  & 0\\
      0  &  V
   \end{array}\right).
\end{align}

\subsubsection{Contraction rules for causally ordered unitary tensors}

For a unitary network, the contractions of its local tensors must satisfy an additional constraint: each contracted incoming leg must be paired with an outgoing leg. By representing each local tensor as a vertex and each leg as a directed edge, the unitary network $U_{\mathrm{Net}}$ defines a directed graph $G[U_{\mathrm{Net}}]=(V,E)$. If $G[U_{\mathrm{Net}}]$ is a directed acyclic graph (DAG), i.e., it contains no directed cycles, then the local unitary matrices admit a well-defined causal ordering.

For two fermionic Gaussian unitary operators with such a causal ordering, their contraction corresponds to the sequential application of the associated unitary gates. Accordingly, the contraction of two single-particle local unitary matrices with a well-defined causal order is given by:
\begin{equation}
\begin{aligned}
    \begin{array}{c}
    \tikzsetnextfilename{contraction_mode_unitary}
    \begin{tikzpicture}
        \node[draw, rectangle, rounded corners=2pt, minimum width=1cm, minimum height=0.6cm, fill=tensorcolor, thick] (U1){$U$};
        \node[draw, rectangle, rounded corners=2pt, minimum width=1cm, minimum height=0.6cm, fill=tensorcolor, thick] (U2) at (0.6,-1.2) {$V$};
        \draw[mid arrow=0.65] ([xshift=-0.3cm]U1.90) -- ++ (0,0.6);
        \draw[mid arrow=0.65] ([xshift=0.3cm]U1.90) -- ++ (0,0.6);
        \draw[mid arrow=0.65] ([xshift=-0.3cm,yshift=-0.6cm]U1.270) --  node[left] {$A$} ++(0,0.6);
        \draw[mid arrow=0.65] ([xshift=0.3cm,yshift=-0.6cm]U1.270) -- node[left] {$B$} ++(0,0.6);
         \draw[mid arrow=0.65] ([xshift=0.9cm,yshift=-0.6cm]U1.270) -- node[left] {$C$} ++(0,0.6);
          \draw[mid arrow=0.65] ([xshift=-0.3cm,yshift=-0.6cm]U2.270) -- ++ (0,0.6);
        \draw[mid arrow=0.65] ([xshift=0.3cm,yshift=-0.6cm]U2.270) -- ++ (0,0.6);
        \draw[mid arrow=0.65, densely dashed] ([shift={(-0.3,-1.8)}]U1.270) -- ++ (0,1.2);
        \draw[mid arrow=0.65, densely dashed] ([shift={(0.3,0.6)}]U2.90) -- ++ (0,1.2);
    \end{tikzpicture}
    \end{array} &= 
  U \times_{B} V \\
  &= (U \oplus I_{C}) \times (I_{A} \oplus V)\\
  &=    \left(
     \begin{array}{c|c|c}
     U_{A}  & U_{B} & 0 \\
     \hline
     0 & 0 & I_{C}
   \end{array} 
\right) \left(
     \begin{array}{c|c}
     I_{A}  & 0  \\
     \hline
     0 & V^B \\
     \hline
     0 & V^{C}
   \end{array} 
\right) \\
&= \left(
     \begin{array}{c|c}
     U_{A}  & U_{B}V^{B}  \\
     \hline
     0 & V^{C} 
   \end{array} 
\right).
\end{aligned}
\end{equation}
In this notation, $U \times_{B} V$ denotes the contraction over the pair of legs $\{B_{U}, B_{V}\}$ and their associated modes. Standard matrix multiplication $UV$ is a special case of this contraction, in which every input mode of $U$ is paired with the corresponding output mode of $V$.

\subsubsection{Self-contraction as a linear fractional transformation}
\label{subsubsec: Contract without causal order}

To handle networks containing directed cycles, we recast their
contraction as a self-contraction of a single enlarged
fermionic Gaussian tensor. Consider a network composed of
local tensors $\{\hat{u}_1,\hat{u}_2,\ldots,\hat{u}_n\}$.
Before contracting any internal legs, we form their graded
tensor product:
\[
\hat{U}_{\mathrm{prod}}
=
\mathop{\bigotimes\nolimits_g}_{i=1}^{n}\hat{u}_i.
\]
The resulting tensor collects all local tensors into a
single enlarged fermionic Gaussian tensor. At the
single-particle level, this construction corresponds to
the direct sum of the associated matrices:
\begin{equation}
 \quad U_{\mathrm{prod}} = \bigoplus_{i} u_{i}.
\end{equation}
The contraction of the original network can be viewed as a sequence of self-contractions of $\hat{U}_{\mathrm{prod}}$. By inserting appropriate fermionic swap gates, each contracted pair of legs, consisting of one incoming and one outgoing leg, can be brought into vertical alignment. As explained in Refs.~\cite{Mortier_2025, Bultinck2017Fermionic, ren2025graphicalcalculusfermionictensors}, a self-contraction over the fermionic Hilbert space $\mathcal{H}_{B}$ is implemented by a partial supertrace rather than an ordinary partial trace:
\[
  \operatorname{STr}_{B}[\hat{U}]
    =
    \operatorname{Tr}_{B}\!\left[\hat{P}_{B}\hat{U}\right],
\]
where $\hat{P}_{B}=(-1)^{\hat{N}_{B}}$
is the fermion-parity operator on $\mathcal{H}_{B}$. The supertrace accounts for the fermionic sign acquired when bringing a bra $\langle y|$ past a ket $|x\rangle$ before contraction. In a $\mathbb{Z}_{2}$-graded tensor product, this exchange gives
\[
 |x\rangle \langle y|
    =
    (-1)^{|x||y|}
    \langle y| |x\rangle ,
\]
where $|x|,|y|\in\{0,1\}$ denote the corresponding fermion parities.

For a fermionic Gaussian unitary operator, the partial supertrace can be evaluated directly from its single-particle unitary matrix. Consider a fermionic Gaussian unitary acting on a bipartite system $A+B$, with single-particle representation
\[
U = \left(
    \begin{array}{cc}
     U_{AA}  & U_{AB} \\
     U_{BA} & U_{BB} 
   \end{array} 
   \right),
\]
and let $\hat{U} = \hat{\rho}(U)$ denote the corresponding many-body operator. As shown in Appendix \ref{Append: Self-contraction rule with fermion number conservation}, the partial supertrace over subsystem $B$ is given by
\begin{equation}
  \mathrm{STr}_{B} \left[ \hat{U} \right] = \mathrm{det} (I-U_{BB}) \cdot \hat{\rho}( \rm{str}_{B} [U]).
\end{equation}
Here, $\rm{str}_{B}[U]$ denotes the corresponding self-contraction at the single-particle level, defined by
\begin{equation}
\label{eq: self contraction}
   \begin{array}{c}
    \begin{tikzpicture}
        \node[draw, rectangle, rounded corners=2pt, minimum width=1cm, minimum height=0.6cm, fill=tensorcolor, thick] (U1){$U$};
        \draw[mid arrow=0.55] ([xshift=0.3cm]U1.90) -- ++ (0,0.4) node[above] {$B$} -- ++(0.4,0) -- ++(0,-1.4) -- ++(-0.4,0) --([xshift=0.3cm]U1.270);
        \draw[mid arrow=0.65] ([xshift=-0.3cm]U1.90) -- ++ (0,0.4) node[above] {$A$} ;
        \draw[mid arrow=0.65] ([xshift=-0.3cm, yshift=-0.4cm]U1.270) -- ([xshift=-0.3cm]U1.270) ;
    \end{tikzpicture}
   \end{array} 
    = \mathrm{str}_{B} [U] \coloneqq U_{AA} + U_{AB}(I_{BB}-U_{BB})^{-1}U_{BA},
\end{equation}
where $I_{BB}$ is the identity matrix on subsystem $B$. The inverse can be expressed as the power series
\begin{equation}
\label{eq: power series}
    (I_{BB}-U_{BB})^{-1} = I_{BB}+U_{BB}+U_{BB}^2 + \cdots = \sum_{n}U_{BB}^{n}.
\end{equation}
The factor $\mathrm{det}(I-U_{BB})$ contributes only an overall normalization and therefore does not affect the single-particle transformation. The resulting matrix $\mathrm{str}_{B}[U]$ acts solely on subsystem $A$ and remains unitary. A proof of this contraction rule and its unitarity is provided in Appendix \ref{Append: Self-contraction rule with fermion number conservation}.

The transformation in Eq.~\eqref{eq: self contraction} is a linear fractional transformation (LFT) of matrices, a structure widely used in control theory \cite{zhou1996robust}. More specifically, it is a special case of the Redheffer star product \cite{redheffer1959difference}, which describes feedback interconnections between systems. This linear fractional transformation admits a simple scattering interpretation. Consider a particle entering through subsystem $A$. After the single-particle unitary matrix $U$ acts, the particle may exit through either subsystem $A$ or subsystem $B$. Since the modes in $B$ are contracted, any amplitude exiting through $B$ is fed back into $U$ and scattered again. This process repeats until the particle eventually exits through $A$. The resulting sequence of repeated scattering processes is encoded in the power-series expansion in Eqs.~\eqref{eq: self contraction} and \eqref{eq: power series}.

Consider, for example, the contraction of two tensors $\hat{U}$ and $\hat{V}$:
\begin{align}
 \begin{array}{c}
    \tikzsetnextfilename{contraction_as_trace0}
    \begin{tikzpicture}
        \node[draw, rectangle, rounded corners=2pt, minimum width=0.6cm, minimum height=0.6cm, fill=tensorcolor, thick] (a) at (0,0) {$\hat{U}$};
        \node[draw, rectangle, rounded corners=2pt, minimum width=0.6cm, minimum height=0.6cm, fill=tensorcolor, thick] (b) at (0,-1.0) {$\hat{V}$};
        \draw[mid arrow=0.65] (a.90) -- ++(0,0.4);
        \draw[mid arrow=0.65] (b.90) -- (a.270);
        \draw[mid arrow=0.65] ([yshift=-0.4cm]b.270) -- (b.270);
    \end{tikzpicture}
    \end{array}
    =
    \begin{array}{c}
    \tikzsetnextfilename{contraction_as_trace}
    \begin{tikzpicture}
        \node[draw, rectangle, rounded corners=2pt, minimum width=0.6cm, minimum height=0.6cm, fill=tensorcolor, thick] (a) at (0,0) {$\hat{U}$};
        \node[draw, rectangle, rounded corners=2pt, minimum width=0.6cm, minimum height=0.6cm, fill=tensorcolor, thick] (b) at (1.0,0) {$\hat{V}$};
        \draw[mid arrow=0.65] (a.90) -- ++(0,0.4);
        \draw[mid arrow=0.55] (b.90) -- ++(0,0.4) -- ++(-0.5, 0) -- ++(0,-1.4) -- ++(-0.5, 0) -- (a.270);
        \draw[mid arrow=0.65] ([yshift=-0.4cm]b.270) -- (b.270);
    \end{tikzpicture}
    \end{array}
    = 
     \begin{array}{c}
    \tikzsetnextfilename{contraction_as_trace2}
    \begin{tikzpicture}
        \node[draw, rectangle, rounded corners=2pt, minimum width=0.6cm, minimum height=0.6cm, fill=tensorcolor, thick] (a) at (0,0) {$\hat{U}$};
        \node[draw, rectangle, rounded corners=2pt, minimum width=0.6cm, minimum height=0.6cm, fill=tensorcolor, thick] (b) at (1.0,0) {$\hat{V}$};
        \draw[mid arrow=0.65] (a.90) -- ++(0,0.4) node[above] {$A$} ;
        \draw[mid arrow=0.65] ([yshift=-1.0cm]a.270) -- ++(0,0.4);
        \draw[rounded corners] ([yshift=-0.6cm]a.270) -- ++(0,0.2) -- ([yshift=-0.2cm]b.270) -- (b.270);
        \draw[rounded corners] ([yshift=-0.6cm]b.270) -- ++(0,0.2) -- ([yshift=-0.2cm]a.270) -- (a.270);
        \draw[mid arrow=0.55] (b.90)  -- ++(0,0.4) node[above] {$B$} -- ++(0.5, 0) -- ++(0,-2.0) -- ++(-0.5, 0) -- ([yshift=-0.6cm]b.270);
    \end{tikzpicture}
    \end{array}
    .
\end{align} 
As illustrated in the diagram above, this contraction can be implemented by first taking the graded tensor product of the two tensors, then applying appropriate fermionic swap gates to align the modes to be contracted \cite{Mortier_2025, ren2025graphicalcalculusfermionictensors}, and finally taking the partial supertrace over those modes:
\begin{equation}
    \hat{U}\hat{V} = \hat{U} \times_{B}\hat{V} = \rm{STr}_{B } [(\hat{U} \otimes_{g} \hat{V}) \cdot \rm{fSWAP}_{AB}], 
\end{equation}
where $B$ denotes the set of contracted modes. If $\hat{U}$ and $\hat{V}$ are fermionic Gaussian unitary operators with single-particle representations $U$ and $V$, respectively, the corresponding contraction rule at the single-particle level reduces to ordinary matrix multiplication. Explicitly,
\begin{equation}
\begin{aligned}
     UV &= \textrm{str}_B [\begin{pmatrix}
        U & 0 \\
        0 & V
    \end{pmatrix}  \begin{pmatrix}
        0 & I \\
        I & 0
    \end{pmatrix}] \\
    &= \textrm{str}_B [\begin{pmatrix}
        0 & U \\
        V & 0
    \end{pmatrix}] \\
    &= 0 + U I^{-1}V \\
    &= UV.
\end{aligned}
\end{equation}
Thus, the single-particle self-contraction rule $\textrm{str}_B$ reproduces the standard composition law for two causally ordered fermionic Gaussian unitaries.

\section{Momentum-space formulation of translation-invariant unitary networks}

We consider a translation-invariant unitary network built from
a uniform local tensor $\hat{u}$. For finite systems, imposing
translation invariance naturally leads to periodic boundary
conditions, which introduce directed cycles and eliminate the
causal ordering of the network. Consequently, global unitarity
is no longer guaranteed for generic unitary networks. In
contrast, fermionic Gaussian unitary networks preserve global
unitarity up to an overall normalization factor, as shown in
Appendices~\ref{Append: Self-contraction rule with fermion number conservation}
and~\ref{Append: Self-contraction rule without fermion number conservation}.

We first consider the particle-number-conserving case and
defer the number-nonconserving case to
Sec.~\ref{sec: Without fermion number conserving}.
The local tensor $\hat{u}$ is represented by a single-particle
unitary matrix $u$, which defines the translation-invariant
network $U_{\mathrm{Net}}[u]$. We partition $u$ into physical
($P$) and bond ($B$) sectors:
\[
u =
\begin{pmatrix}
u_{PP} & u_{PB} \\
u_{BP} & u_{BB}
\end{pmatrix}.
\]
Here, $P$ and $B$ denote the physical and bond modes,
respectively. Each bond leg defines a directed edge in
$U_{\mathrm{Net}}[u]$, labeled by
$(\mathrm{in/out},\mathbf{b})$, where $\mathrm{in/out}$
specifies the leg orientation and $\mathbf{b}$ is the
displacement vector from the tail to the head of the edge.
The diagram below illustrates this convention, with $a$
denoting the lattice spacing:
\begin{equation}
   \begin{array}{c}
    \begin{tikzpicture}
        \node[draw, rectangle, rounded corners=2pt, minimum width=0.6cm, minimum height=0.6cm, fill=tensorcolor, thick] (U1){$u$};
        \draw[mid arrow=0.65] (U1.90) -- ++ (0,0.6) node[above] {$(\mathrm{out},P)$} ;
        \draw[mid arrow=0.65] ([yshift=-0.6cm]U1.270) node[below] {$(\mathrm{in},P)$} -- (U1.270) ;
        \draw[mid arrow=0.65] ([yshift=0.1cm]U1.0)  -- ++(0.6,0) node[above] {$(\mathrm{out}, a\mathbf{\hat{i}})$} ;
        \draw[mid arrow=0.65] ([xshift=0.6cm, yshift=-0.1cm]U1.0) node[below] {$(\mathrm{in},-a\mathbf{\hat{i}})$} -- ([yshift=-0.1cm]U1.0)  ;
        \draw[mid arrow=0.65] ([xshift=-0.6cm, yshift=0.1cm]U1.180) node[above] {$(\mathrm{in}, a\mathbf{\hat{i}})$} -- ([yshift=0.1cm]U1.180)  ;
        \draw[mid arrow=0.65] ([yshift=-0.1cm]U1.180)  -- ++(-0.6, 0)  node[below] {$(\mathrm{out},-a\mathbf{\hat{i}})$};
    \end{tikzpicture}
   \end{array}.
\end{equation}
We assume that each $(\mathrm{in},\mathbf{b})$ leg is paired with a unique $(\mathrm{out},\mathbf{b})$ leg carrying the same number of modes. In the matrix representation of $u$, rows and columns correspond to outgoing and incoming legs, respectively. When the orientation is clear from context, we label a leg simply by $\mathbf{b}$. For a two-dimensional square lattice, for example, the bond legs can be labeled as
\[
\begin{aligned}
    B &= (\mathbf{b}_{1}, \mathbf{b}_{2}, \mathbf{b}_{3}, \mathbf{b}_{4})\\
      &= (a\mathbf{\hat{i}}, -a\mathbf{\hat{i}}, a\mathbf{\hat{j}}, -a\mathbf{\hat{j}}),
\end{aligned}
\]
where $a$ is the lattice spacing, and $\mathbf{\hat{i}}$ and $\mathbf{\hat{j}}$ are unit vectors along the $x$- and $y$-axes, respectively.

Because the unitary network is translation invariant, it satisfies $[U_{\mathrm{Net}},T_i]=0$ for $i=x,y,z,\ldots$. The global single-particle unitary matrix therefore decomposes into independent momentum sectors:
\[
 U_{\mathrm{Net}}[u]
    =
    \bigoplus_{\mathbf{k}\in\mathrm{BZ}}
    U_{\mathrm{Net}}[u](\mathbf{k}).
\]
Using the plane-wave basis
$\psi_{\mathbf{k},\alpha}
=
\frac{1}{\sqrt{N}}
\sum_{\mathbf{x}\in\Gamma}
e^{i\mathbf{k}\cdot\mathbf{x}}
\phi_{\mathbf{x},\alpha}$,
the matrix elements of $U_{\mathrm{Net}}[u](\mathbf{k})$ are
\begin{equation}
\label{eq: enrty of Unet}
\begin{aligned}
U_{\mathrm{Net}}[u](\mathbf{k})_{\alpha\beta}
&=
\langle
\psi_{\mathbf{k},\alpha}
|
U_{\mathrm{Net}}[u]
|
\psi_{\mathbf{k},\beta}
\rangle
\\
&=
\frac{1}{N}
\sum_{\mathbf{x},\mathbf{y}\in\Gamma}
e^{i\mathbf{k}\cdot(\mathbf{y}-\mathbf{x})}
\langle
\phi_{\mathbf{x},\alpha}
|
U_{\mathrm{Net}}[u]
|
\phi_{\mathbf{y},\beta}
\rangle
\\
&=
\sum_{\mathbf{x}\in\Gamma}
e^{-i\mathbf{k}\cdot\mathbf{x}}
\langle
\phi_{\mathbf{x},\alpha}
|
U_{\mathrm{Net}}[u]
|
\phi_{\mathbf{0},\beta}
\rangle .
\end{aligned}
\end{equation}

Equation~\eqref{eq: enrty of Unet} admits a simple propagation interpretation. A state $\phi_{\mathbf{0},\beta}$ is injected into the network at the origin and propagates through successive applications of the local tensor $u$. Each time the state propagates along a bond vector $\mathbf{b}_i$, it acquires a phase factor $e^{-i\mathbf{k}\cdot\mathbf{b}_i}$. Summing over all possible numbers of internal scattering events therefore gives
\begin{equation}
\label{eq: k-space formula}
\begin{aligned}
   U_{\mathrm{Net}}[u](\mathbf{k})
   &= u_{PP} + u_{PB} \Phi_{B} \left[\sum_{m=0}^{\infty} (u_{BB}\Phi_{B})^{m} \right] u_{BP}
   \\
   &= u_{PP} +  u_{PB}\Phi_{B}(1- u_{BB}\Phi_{B})^{-1} u_{BP}
   \\
   &= u_{PP} +  u_{PB}(1- \Phi_{B}u_{BB})^{-1} \Phi_{B}u_{BP}
   \\
   &= \mathrm{str}_{B}[u \times_{B} \Phi_{B}( \mathbf{k})]
   \\
   &= \begin{array}{c}
   \begin{tikzpicture}
   \node[draw, rectangle, rounded corners=2pt, minimum width=0.6cm, minimum height=0.6cm, fill=tensorcolor, thick] (a) at (0,0) {$u$};
   \node[draw, rectangle, rounded corners=2pt, minimum width=0.6cm, minimum height=0.6cm, fill=tensorcolor, thick] (b) at (1.6,0) {$\Phi_{B}(\mathbf{k})$};
   \draw[mid arrow=0.65] (a.90) -- ++(0,0.6);
   \draw[mid arrow=0.65] ([yshift=-0.6cm]a.270) -- ++(0,0.6);
   \draw[mid arrow=0.65] (a.0) -- (b.180);
   \draw[mid arrow=0.5] (b.0) --  ++(0.3,0)-- ++(0,-0.5) -- ++(-3,0) -- ++(0,0.5) --(a.180);
\end{tikzpicture}
\end{array}.
\end{aligned}
\end{equation}
Here, $\Phi_{B}(\mathbf{k})$ is the diagonal phase matrix associated with propagation along the bond legs:
\[
\Phi_{B}(\mathbf{k})
   =
   \mathrm{diag}
   \left(
   e^{-i \mathbf{k} \cdot \mathbf{b}_{1}} I,
   e^{-i \mathbf{k} \cdot \mathbf{b}_{2}} I,
   e^{-i \mathbf{k} \cdot \mathbf{b}_{3}} I,
   e^{-i \mathbf{k} \cdot \mathbf{b}_{4}} I
   \right).
\]

\section{Particle-number-nonconserving case}
\label{sec: Without fermion number conserving}

For a generic fermionic Gaussian unitary, fermion number need not be conserved. In this case, the unitary induces a Bogoliubov transformation on the Majorana operators, represented at the single-particle level by an orthogonal matrix. This defines a projective representation
$\hat{\rho}: \mathrm{O}(2N) \rightarrow \mathrm{U}(2^N)$,
\begin{equation}
\begin{aligned}
  & \hat{\rho}(e^{K}) = \pm e^{\hat{K}},  \\
  & \hat{K} = \frac{1}{2} g_{ac}K^{c}_{b}\hat{\xi}^{a}\hat{\xi}^{b}.
\end{aligned}
\end{equation}
A more detailed discussion of fermionic Gaussian unitary operators is given in Appendix~\ref{Append: Fermionic Gaussian unitary operations}.

Accordingly, in a fermionic Gaussian unitary network, we represent each local unitary tensor by an orthogonal matrix $o \in \mathrm{O}(2N)$, whose rows and columns correspond to output and input Majorana modes, respectively. Since both the local tensors and the global transformation are represented at the single-particle level by orthogonal matrices, we denote the single-particle representation of the network $\hat{U}_{\mathrm{Net}}[\hat{u}]$ by $U_{\mathrm{Net}}[o]$, where $o$ is chosen such that $\hat{\rho}(o)=\pm\hat{u}$.

\subsection{Computation Rules}

The computational rules for fermionic Gaussian unitary networks without fermion-number conservation follow directly from those of the number-conserving case, with two modifications:
\begin{itemize}
    \item Each local tensor is represented by an orthogonal matrix $O \in \mathrm{O}(2N)$ rather than a unitary matrix $U \in \mathrm{U}(N)$.
    \item Each outgoing (incoming) leg must correspond to an even number of rows (columns) of the orthogonal matrix. This is because a Majorana mode carries half the degrees of freedom of a Dirac mode, so each tensor leg must contain an even number of Majorana modes.
\end{itemize}

Apart from these modifications, the contraction rules are identical to those in the fermion-number-conserving case. For example, the self-contraction of a fermionic Gaussian unitary operator is given by
\begin{equation}
\label{eq: self contraction2}
   \begin{array}{c}
    \begin{tikzpicture}
        \node[draw, rectangle, rounded corners=2pt, minimum width=1cm, minimum height=0.6cm, fill=tensorcolor, thick] (U1){$O$};
        \draw[mid arrow=0.55] ([xshift=0.3cm]U1.90) -- ++ (0,0.4) node[above] {$B$} -- ++(0.4,0) -- ++(0,-1.4) -- ++(-0.4,0) --([xshift=0.3cm]U1.270);
        \draw[mid arrow=0.65] ([xshift=-0.3cm]U1.90) -- ++ (0,0.4) node[above] {$A$} ;
        \draw[mid arrow=0.65] ([xshift=-0.3cm, yshift=-0.4cm]U1.270) -- ([xshift=-0.3cm]U1.270) ;
    \end{tikzpicture}
   \end{array} 
    = \mathrm{str}_{B} [O] \coloneqq O_{AA} + O_{AB}(I_{BB}-O_{BB})^{-1}O_{BA},
\end{equation}
A derivation of this self-contraction rule for fermionic Gaussian unitary operators without fermion-number conservation is given in Appendix \ref{Append: Self-contraction rule without fermion number conservation}.

\subsection{Momentum-Space Formulation of Translation-Invariant Networks}

Analogous to the fermion-number-conserving case, translational invariance allows the global unitary operator to be decomposed into independent momentum sectors, with each block acting on the mode subspace associated with a fixed crystal momentum $\mathbf{k}$:
\begin{equation}
\begin{aligned}
   U_{\text{Net}}[o](\mathbf{k}) &= \mathrm{str}_{B}[o \times_{B} \Phi_{B}( \mathbf{k})] \\
   &= \begin{array}{c}
    \begin{tikzpicture} 
    \node[draw, rectangle, rounded corners=2pt, minimum width=0.6cm, minimum height=0.6cm, fill=tensorcolor, thick] (a) at (0,0) {$o$};
    \node[draw, rectangle, rounded corners=2pt, minimum width=0.6cm, minimum height=0.6cm, fill=tensorcolor, thick] (b) at (1.6,0) {$\Phi_{B}(\mathbf{k})$};
    \draw[mid arrow=0.65] (a.90) -- ++(0,0.6);
    \draw[mid arrow=0.65] ([yshift=-0.6cm]a.270) -- ++(0,0.6);
    \draw[mid arrow=0.65] (a.0) -- (b.180);
    \draw[mid arrow=0.5] (b.0) --  ++(0.3,0)-- ++(0,-0.5) -- ++(-3,0) -- ++(0,0.5) --(a.180);
\end{tikzpicture}
\end{array}.
\end{aligned}
\end{equation}
where $\Phi_{B}$ denotes a diagonal matrix that is a function of $\mathbf{k}$:
\[
 \Phi_{B}(\mathbf{k}) =  \mathrm{diag}(e^{-i \mathbf{k} \cdot \mathbf{b}_{1}} I, e^{-i \mathbf{k} \cdot \mathbf{b}_{2}} I, e^{-i \mathbf{k} \cdot \mathbf{b}_{2}} I, e^{-i \mathbf{k} \cdot \mathbf{b}_{4}} I). 
\]
We remark that, in general, $U_{\mathrm{Net}}[o](\mathbf{k})$ is unitary rather than orthogonal. This distinction arises because the Fourier transform maps the real Majorana basis to a complex momentum-space basis. In particular,
\[
  \hat{q}_{k} = \frac{1}{\sqrt{N}} \sum_{x} e^{-ikx} \hat{q}_{x} ,
    \quad  \hat{p}_{k} = \frac{1}{\sqrt{N}} \sum_{x} e^{-ikx} \hat{p}_{x} ,
\]
are not Hermitian operators. Instead, they satisfy
\[
 \hat{q}_{k}^{\dagger} = \hat{q}_{-k}, \quad \hat{p}_{k}^{\dagger} = \hat{p}_{-k}.
\]
Consequently, orthogonality in real space becomes a constraint relating the $\mathbf{k}$ and $-\mathbf{k}$ sectors, rather than requiring each individual block $U_{\mathrm{Net}}[o](\mathbf{k})$ to be orthogonal.

\section{Variational optimization}

We use a fermionic Gaussian unitary network as a variational ansatz for a target fermionic Gaussian unitary transformation. The variational parameters are optimized by minimizing the discrepancy between the unitary represented by the network and the target transformation. In general, the local tensors may be chosen independently and optimized separately. Here, for concreteness, we restrict to translation-invariant unitary networks, for which all local tensors are identical. The variational parameters are therefore reduced to the matrix entries of a single unitary matrix $u$, corresponding to the uniform local tensor $\hat{u}$.

Translational invariance implies that the contracted unitary matrix decomposes into independent momentum sectors,
\[ 
U_{\mathrm{Net}}[u]
    =
    \bigoplus_{\mathbf{k} \in \mathrm{BZ}}
    U_{\mathrm{Net}}[u](\mathbf{k}).
\]
This decomposition naturally suggests a machine-learning formulation. We regard the unitary network as a parametrized map from crystal momentum $\mathbf{k}$ to a momentum-resolved unitary matrix $U_{\mathbf{k}}$, with the local unitary matrix $u$ serving as the trainable parameter:
\begin{equation}
    U_{\mathrm{Net}}[u]: \mathbf{k} \rightarrow U_{\mathbf{k}} = U_{\mathrm{Net}}[u](\mathbf{k}).
\end{equation}
Each $\mathbf{k}$ point in the Brillouin zone can then be treated as an independent data sample, with a corresponding loss evaluated separately. This viewpoint offers two practical advantages:
\begin{itemize}
    \item \textit{Optimization.} Since the loss decomposes over momentum points, the parameters can be optimized using stochastic or mini-batch gradient descent. At each update step, only a subset of $\mathbf{k}$ points needs to be evaluated, reducing the computational cost per iteration. The stochasticity of mini-batch updates can also improve exploration of the optimization landscape.

    \item \textit{Generalization.} Consider a family of systems described by the same momentum-space Hamiltonian $h(\mathbf{k})$ but with different linear sizes $L$. Smaller systems sample the Brillouin zone on a coarser momentum grid, whereas larger systems provide a finer sampling. Since the same local tensor $u$ defines $U_{\mathrm{Net}}[u](\mathbf{k})$ for all system sizes, the network can be trained using the momentum points of a smaller system and then evaluated on the denser momentum grid of a larger system. In this sense, the learned network performs an interpolation across momentum space.
\end{itemize}

\section{State preparation}

For a number-conserving quadratic fermion Hamiltonian
$\hat{H}=\sum_{ij} h_{ij}\hat{c}^{\dagger}_{i}\hat{c}_{j}$,
the ground state is a fermionic Gaussian state determined by the occupied single-particle eigenmodes of $h$. We therefore use a fermionic Gaussian unitary network as a variational ansatz for the ground state. Specifically, we parameterize the trial state as
\begin{equation}
\label{eq: UN state ansatz}
     \Psi_{\mathrm{var}} = \hat{U}_{\mathrm{Net}} \Phi_0,
\end{equation}
where $\Phi_0 = \bigotimes_g \Phi_{0,i}$ is a simple real-space product state and $\hat{U}_{\mathrm{Net}}$ is a variational fermionic Gaussian unitary network. The architecture is illustrated below for a one-dimensional system; the same construction extends straightforwardly to higher spatial dimensions:
\begin{align} \label{fig:1D_tensor_network_state}
    \begin{array}{c}
    \tikzsetnextfilename{1D_tensor_network_state}
    \begin{tikzpicture} 
    \node[draw, rectangle, rounded corners=2pt, minimum width=0.6cm, minimum height=0.6cm, fill=tensorcolor, thick] (a) at (0,0) {$\hat{u}$};
    \node[draw, rectangle, rounded corners=2pt, minimum width=0.6cm, minimum height=0.6cm, fill=tensorcolor, thick] (b) at (1.2,0) {$\hat{u}$};
    \node[draw, rectangle, rounded corners=2pt, minimum width=0.6cm, minimum height=0.6cm, fill=tensorcolor, thick] (c) at (2.4,0) {$\hat{u}$};
    \node[draw, rectangle, rounded corners=2pt, minimum width=0.6cm, minimum height=0.6cm, fill=tensorcolor, thick] (d) at (3.6,0) {$\hat{u}$};
    \node[draw, rectangle, rounded corners=2pt, minimum width=0.6cm, minimum height=0.6cm, fill=tensorcolor, thick] (a0) at (0,-1.2) {$\Phi_{0}$};
    \node[draw, rectangle, rounded corners=2pt, minimum width=0.6cm, minimum height=0.6cm, fill=tensorcolor, thick] (b0) at (1.2,-1.2) {$\Phi_{0}$};
    \node[draw, rectangle, rounded corners=2pt, minimum width=0.6cm, minimum height=0.6cm, fill=tensorcolor, thick] (c0) at (2.4,-1.2) {$\Phi_{0}$};
    \node[draw, rectangle, rounded corners=2pt, minimum width=0.6cm, minimum height=0.6cm, fill=tensorcolor, thick] (d0) at (3.6,-1.2) {$\Phi_{0}$};
    \draw[mid arrow=0.65] (a.90) -- ++(0,0.6);
    \draw[mid arrow=0.65] (a0.90) -- (a.270);
    \draw[mid arrow=0.65] (b.90) -- ++(0,0.6);
    \draw[mid arrow=0.65] (b0.90) -- (b.270);
    \draw[mid arrow=0.65] (c.90) -- ++(0,0.6);
    \draw[mid arrow=0.65] (c0.90) -- (c.270);
    \draw[mid arrow=0.65] (d.90) -- ++(0,0.6);
    \draw[mid arrow=0.65] (d0.90) -- (d.270);
    \draw[mid arrow=0.65] ([yshift= 2pt]a.0) -- ([yshift= 2pt]b.180);
    \draw[mid arrow=0.65] ([yshift=-2pt]b.180) -- ([yshift=-2pt]a.0);
    \draw[mid arrow=0.65] ([yshift= 2pt]b.0) -- ([yshift= 2pt]c.180);
    \draw[mid arrow=0.65] ([yshift=-2pt]c.180) -- ([yshift=-2pt]b.0);
    \draw[mid arrow=0.65] ([yshift= 2pt]c.0) -- ([yshift= 2pt]d.180);
    \draw[mid arrow=0.65] ([yshift=-2pt]d.180) -- ([yshift=-2pt]c.0);
    \draw[mid arrow=0.5]
        ([yshift= 2pt]d.0) -- ++(0.4,0)
        -- ++(0,-0.5) -- ++(-5.0,0)
        -- ++(0,0.5) -- ([yshift= 2pt]a.180);
    \draw[mid arrow=0.5]
        ([yshift=-2pt]a.180) -- ++(-0.3,0)
        -- ++(0,-0.5) -- ++(4.8,0)
        -- ++(0,0.5) -- ([yshift=-2pt]d.0);
\end{tikzpicture}
\end{array}.
\end{align}
The network acts as an entangling transformation that maps $\Phi_0$ to the variational fermionic Gaussian state $\Psi_{\mathrm{var}}$, whose parameters are optimized to approximate the target ground state $\Psi_{\mathrm{tar}}$.

Because both the target and variational states are fermionic Gaussian, the optimization can be performed entirely at the single-particle level. We represent the target state by its projector in each momentum sector $\mathbf{k}$:
\[
P_{\mathrm{tar, \mathbf{k}}}=\sum_{n\in \mathrm{occ}}
    |\phi_{n,\mathbf{k}}\rangle\langle \phi_{n,\mathbf{k}}| ,
\]
where $\{|\phi_{n,\mathbf{k}}\rangle\}_{n\in \mathrm{occ}}$ are the occupied single-particle eigenstates at momentum $\mathbf{k}$. Similarly, let $P_{0,\mathbf{k}}$ denote the single-particle projector of the reference state $\Phi_0$ in the momentum sector $\mathbf{k}$. The unitary-network ansatz transforms it into
$U_\mathbf{k}P_{0,\mathbf{k}}U_\mathbf{k}^{\dagger}$,
where $U_\mathbf{k}$ is the momentum-resolved single-particle transformation generated by $\hat{U}_{\mathrm{Net}}$. We optimize the local parameters of the unitary network by minimizing the momentum-resolved fidelity loss
\begin{equation}
\label{eq: fidelity loss}
   L_{\mathrm{fidelity}}[U_\mathbf{k}]
   =
   -\mathrm{Tr}\!\left[
   P_{\mathrm{tar},\mathbf{k}}\,
   U_\mathbf{k} P_{0,\mathbf{k}} U_\mathbf{k}^{\dagger}
   \right].
\end{equation}
Maximizing the overlap between the transformed and target projectors therefore provides a single-particle variational principle for preparing the target fermionic Gaussian ground state with the unitary-network ansatz.

\subsection{One-Layer Unitary Network}

\begin{figure}
    \centering
    \includegraphics[width=\linewidth]{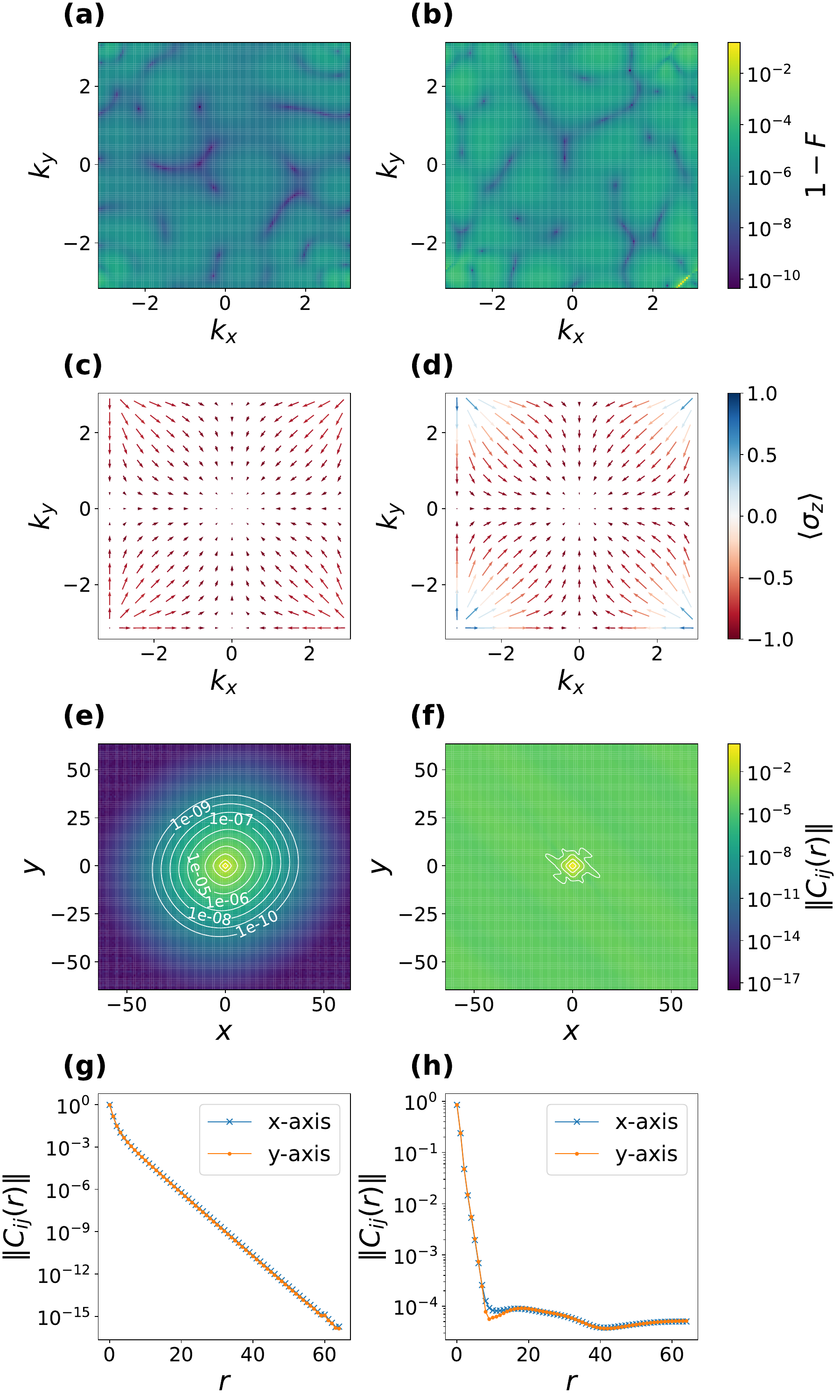}
    \caption{
    Tensor-network approximation of the ground state using entangling unitary network transformations.
    The unitary network maps a product state to an approximation of the ground state of the Qi--Wu--Zhang Hamiltonian $H_{\mathrm{QWZ}}$, comparing the trivial phase ($m=2.5$, left column) and the topological phase ($m=1.5$, right column).
    (a,b) Momentum-resolved projector infidelity $1-F$, shown on a shared logarithmic color scale.
    (c,d) Spin texture in the Brillouin zone: arrows represent the in-plane components $\langle \sigma_x \rangle$ and $\langle \sigma_y \rangle$, while the shared color scale indicates $\langle \sigma_z \rangle$.
    (e,f) Frobenius norm of the real-space correlation matrix $C_{\mathbf{r}}$, shown on a shared logarithmic color scale.
    (g,h) Correlation norms along the $x$ and $y$ axes, plotted on semilogarithmic axes to compare their spatial decay.
    }
    \label{fig:ground state}
\end{figure}

As a demonstration, we consider the Qi--Wu--Zhang model~\cite{Qi2006Topological}, whose Bloch Hamiltonian is
\begin{equation}
\label{eq:QWZ hamiltonian}
H_{\mathrm{QWZ}}(\mathbf{k})
=
\sin k_x \, \sigma_x
+
\sin k_y \, \sigma_y
+
\bigl(m+\cos k_x+\cos k_y\bigr)\sigma_z .
\end{equation}
The parameter $m$ controls the band topology: the system is topological for $0<|m|<2$ and trivial for $|m|>2$. We apply the fermionic Gaussian unitary-network ansatz in Eq.~\eqref{eq: UN state ansatz} to two representative values, $m=1.5$ and $m=2.5$, corresponding to the topological and trivial phases, respectively.

The numerical results are shown in Fig.~\ref{fig:ground state}. In both phases, the states generated by the unitary-network ansatz achieve high fidelity with the target ground states. For $m=1.5$, the spin texture wraps once around the Bloch sphere, consistent with a nonzero Chern number. However, despite accurately reproducing the target state in momentum space, the resulting tensor-network state does not exhibit exponentially decaying real-space correlations. This behavior is consistent with, and can be understood in light of, previous studies of fermionic Gaussian tensor-network representations of Chern insulators~\cite{Wahl2013Projected, Dubail2015Tensor}.

\subsection{Scale Invariant Layered Unitary Network for Metal}

For gapless fermionic systems with a Fermi surface---consisting of Fermi points in one dimension and an extended Fermi surface in higher dimensions---the entanglement entropy generally exhibits a logarithmic violation of the area law \cite{Pasquale_Calabrese_2004,Wolf2006Violation}. In contrast, fixed-bond-dimension tensor-network states, such
as MPS and PEPS, obey an area-law bound on entanglement
\cite{Verstraete_2006,Eisert2010Colloquium}. However, metallic
ground states with a finite Fermi surface exhibit  entanglement scaling  $S_A\sim L^{d-1}\log L$, violating the
area law \cite{Wolf_2006,Gioev_2006, Swingle_2010}. Consequently, fixed-bond-
dimension tensor networks cannot capture their asymptotic
entanglement scaling in the thermodynamic limit. To overcome this
limitation, we employ the \textit{zipper entanglement renormalization}
(ZER) scheme \cite{wong2022zipperentanglementrenormalizationfree},
which organizes entanglement renormalization into successive
layers of quasi-local disentangling unitaries. Our unitary-network
ansatz provides a natural representation of these
scale-dependent disentangling unitaries. We adopt the scale-invariant layered architecture illustrated below.
\begin{align}\label{fig:hierarchical_tensor_network}
\begin{array}{c}
\tikzsetnextfilename{hierarchical_tensor_network}
\begin{tikzpicture}[x=0.5cm,y=1cm]
\tikzset{
tensor/.style={draw,rectangle,rounded corners=2pt,minimum height=0.6cm,fill=tensorcolor,thick},
state/.style={draw,rectangle,minimum width=0.9cm,minimum height=0.5cm,thick},
bridge/.style={preaction={draw=white,line width=2.6pt},mid arrow=0.65},
pbc/.style={rounded corners=2pt,mid arrow=0.55}
}
\foreach \i in {0,2,...,14}{
\coordinate (p\i) at (\i,0);
}
\node[tensor,minimum width=1.5cm] (u1) at (1,0) {$\hat{u}$};
\node[tensor,minimum width=1.5cm] (u2) at (5,0) {$\hat{u}$};
\node[tensor,minimum width=1.5cm] (u3) at (9,0) {$\hat{u}$};
\node[tensor,minimum width=1.5cm] (u4) at (13,0) {$\hat{u}$};
\node[tensor,minimum width=2.8cm,minimum height=0.7cm] (v1) at (2,-2) {$\hat{u}$};
\node[tensor,minimum width=2.8cm,minimum height=0.7cm] (v2) at (10,-2) {$\hat{u}$};
\node[tensor,minimum width=5.2cm,minimum height=0.75cm] (w) at (4,-4.2) {$\hat{u}$};
\node[state] (a1) at ([yshift=-1.05cm]p2) {$\Phi_0$};
\node[state] (a2) at ([yshift=-1.05cm]p6) {$\Phi_0$};
\node[state] (a3) at ([yshift=-1.05cm]p10) {$\Phi_0$};
\node[state] (a4) at ([yshift=-1.05cm]p14) {$\Phi_0$};
\node[state] (a5) at ([yshift=-3.10cm]p4) {$\Phi_0$};
\node[state] (a6) at ([yshift=-3.10cm]p12) {$\Phi_0$};
\node[state,minimum width=1.8cm,minimum height=0.55cm] (pk) at ([yshift=-5.45cm]p0) {$\Psi_{\bm{k}=0}$};
\node[state,minimum height=0.55cm] (a7) at ([yshift=-5.45cm]p8) {$\Phi_0$};
\draw[mid arrow=0.65] ([yshift=0.1cm]u1.east) -- ([yshift=0.1cm]u2.west);
\draw[mid arrow=0.65] ([yshift=-0.1cm]u2.west) -- ([yshift=-0.1cm]u1.east);
\draw[mid arrow=0.65] ([yshift=0.1cm]u2.east) -- ([yshift=0.1cm]u3.west);
\draw[mid arrow=0.65] ([yshift=-0.1cm]u3.west) -- ([yshift=-0.1cm]u2.east);
\draw[mid arrow=0.65] ([yshift=0.1cm]u3.east) -- ([yshift=0.1cm]u4.west);
\draw[mid arrow=0.65] ([yshift=-0.1cm]u4.west) -- ([yshift=-0.1cm]u3.east);
\draw[mid arrow=0.65] ([yshift=0.1cm]v1.east) -- ([yshift=0.1cm]v2.west);
\draw[mid arrow=0.65] ([yshift=-0.1cm]v2.west) -- ([yshift=-0.1cm]v1.east);
\coordinate (uLtop) at ([xshift=-0.28cm,yshift=0.1cm]u1.west);
\coordinate (uLbot) at ([xshift=-0.42cm,yshift=-0.1cm]u1.west);
\coordinate (uRtop) at ([xshift=0.28cm,yshift=0.1cm]u4.east);
\coordinate (uRbot) at ([xshift=0.42cm,yshift=-0.1cm]u4.east);
\coordinate (uLOutTop) at ([xshift=-0.08cm]uLtop);
\coordinate (uROutTop) at ([xshift=0.08cm]uRtop);
\coordinate (uLOutBot) at ([xshift=-0.08cm]uLbot);
\coordinate (uROutBot) at ([xshift=0.08cm]uRbot);
\coordinate (uHigh) at ([yshift=-0.18cm]u1.south);
\coordinate (uLow) at ([yshift=-0.38cm]u1.south);
\coordinate (vLtop) at ([xshift=-0.28cm,yshift=0.1cm]v1.west);
\coordinate (vLbot) at ([xshift=-0.42cm,yshift=-0.1cm]v1.west);
\coordinate (vRtop) at ([xshift=0.28cm,yshift=0.1cm]v2.east);
\coordinate (vRbot) at ([xshift=0.42cm,yshift=-0.1cm]v2.east);
\coordinate (vLOutTop) at ([xshift=-0.08cm]vLtop);
\coordinate (vROutTop) at ([xshift=0.08cm]vRtop);
\coordinate (vLOutBot) at ([xshift=-0.08cm]vLbot);
\coordinate (vROutBot) at ([xshift=0.08cm]vRbot);
\coordinate (vHigh) at ([yshift=-0.18cm]v1.south);
\coordinate (vLow) at ([yshift=-0.38cm]v1.south);
\coordinate (wLtop) at ([xshift=-0.28cm,yshift=0.1cm]w.west);
\coordinate (wLbot) at ([xshift=-0.42cm,yshift=-0.1cm]w.west);
\coordinate (wRtop) at ([xshift=0.28cm,yshift=0.1cm]w.east);
\coordinate (wRbot) at ([xshift=0.42cm,yshift=-0.1cm]w.east);
\coordinate (wLOutTop) at ([xshift=-0.08cm]wLtop);
\coordinate (wROutTop) at ([xshift=0.08cm]wRtop);
\coordinate (wLOutBot) at ([xshift=-0.08cm]wLbot);
\coordinate (wROutBot) at ([xshift=0.08cm]wRbot);
\coordinate (wHigh) at ([yshift=-0.18cm]w.south);
\coordinate (wLow) at ([yshift=-0.38cm]w.south);
\draw[mid arrow=0.65] ([yshift=0.1cm]u4.east) -- (uRtop);
\draw[pbc] (uRtop) -- (uROutTop) -- (uROutTop |- uHigh) -- (uLOutTop |- uHigh) -- (uLOutTop) -- (uLtop);
\draw[mid arrow=0.65] (uLtop) -- ([yshift=0.1cm]u1.west);
\draw[mid arrow=0.65] ([yshift=-0.1cm]u1.west) -- (uLbot);
\draw[pbc] (uLbot) -- (uLOutBot) -- (uLOutBot |- uLow) -- (uROutBot |- uLow) -- (uROutBot) -- (uRbot);
\draw[mid arrow=0.65] (uRbot) -- ([yshift=-0.1cm]u4.east);
\draw[mid arrow=0.65] ([yshift=0.1cm]v2.east) -- (vRtop);
\draw[pbc] (vRtop) -- (vROutTop) -- (vROutTop |- vHigh) -- (vLOutTop |- vHigh) -- (vLOutTop) -- (vLtop);
\draw[mid arrow=0.65] (vLtop) -- ([yshift=0.1cm]v1.west);
\draw[mid arrow=0.65] ([yshift=-0.1cm]v1.west) -- (vLbot);
\draw[pbc] (vLbot) -- (vLOutBot) -- (vLOutBot |- vLow) -- (vROutBot |- vLow) -- (vROutBot) -- (vRbot);
\draw[mid arrow=0.65] (vRbot) -- ([yshift=-0.1cm]v2.east);
\draw[mid arrow=0.65] ([yshift=0.1cm]w.east) -- (wRtop);
\draw[pbc] (wRtop) -- (wROutTop) -- (wROutTop |- wHigh) -- (wLOutTop |- wHigh) -- (wLOutTop) -- (wLtop);
\draw[mid arrow=0.65] (wLtop) -- ([yshift=0.1cm]w.west);
\draw[mid arrow=0.65] ([yshift=-0.1cm]w.west) -- (wLbot);
\draw[pbc] (wLbot) -- (wLOutBot) -- (wLOutBot |- wLow) -- (wROutBot |- wLow) -- (wROutBot) -- (wRbot);
\draw[mid arrow=0.65] (wRbot) -- ([yshift=-0.1cm]w.east);
\draw[mid arrow=0.65] (p0 |- u1.north) -- ++(0,0.65);
\draw[mid arrow=0.65] (p2 |- u1.north) -- ++(0,0.65);
\draw[mid arrow=0.65] (p4 |- u2.north) -- ++(0,0.65);
\draw[mid arrow=0.65] (p6 |- u2.north) -- ++(0,0.65);
\draw[mid arrow=0.65] (p8 |- u3.north) -- ++(0,0.65);
\draw[mid arrow=0.65] (p10 |- u3.north) -- ++(0,0.65);
\draw[mid arrow=0.65] (p12 |- u4.north) -- ++(0,0.65);
\draw[mid arrow=0.65] (p14 |- u4.north) -- ++(0,0.65);
\draw[bridge] (p0 |- v1.north) -- (p0 |- u1.south);
\draw[bridge] (p4 |- v1.north) -- (p4 |- u2.south);
\draw[bridge] (p8 |- v2.north) -- (p8 |- u3.south);
\draw[bridge] (p12 |- v2.north) -- (p12 |- u4.south);
\draw[bridge] (a1.north) -- (p2 |- u1.south);
\draw[bridge] (a2.north) -- (p6 |- u2.south);
\draw[bridge] (a3.north) -- (p10 |- u3.south);
\draw[bridge] (a4.north) -- (p14 |- u4.south);
\draw[bridge] (p0 |- w.north) -- (p0 |- v1.south);
\draw[bridge] (p8 |- w.north) -- (p8 |- v2.south);
\draw[bridge] (a5.north) -- (p4 |- v1.south);
\draw[bridge] (a6.north) -- (p12 |- v2.south);
\draw[bridge] (pk.north) -- (p0 |- w.south);
\draw[bridge] (a7.north) -- (p8 |- w.south);
\end{tikzpicture}
\end{array}.
\end{align}
We initialize the construction from a single-site state,
$\Psi(\bm{k}=\bm{0})=\Psi(\bm{r}=\bm{0})$. At each layer, additional product-state degrees of freedom $\Phi_{0}$, consisting of fully localized empty and occupied modes, are introduced and entangled with the existing state by the unitary network $\hat{U}_{\mathrm{Net}}[\hat{u}]$. 

At each layer, we apply a translation-invariant unitary network constructed from the same local tensor $\hat{u}$. As follows from Eq.~\eqref{eq: k-space formula}, its action is described by the same momentum-dependent unitary function at every layer. We therefore optimize $u$ by minimizing the fidelity loss in Eq.~\eqref{eq: fidelity loss} so that the network realizes the transformation
\begin{equation}
\begin{aligned}
 P_{\mathrm{tar}}(\bm{k}) \oplus P_{\mathrm{prod}}(\bm{k}) &\overset{U_{\mathrm{Net}}[u](\mathbf{k})}{\longrightarrow}
   \mathcal{F}^{-1}_{\mathrm{block}}
   \left[
   \bigoplus_{\alpha=1}^{2}
   P_{\mathrm{tar}}(\bm{k}/2+{G}_{\alpha})
   \right],
   \\
   \text{for all } \bm{k}
   &\in [0,2\pi).
\end{aligned}
\end{equation}
Here, $\mathcal{F}^{-1}_{\mathrm{block}}$ denotes the inverse Fourier transformation from the folded momentum basis to the real-space basis of the two sites within each block. The shifts $G_{1}=0$ and $G_{2}=\pi$ label the two fine-grid momentum sectors, $\bm{k}/2$ and $\bm{k}/2+\pi$, that fold onto the same coarse momentum $\bm{k}$ under factor-two blocking. Thus, each layer combines the coarse-grid state with localized product modes and maps it to the corresponding state on a refined momentum grid.

Repeated application of the same scale-independent transformation generates approximations to the target free-fermion state at progressively larger system sizes while preserving the same momentum-space structure. After the $\tau$-th layer, we require the resulting Gaussian state to approximate the target projector
\begin{equation}
    P_{\tau} = \bigoplus_{\bm{k} \in \mathrm{BZ}_{\tau}} P_{\mathrm{tar}}(\bm{k}),
\end{equation}
where
\[
  \mathrm{BZ}_{0} \subset \mathrm{BZ}_{1} \subset \cdots \subset \mathrm{BZ}_{T}
\]
is a sequence of progressively denser dimensionless momentum grids. In one dimension, these grids sample the interval $[0,2\pi)$. Thus, each additional layer increases the system size and refines the momentum-space resolution while sampling the same target projector $P_{\mathrm{tar}}(\bm{k})$.

We apply the above ansatz to the half-filled one-dimensional nearest-neighbor free-fermion model,
\[
H(k) = -\cos(k).
\]
To construct a translation-invariant unitary network, we group
pairs of physical sites into blocks and choose the local tensor $u$ to act on four physical sites. Accordingly, $P_{\mathrm{tar}}(k)$ denotes the target projector in the blocked representation. The auxiliary product state
$\Phi_0$ contains one occupied and one empty mode per block.
To preserve time-reversal symmetry, we keep both degenerate
modes at the Fermi surface $k=\pm\pi/2$ empty. In the reconstruction, we shift the Brillouin-zone convention so that the two Fermi points fold onto $k=0$ in the blocked representation, and input to the network with the fully
empty initial state $\Psi(\bm{k}=\bm{0})$. After applying all layers
of the unitary network, we restore the original
Brillouin-zone convention.

\begin{figure}
    \centering
    \includegraphics[width=1\linewidth]{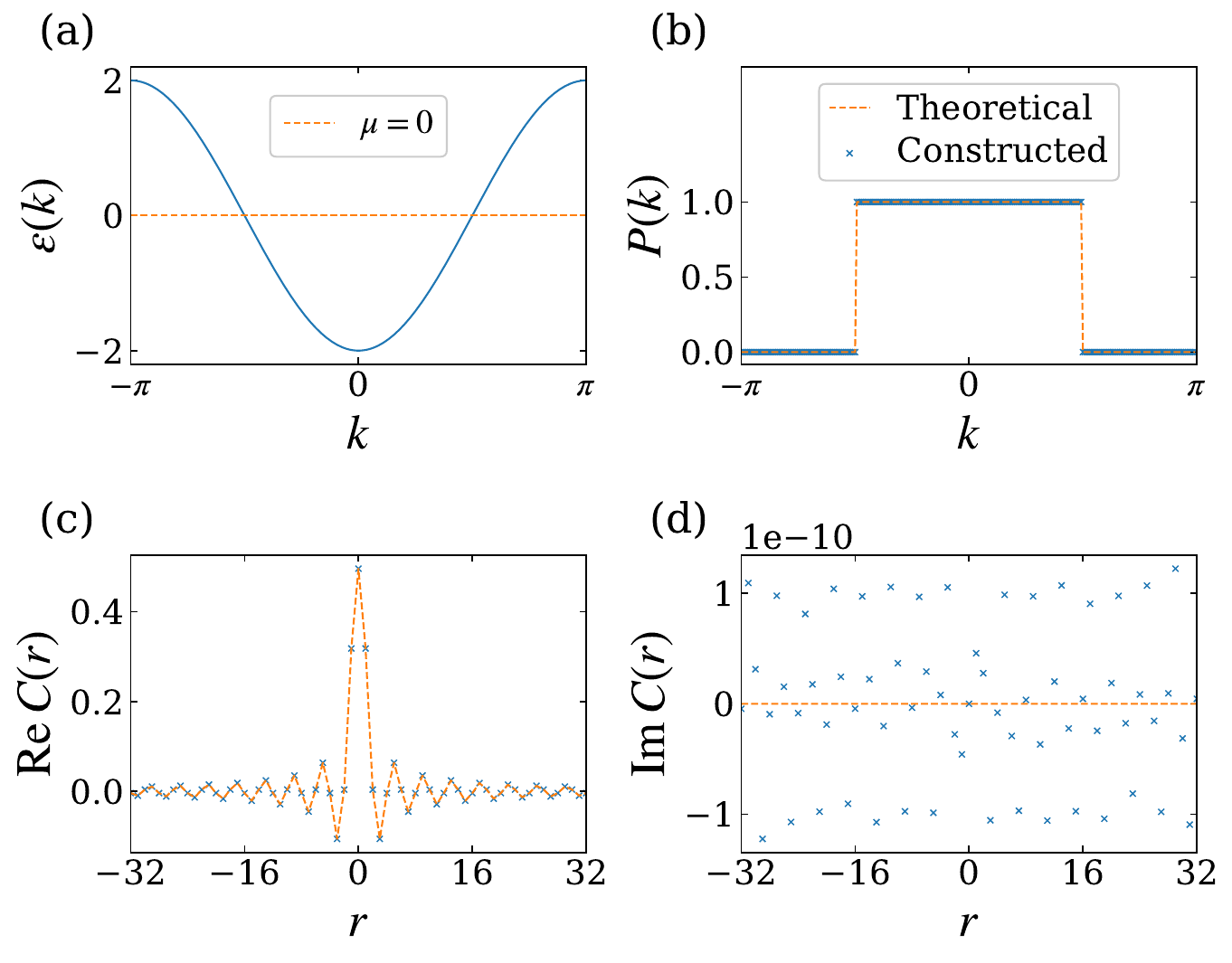}
    \caption{
Ground-state reconstruction of a half-filled one-dimensional metal using the fermionic Gaussian unitary-network ansatz. 
(a) Single-particle dispersion $\epsilon(k)$, with the dashed line indicating the Fermi energy $\mu=0$. 
(b) Momentum distribution $P(k)=\langle c_k^\dagger c_k\rangle$ of the exact ground state and the state generated by the unitary-network ansatz. 
(c,d) Real and imaginary parts of the equal-time correlation
function, $C(r)=\langle c_0^\dagger c_r\rangle
=L_x^{-1}\sum_k e^{ikr}P(k)$, respectively. Time-reversal
symmetry requires $C(r)$ to be real, so its imaginary part shall
vanish. The agreement with the exact results demonstrates
that the unitary-network ansatz accurately captures both the
Fermi-surface structure and the long-range correlations of
the metallic ground state.
}
\label{fig:metal}
\end{figure}

The numerical results are shown in Fig.~\ref{fig:metal}. The unitary-network ansatz accurately reproduces both the Fermi points in momentum space and the long-range correlations in real space.

The scale-invariant layered unitary-network ansatz introduced 
above extends naturally to higher dimensions and can describe 
both metallic and gapped states. Examples of these extensions are presented in Ref.~\cite{chan2026cherninsulatorsfixedpoints}. However, reproducing a finite, nonzero correlation length is sensitive to small approximation 
errors and remains challenging.

\section{Time evolution}

Another immediate application , we use fermionic Gaussian unitary networks to approximate real-time evolution under quadratic fermionic Hamiltonians $\hat{U}_0(\delta t)=e^{-i\hat{H}_0\delta t}$.

\subsection{Free-fermion time evolution}

We first consider free-fermion time evolution and examine the
capabilities and limitations of unitary networks in representing
it.

For a generic dispersive Hamiltonian, a quantum quench from a low-entanglement initial state generates quasi-particle
excitations with nonzero group velocities
\[
\mathbf{v}_{\alpha}(\mathbf{k})
    = \nabla_{\mathbf{k}}\varepsilon_{\alpha}(\mathbf{k}),
\]
leading to ballistic correlation spreading and, generically,
linear entanglement growth
\cite{Calabrese_2005, Fagotti2008Evolution, Alba_2017}.
A tensor network with fixed bond dimension $\chi$
can support at most
$S_A \lesssim |\partial A|\log\chi$
for a bipartition with $O(|\partial A|)$ bonds crossing its
boundary. Consequently, a fixed-bond-
dimension unitary network is naturally suited
to short-time evolution but generally fails to capture
long-time dynamics as entanglement grows.

The situation is different for flat-band Hamiltonians, for
which $\varepsilon_{\alpha}(\mathbf{k})=\mathrm{const.}$ and
hence
\[
\mathbf{v}_{\alpha}(\mathbf{k})
=\nabla_{\mathbf{k}}\varepsilon_{\alpha}(\mathbf{k})=0.
\]
The usual dispersive quasiparticle mechanism for ballistic
propagation and linear entanglement growth is therefore absent,
although nontrivial real-space correlations can still develop.
Moreover, when the flat-band energies are commensurate,
the time evolution is periodic (up to a global phase).
These properties make flat-band dynamics particularly
suitable for approximation by finite-bond-dimension
fermionic Gaussian unitary networks.

\begin{figure}
    \centering
    \includegraphics[width=1\linewidth]{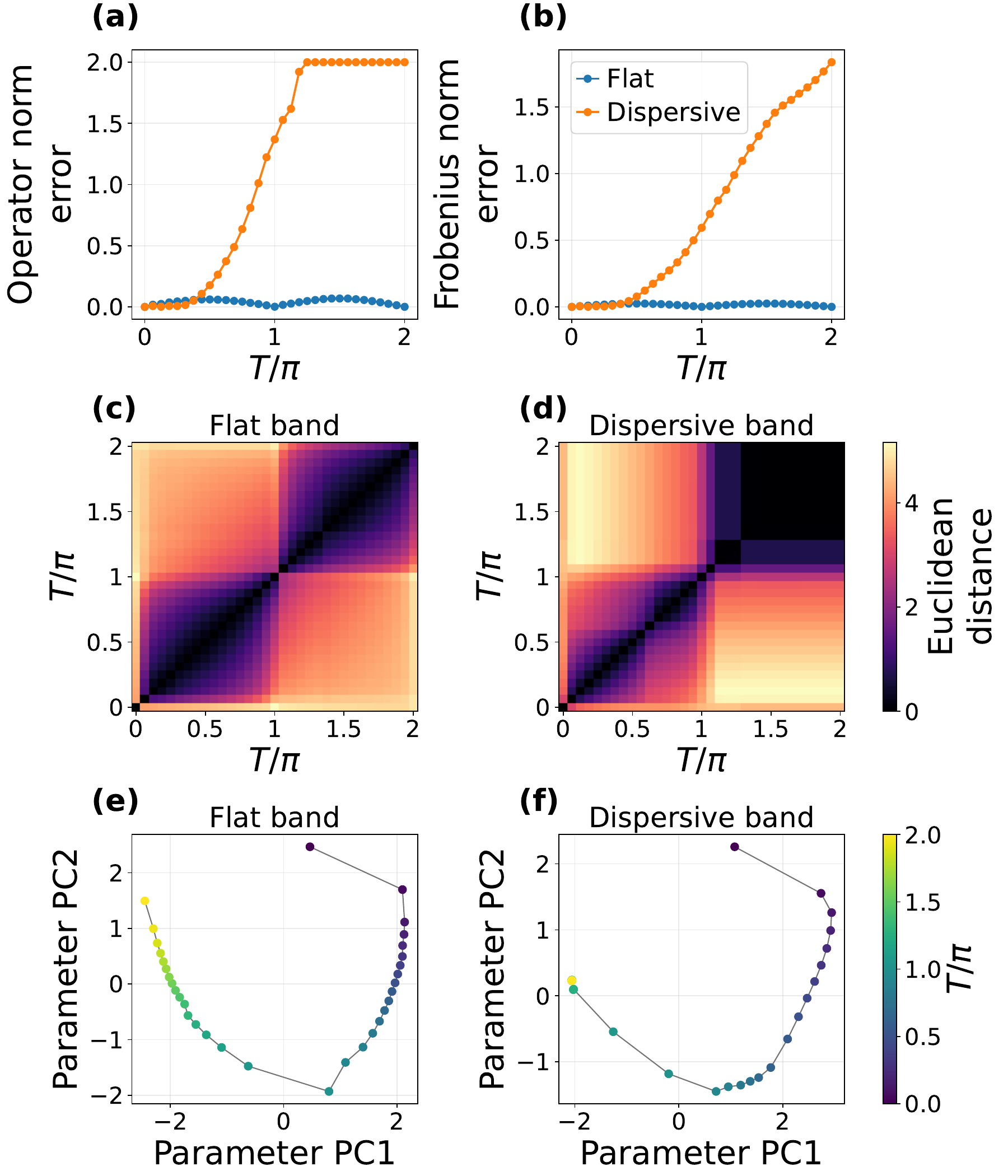}
    \caption{Unitary-network approximation of real-time evolution.
    We compare the dispersive Qi--Wu--Zhang Hamiltonian
    $H_{\mathrm{QWZ}}$ ($m=1.5$) with the flat-band Hamiltonian
    $H_{\mathrm{flat}}=I-2P_{\mathrm{QWZ}}$.
    (a,b) Maximum operator-norm and mean Frobenius-norm errors,
    respectively, over lattice momenta as functions of $T/\pi$.
    (c,d) Pairwise Euclidean distances between local-unitary
    parameter vectors for flat and dispersive bands, respectively,
    using a shared color scale. Parameters at adjacent times
    generally remain close. For flat-band evolution, deviations
    near $T=0$ and $T=\pi$ arise because the evolution is
    proportional to the identity, allowing physical and bond
    degrees of freedom to decouple and redundant bond parameters
    to represent the same physical operation.
    (e,f) Parameter trajectories for flat and dispersive bands,
    respectively, projected onto the first two principal
    components, with color indicating $T/\pi$.}
    \label{fig:time_evo}
\end{figure}

Fig.~\ref{fig:time_evo} (a,b) compares unitary-network
approximations of dispersive and flat-band time evolution.
For the dispersive case, we use the Qi--Wu--Zhang Hamiltonian
$H_{\mathrm{QWZ}}$ in the topological phase with $m=1.5$,
defined in Eq.~\eqref{eq:QWZ hamiltonian}. We obtain the
corresponding flat-band Hamiltonian by spectrally flattening
the same model:
\[
H_{\mathrm{flat}}=I-2P_{\mathrm{QWZ}},
\]
where $P_{\mathrm{QWZ}}$ is the occupied-band projector.
We evaluate the approximation errors using the operator and
Frobenius norms over lattice momenta. As shown in
Figs.~\ref{fig:time_evo}(a) and (b), the error for the
dispersive Hamiltonian increases with time, whereas the
flat-band evolution remains accurately represented
throughout the full period $T=2\pi$.

The unitary-network ansatz can also accurately represent
dynamics restricted to a flat-band subspace, even when
other bands are dispersive. We therefore consider the
projected time-evolution operator
\[
P_{\mathrm{flat}}e^{-iHt}P_{\mathrm{flat}},
\]
which is unitary within the flat-band subspace.

\begin{figure}
    \centering
    \includegraphics[width=1\linewidth]{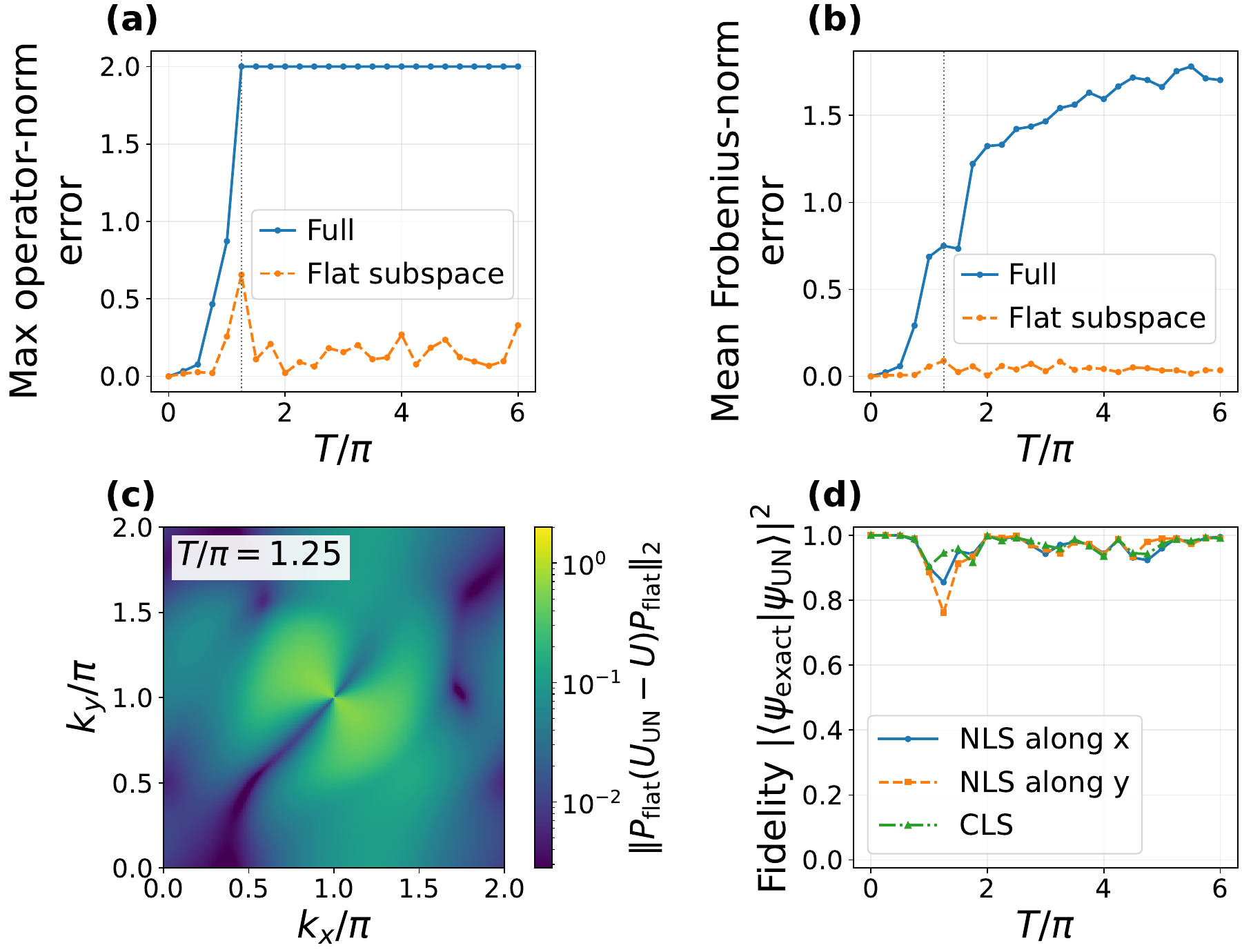}
    \caption{Unitary-network approximation of real-time evolution
on the Lieb lattice.
(a) Maximum operator-norm errors of the full evolution,
$D_{\mathrm{full}}=U_{\mathrm{UN}}-U$, and its projection onto
the flat-band subspace,
$D_{\mathrm{flat}}=P_{\mathrm{flat}}D_{\mathrm{full}}P_{\mathrm{flat}}$,
as functions of time. At $M=(\pi,\pi)$, all three bands are degenerate and
$P_{\mathrm{flat}}$ is chosen as the projector onto the two rim orbitals.
(b) Mean Frobenius-norm errors of the same operators.
The error remains small within the flat-band subspace.
(c) Momentum-resolved operator-norm error of
$D_{\mathrm{flat}}$ at $T=5\pi/4$, marked by the vertical
dotted lines in (a,b), on a logarithmic color scale.
The maximum error, $0.656$, occurs at $M=(\pi,\pi)$.
(d) State fidelities between unitary network approximation and exact evolution for a compact localized state (CLS, green dash-dotted) and two noncontractible loop states extending along the $x$ (blue solid) and $y$ (orange dashed) directions.
}
    \label{fig:Lieb}
\end{figure}

As an example, we consider the Lieb
lattice \cite{Lieb1989Two}, which hosts one flat band and two dispersive bands
that become degenerate at $\bm{k}=(\pi,\pi)$. We use a
fermionic Gaussian unitary network to approximate its time
evolution. As shown in Fig.~\ref{fig:Lieb}, although the error in the full
evolution increases with time, the error projected onto the
flat-band subspace remains small, except in the vicinity of the
three-band touching point. This result suggests that the unitary
network can capture flat-band dynamics in systems with
interesting quantum geometry
\cite{Julku_2016}, including singular flat bands \cite{Rhim_2019,Rhim01012021}.

The continuity of time evolution can be exploited to improve
the efficiency of unitary-network optimization. Since
$U(t+\Delta t)=e^{-iH\Delta t}U(t)$, the target unitaries
at nearby times are close. We therefore expect their optimal
network parameters to vary smoothly, provided the optimization
follows the same solution branch. Rather than optimizing each
time independently, we adopt a time-continuation scheme:
the optimized parameters at time $t$ are used to initialize
the optimization at $t+\Delta t$. This warm-start strategy
exploits temporal correlations between successive solutions
and typically accelerates convergence compared with random
initialization.

We parameterize each local unitary as $u=e^{ih}$, with $h$
Hermitian, and quantify the distance between two parameter
sets by the Frobenius norm
\[
    \|h_1-h_2\|_F
    =\sqrt{\sum_{i,j}|(h_1-h_2)_{ij}|^2}.
\]
Fig.~\ref{fig:time_evo} (c,d) show that optimized
parameters at neighboring times are generally close,
supporting the time-continuation scheme. However, large
parameter variations occur near $t=0$ and $t=\pi$ for the
flat-band Hamiltonian, where the evolution is proportional
to the identity.

This apparent discontinuity originates from the
nonuniqueness of the local-tensor representation. When the
physical and bond degrees of freedom decouple, a local tensor
can take the block-diagonal form
\[
h=
\begin{pmatrix}
h_{\mathrm{phys}} & 0\\
0 & h_{\mathrm{bond}}
\end{pmatrix},
\qquad
u=e^{ih}=
\begin{pmatrix}
u_{\mathrm{phys}} & 0\\
0 & u_{\mathrm{bond}}
\end{pmatrix}.
\]
In this case, $u_{\mathrm{bond}}$ can vary without affecting
the physical operation represented by the network. Thus,
the same global unitary admits a continuous family of local
parameterizations. Near these highly redundant points, the
optimization can move substantially in parameter space while
leaving the represented physical operation nearly unchanged.
This behavior is also reflected in the optimization
trajectories in Figs.~\ref{fig:time_evo} (e,f).

\subsection{With on-site interaction}

Although fermionic Gaussian unitary networks represent only
fermionic Gaussian unitaries, they can serve as building blocks
for tensor-network representations of interacting many-body
dynamics. Consider a Hamiltonian
$\hat{H}=\hat{H}_0+\hat{V}$, where $\hat{H}_0$ is quadratic
and $\hat{V}=\sum_i v_i$ consists of on-site interactions.
The time-evolution operator over time $T$ can be approximated
using a second-order Trotter decomposition \cite{strang1968construction, suzuki1976generalized} with
$N$ steps of size $\delta t=T/N$:
\begin{equation}
\begin{aligned}
\hat{U}(T)
&=e^{-i\hat{H}T}\\
&\approx
\left[
e^{-i\hat{V}\delta t/2}
e^{-i\hat{H}_0\delta t}
e^{-i\hat{V}\delta t/2}
\right]^N.
\end{aligned}
\end{equation}
The quadratic evolution
$\hat{U}_0(\delta t)=e^{-i\hat{H}_0\delta t}$
is a fermionic Gaussian unitary and can thus be represented
by a fermionic Gaussian unitary network. Since $\hat{V}$
contains only on-site interactions, its evolution $e^{-i\hat{V}\delta t/2}$ factorizes
into local many-body gates, each represented by a tensor with
bond dimension one. Combining these two components yields a
tensor-network representation of the full interacting
dynamics.

After optimizing the single-particle matrix $u_0$ of the local
Gaussian tensor, its Fock-space representation
$\hat{u}_0=\hat{\rho}(u_0)$ can be constructed explicitly.
In the number-conserving case, its matrix elements are given
by the determinant formula
\cite{rajabpour2025matrixelementsfermionicgaussian}:
\begin{equation}
    \hat{\rho}(u_0)_{MN}
    = \det u_0[M,N],
\end{equation}
where $M$ and $N$ denote the occupied output and input modes,
respectively, with $|M|=|N|$, and $u_0[M,N]$ is the
corresponding submatrix. This allows the optimized Gaussian
tensor to be converted into a many-body tensor and combined
with local interaction gates.

Since $\hat{V}$ consists of on-site terms, its evolution
factorizes into a product of local unitaries:
\begin{equation}
    \hat{U}_I(\delta t/2)
    = e^{-i\hat{V}\delta t/2}
    = \bigotimes_i e^{-i\hat{v}^{(i)}\delta t/2}
    \equiv \bigotimes_i u_I^{(i)},
\end{equation}
where $i$ labels the lattice sites. This product operator
has bond dimension one. Consequently, a symmetric Trotter
step is represented by composing the Gaussian unitary
network with the local interaction layers:
\begin{equation}
    \hat{U}(\delta t)
    \approx \hat{U}_I(\delta t/2) \hat{U}_0(\delta t) \hat{U}_I(\delta t/2),
\end{equation}
\begin{equation}
   \begin{array}{c}
    \begin{tikzpicture}
        \node[draw, rectangle, rounded corners=2pt, minimum width=0.6cm, minimum height=0.6cm, fill=tensorcolor, thick] (U1){$\hat{u}$};
        \draw[mid arrow=0.65] (U1.90) -- ++ (0,0.6) ;
        \draw[mid arrow=0.65] ([yshift=-0.6cm]U1.270) -- (U1.270);
        \draw[mid arrow=0.65] ([yshift=0.1cm]U1.0)  -- ++(0.6,0)  ;
        \draw[mid arrow=0.65] ([xshift=0.6cm, yshift=-0.1cm]U1.0)  -- ([yshift=-0.1cm]U1.0)  ;
        \draw[mid arrow=0.65] ([xshift=-0.6cm, yshift=0.1cm]U1.180) -- ([yshift=0.1cm]U1.180)  ;
        \draw[mid arrow=0.65] ([yshift=-0.1cm]U1.180)  -- ++(-0.6, 0) ;
    \end{tikzpicture}
   \end{array} = 
   \begin{array}{c}
    \begin{tikzpicture}
        \node[draw, rectangle, rounded corners=2pt, minimum width=0.6cm, minimum height=0.6cm, fill=tensorcolor, thick] (U1) at (0,0) {$\hat{u}_{0}$};
         \node[draw, rectangle, rounded corners=2pt, minimum width=0.6cm, minimum height=0.6cm, fill=tensorcolor, thick] (U2) at (0,1.0) {$\hat{u}_{I}$};
         \node[draw, rectangle, rounded corners=2pt, minimum width=0.6cm, minimum height=0.6cm, fill=tensorcolor, thick] (U0) at (0,-1.0) {$\hat{u}_{I}$};
        \draw[mid arrow=0.65] (U1.90) -- (U2.270) ;
        \draw[mid arrow=0.65] ([yshift=-0.4cm]U1.270) -- (U1.270);
        \draw[mid arrow=0.65] ([yshift=0.1cm]U1.0)  -- ++(0.6,0)  ;
        \draw[mid arrow=0.65] ([xshift=0.6cm, yshift=-0.1cm]U1.0)  -- ([yshift=-0.1cm]U1.0)  ;
        \draw[mid arrow=0.65] ([xshift=-0.6cm, yshift=0.1cm]U1.180) -- ([yshift=0.1cm]U1.180)  ;
        \draw[mid arrow=0.65] ([yshift=-0.1cm]U1.180)  -- ++(-0.6, 0) ;
        \draw[mid arrow=0.65] (U2.90) -- ++(0,0.4) ;
         \draw[mid arrow=0.65] (U0.90) -- (U1.270) ;
        \draw[mid arrow=0.65] ([yshift=-0.4cm]U0.270) -- (U0.270) ;
    \end{tikzpicture}
   \end{array}.
\end{equation}
We note that, because $\hat{u}$ is a fermionic tensor, permuting its legs or contracting them can introduce additional fermionic sign factors. The contraction rules therefore differ from their bosonic counterparts and must account for the $\mathbb{Z}_2$ grading. The corresponding sign conventions are described in Refs.~\cite{Barthel_2009, Mortier_2025}.

\begin{figure}
    \centering
    \includegraphics[width=1\linewidth]{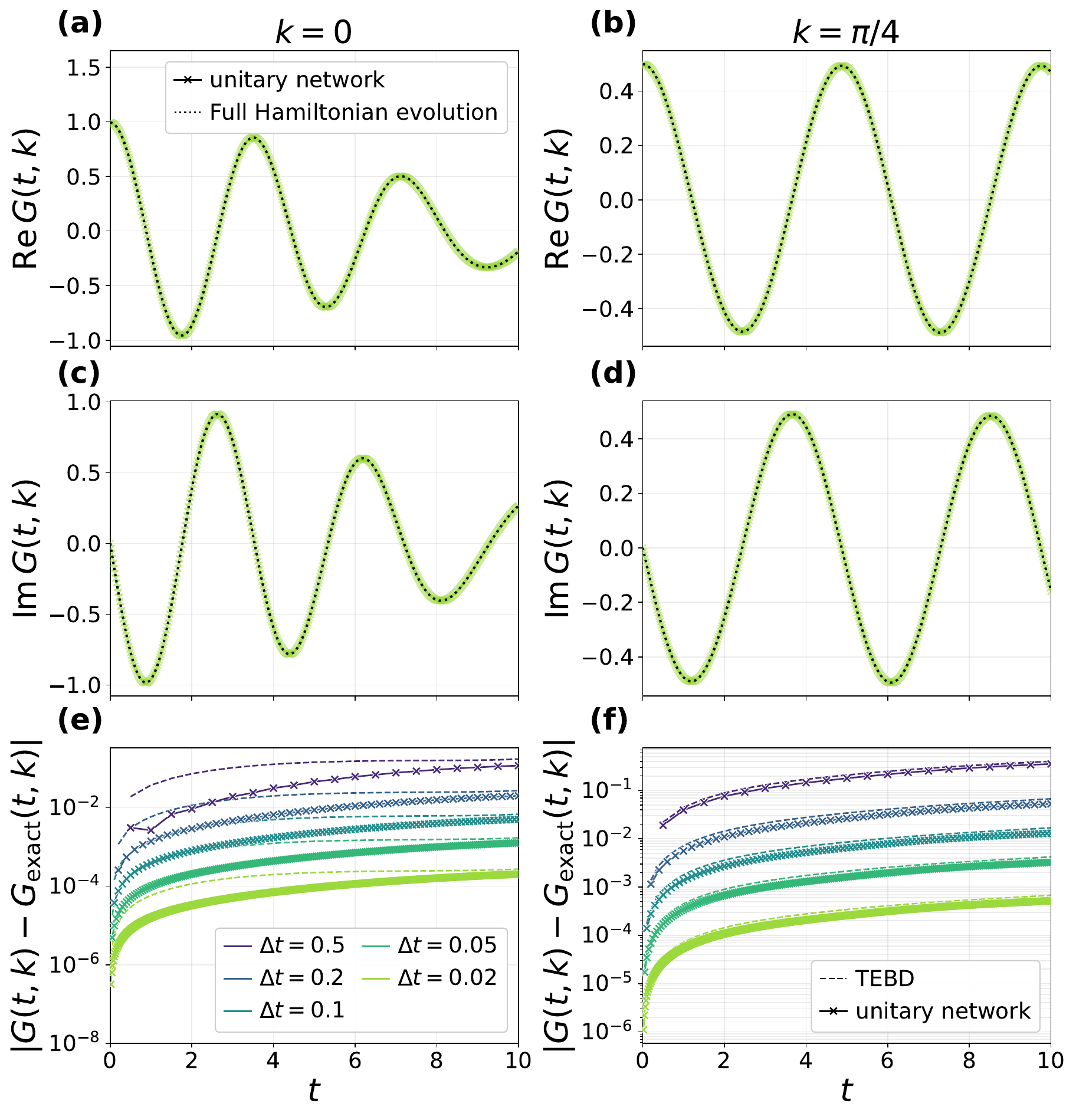}
    \caption{
Real-time hole Green's functions and evolution errors for the periodic
Hubbard chain with $L=8$, at quarter filling
$N_\uparrow=N_\downarrow=2$, and $U=1$.
Panels \textbf{(a,b)} and \textbf{(c,d)} show the real and imaginary
parts of $G(t,k)$, respectively, comparing unitary-network evolution
at $\Delta t=0.02$ with the full-Hamiltonian evolution.
Panels \textbf{(e,f)} compare the absolute errors
$|G(t,k)-G_{\mathrm{exact}}(t,k)|$ of second-order TEBD
(dashed curves) and unitary-network evolution
(solid curves) for
$\Delta t=0.02,0.05,0.1,0.2,0.5$, distinguished by color.
The left and right columns correspond to $k=0$ and $k=\pi/4$,
respectively.
The unitary network uses two bond modes per leg.
All Green's functions are evaluated from the same initial DMRG ground state.
}
    \label{fig:hubbard1}
\end{figure}

We benchmark the unitary-network method above using the one-dimensional
Hubbard model with periodic boundary conditions:
\begin{equation}
\hat{H}
=
-J\sum_{j,\sigma}
\left(
\hat{c}^{\dagger}_{j\sigma}\hat{c}_{j+1,\sigma}
+
\hat{c}^{\dagger}_{j+1,\sigma}\hat{c}_{j\sigma}
\right)
+
U\sum_j
\hat{n}_{j\uparrow}\hat{n}_{j\downarrow},
\end{equation}
where $J$ and $U$ denote the nearest-neighbor hopping
amplitude and on-site interaction strength, respectively,
and $\hat{n}_{j\sigma}=\hat{c}^{\dagger}_{j\sigma}\hat{c}_{j\sigma}$.
We compare time evolution using TEBD (dashed lines) and
the unitary-network method. Starting from the ground state $|\psi_0\rangle$,
obtained using DMRG, we evaluate the hole Green's function
\begin{equation}
\begin{aligned}
G^{-}(t,k) &= \sum_j e^{-ikj}G^{-}(t,j),\\
G^{-}(t,j) &= e^{iE_0t}
\langle\psi_0|
\hat{c}^{\dagger}_{j\uparrow}
e^{-i\hat{H}t}
\hat{c}_{0\uparrow}
|\psi_0\rangle,
\end{aligned}
\end{equation}
where $E_0$ is the ground-state energy. The results are shown in Fig.~\ref{fig:hubbard1}. Both methods
capture the dynamics, but the unitary-network ansatz achieves
smaller errors at a higher computational cost, as shown in
Figs.~\ref{fig:hubbard1}(e) and (f).

In our implementation, the additional computational cost arises
primarily from handling periodic boundary conditions. Although
the physical chain is periodic, the state is represented as an
open-boundary MPS, allowing efficient Schmidt truncation in
mixed-canonical form. The unitary network, however, is naturally
periodic. Converting it into an open-boundary MPO requires
carrying the closing virtual index along the chain, increasing
the bond dimension from $D$ to $D^2$ before compression.
In contrast, TEBD handles the boundary interaction using
fermionic swap gates, avoiding this increase in bond dimension.
The resulting larger contractions and singular value
decompositions make the unitary-network evolution more
computationally expensive in our implementation, despite
its higher accuracy in this benchmark.

\subsection{From unitary network to uniform matrix product operator}

The unitary-network construction also provides a natural route
to uniform matrix product operator (MPO) representations~\cite{Pirvu_2010} of
fermionic time evolution. In particular, the local unitary tensor $\hat{u}$ can serve
directly as the local tensor of a uniform MPO, explicitly preserving translation invariance
at each Trotter step. Following Ref.~\cite{Pirvu_2010}, time evolution is implemented
by stacking local tensors from successive time steps and
compressing the resulting MPO through bond-space truncation.
Details of the compression procedure are given in
Appendix~\ref{app:mpo-compression}. 

A key advantage of the uniform MPO representation is that
it preserves the translation invariance of the time-evolution
operator, even when the initial state is not translation
invariant. This is particularly useful for computing Green's
functions, where a local excitation breaks the translation
invariance of the state without affecting that of the
evolution operator.

However, bond-dimension
truncation generally breaks exact unitarity. Moreover, since
the compressed MPO approximates the full time-evolution
operator rather than a particular evolved state, accurately
capturing generic long-time dynamics requires an increasing
bond dimension due to operator-entanglement growth
\cite{Dubail_2017,Pal_2018}. Consequently, one must either
increase the bond dimension with time or accept potentially
substantial compression errors.

\begin{figure}
    \centering
    \includegraphics[width=0.8\linewidth]{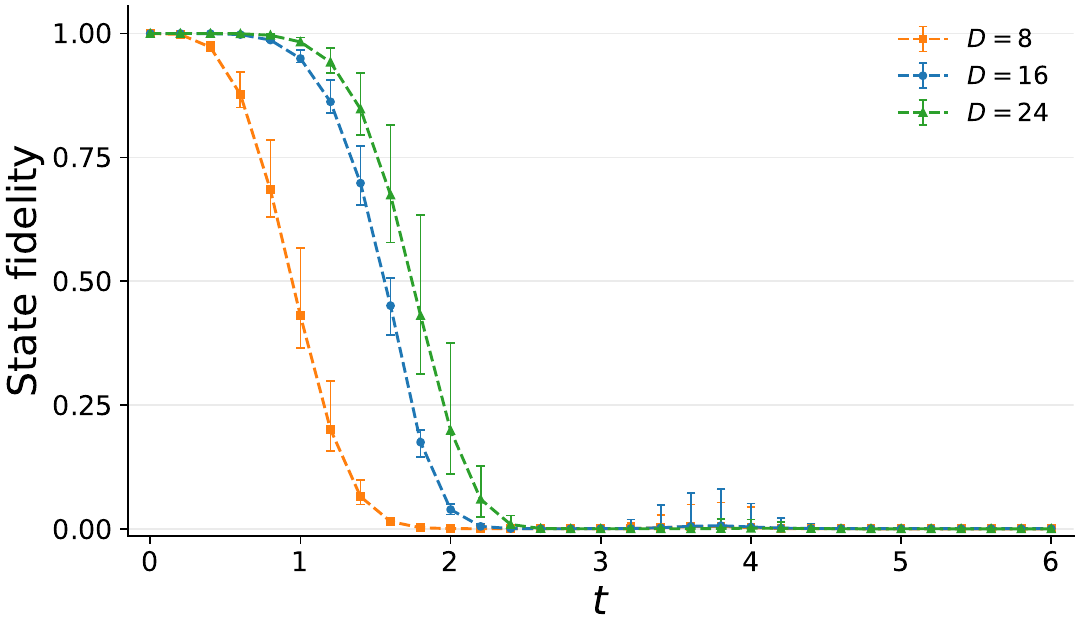}
    \caption{State fidelity between compressed MPO evolution and exact
    full-Hamiltonian evolution for the periodic Hubbard chain with
    $L=8$, $U=1$, and $\mu=0$. The unitary network uses time
    step $\Delta t=0.02$ and maximum MPO bond dimensions $D=8,16,24$,
    with evolution shown up to $t=6$. Each marker shows the median
    squared normalized overlap over the same $302$ equally weighted
    input states: $20$ random states in each particle-number sector
    $N=1,\ldots,15$, and one state each in the $N=0$ and $N=16$ sectors.
    Identical input states are used across all bond dimensions and
    times. Error bars span the 25th--75th percentiles.}
    \label{fig:interaction_un_fidelity}
\end{figure}

To illustrate this limitation, we compare the fidelities of
states evolved using uniform MPOs with fixed bond dimensions
$D=8,16,24$ against exact full-Hamiltonian evolution for an
ensemble of initial states, as shown in
Fig.~\ref{fig:interaction_un_fidelity}. The fidelities remain
high at short times but decrease as time progresses,
reflecting the accumulation of MPO compression errors.
Larger bond dimensions maintain high fidelity over longer times.

\begin{figure}
    \centering
    \includegraphics[width=1\linewidth]{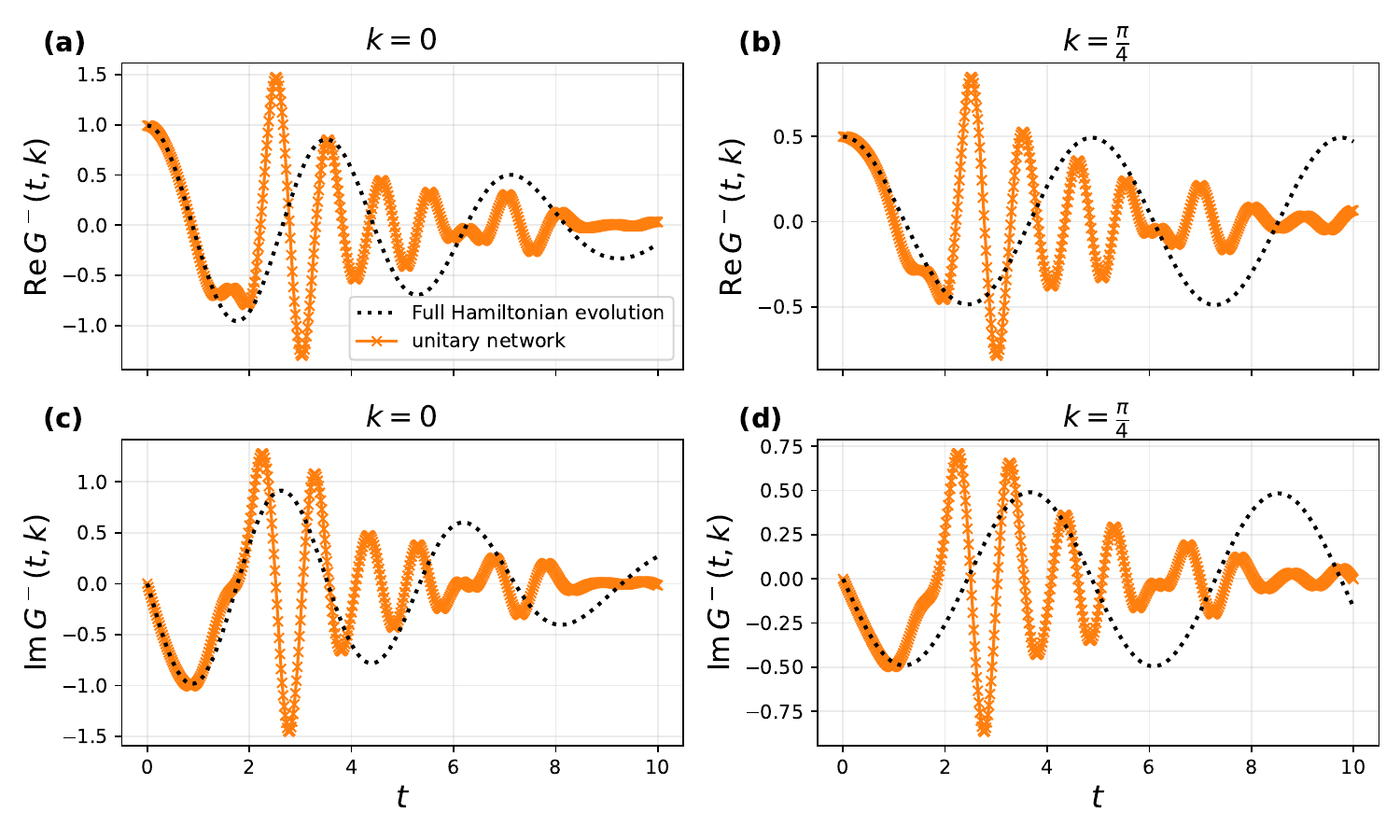}
    \caption{
Hole Green's functions for the periodic Hubbard chain with $L=8$,
$N_\uparrow=N_\downarrow=2$ (quarter filling), $U=1$, and $\mu=0$.
Panels (a,b) show the real part at $k=0$ and $k=\pi/4$,
respectively; panels (c,d) show the imaginary part.
Black dotted curves denote evolution under the full Hamiltonian. Orange solid curves denote MPO evolution with $\Delta t=0.02$, and successive
operator-layer compression to MPO bond dimension $16$.
Both calculations use the same initial DMRG state.
}
    \label{fig:MPO}
\end{figure}

We also compare the hole Green's functions obtained using the
uniform MPO with those from exact full-Hamiltonian evolution
for the periodic Hubbard chain. As shown in Fig.~\ref{fig:MPO},
the results agree well at short times, but deviations increase
at later times.

\section{Conclusion}

In this work, we introduced fermionic Gaussian unitary networks as variational ansatzes for fermionic Gaussian unitary transformations. We showed that, even in the presence of directed loops, the contracted network preserves global unitarity up to an overall normalization factor, thereby extending unitary tensor-network constructions beyond directed acyclic geometries.

We derived explicit contraction rules that reduce the full tensor-network contraction to the single-particle level. Local tensors are represented by orthogonal matrices in the Majorana basis, or by unitary matrices in the particle-number-conserving case. For translation-invariant networks, the contraction further decomposes into independent momentum sectors, so that the global transformation at each momentum $\mathbf{k}$ is determined directly from the local tensor data. This momentum-space formulation recasts variational optimization as a data-driven problem over the Brillouin zone, with different momentum points serving as training samples and allowing stochastic optimization.

We demonstrated the expressive power of the ansatz using the Qi--Wu--Zhang model, where a finite-bond-dimension network maps a simple product state to an accurate approximation of the free-fermion ground state. For gapless states, exemplified by the half-filled one-dimensional nearest-neighbor metal, we introduced a scale-invariant layered architecture. By repeatedly applying the same unitary network across successive scales, this construction generates approximations to target states at progressively larger system sizes. Such a scale-invariant structure provides a natural framework for representing gapless free-fermion states with correlations extending over arbitrarily long length scales.

We also applied the framework to real-time evolution generated by quadratic fermionic Hamiltonians. For generic dispersive bands, the approximation error increases at long times as entanglement grows beyond the capacity of a fixed-bond-dimension network. In contrast, flat-band Hamiltonians, characterized by vanishing group velocity and periodic dynamics, are naturally suited to finite-bond-dimension fermionic Gaussian unitary networks. We showed that an optimized Gaussian network can be converted into a many-body tensor-network operator and composed with on-site interaction gates, providing a systematic extension toward interacting dynamics. We demonstrated this construction for the one-dimensional Hubbard model and benchmarked the resulting time evolution against TEBD method.

These results establish fermionic Gaussian unitary networks as a flexible framework for representing Gaussian states and dynamics while preserving unitarity at the network level. Their compatibility with momentum-space optimization, scale-invariant constructions, and local interacting gates provides several directions for extending the approach to more general fermionic many-body problems.

\begin{acknowledgments}
WX and HCP were supported by the Hong Kong Research Grants Council (GRF grants 16308822 and 16310625), the Croucher Foundation (CIA23SC01), and the Future Science Awards Foundation through an Asian Young Scientist Fellowship.
SO was supported by JST, PRESTO Grant No.~JPMJPR26A5, JSPS KAKENHI Grant No.~JP25K17322.

ChatGPT (OpenAI; GPT-6) was used to assist with mathematical derivations in Appendix~\ref{Append: Self-contraction rule without fermion number conservation}, the development of numerical codes, and literature review. All AI-assisted derivations and numerical implementations were critically reviewed, revised where necessary, and independently verified by the authors. The authors take full responsibility for the overall content of the manuscript.

\end{acknowledgments}

\appendix

\section{Block-Matrix Identities}

In many calculations, we make use of standard identities for block matrices. Consider a block matrix
\begin{equation}
    M = \left(
    \begin{array}{cc}
     A  & B  \\
     C & D
   \end{array}
   \right).
\end{equation}
If $D$ is invertible, the Schur complement of $D$ is defined as
\begin{equation}
    \Sigma_{D} = A-BD^{-1}C.
\end{equation}
Alternatively, if $A$ is invertible, the Schur complement of $A$ is
\begin{equation}
    \Sigma_{A} = D-CA^{-1}B.
\end{equation}
Under the corresponding invertibility conditions, the block matrix admits the factorizations
\begin{equation}
\begin{aligned}
    M
    &= \left(
    \begin{array}{cc}
     I  & BD^{-1}  \\
     0 & I
   \end{array}
   \right)
   \left(
    \begin{array}{cc}
     \Sigma_{D}  & 0  \\
     0 & D
   \end{array}
   \right)
   \left(
    \begin{array}{cc}
     I  & 0  \\
     D^{-1}C & I
   \end{array}
   \right)
   \\
   &=
   \left(
    \begin{array}{cc}
     I  & 0  \\
     CA^{-1} & I
   \end{array}
   \right)
   \left(
    \begin{array}{cc}
     A  & 0  \\
     0 & \Sigma_{A}
   \end{array}
   \right)
   \left(
    \begin{array}{cc}
     I  & A^{-1}B  \\
     0 & I
   \end{array}
   \right).
\end{aligned}
\end{equation}

\subsection{Determinant of a Block Matrix}

The above factorizations immediately give the determinant identities
\begin{equation}
\begin{aligned}
    \mathrm{det}\,M
    &= \mathrm{det}\,\Sigma_{D}\,\mathrm{det}\,D
    = \mathrm{det}(A-BD^{-1}C)\,\mathrm{det}\,D \\
    &= \mathrm{det}\,\Sigma_{A}\,\mathrm{det}\,A
    = \mathrm{det}(D-CA^{-1}B)\,\mathrm{det}\,A,
\end{aligned}
\end{equation}
where the first line assumes that $D$ is invertible, while the second assumes that $A$ is invertible.

\subsection{Inverse of a Block Matrix}

If $D$ is invertible, then $M$ is invertible if and only if $\Sigma_{D}$ is invertible. Similarly, if $A$ is invertible, then $M$ is invertible if and only if $\Sigma_{A}$ is invertible. The inverse of $M$ can therefore be written as
\begin{equation}
\label{eq: block inverse}
\begin{aligned}
      M^{-1}
      &= \left(
    \begin{array}{c|c}
     \Sigma_{D}^{-1}
     &
     -\Sigma_{D}^{-1}BD^{-1}
     \\
     \hline
     -D^{-1}C\Sigma_{D}^{-1}
     &
     D^{-1}+D^{-1}C\Sigma_{D}^{-1}BD^{-1}
   \end{array}
   \right)
   \\
   &=
   \left(
    \begin{array}{c|c}
     A^{-1}+A^{-1}B\Sigma_{A}^{-1}CA^{-1}
     &
     -A^{-1}B\Sigma_{A}^{-1}
     \\
     \hline
     -\Sigma_{A}^{-1}CA^{-1}
     &
     \Sigma_{A}^{-1}
   \end{array}
   \right).
\end{aligned}
\end{equation}

\section{Fermionic Gaussian unitary operations}
\label{Append: Fermionic Gaussian unitary operations}

In this appendix, we summarize the different representations of fermionic Gaussian unitary operations and the relations among them. A fermionic Gaussian unitary $\hat{U}$ may be described in the many-body Fock basis through its matrix elements $U_{MN}$, or at the single-particle level through a Bogoliubov transformation or an orthogonal matrix $O\in\mathrm{O}(2N)$. We also introduce the associated covariance-matrix representation $\Gamma$. These descriptions are equivalent in the sense relevant here and can be converted into one another. Establishing these correspondences will be essential for deriving the fermionic Gaussian tensor-contraction rules used later.

\subsection{Overview}

Consider a system of $N$ fermionic modes, with Hilbert space
\begin{equation}
    \mathcal{H}_{f}^{\otimes_{g} N}
    =
    \underbrace{
    \mathcal{H}_{f}\otimes_{g}\mathcal{H}_{f}\otimes_{g}\cdots\otimes_{g}\mathcal{H}_{f}
    }_{N\ \text{factors}},
\end{equation}
where $\mathcal{H}_{f}\cong\mathbb{C}^{2}$ is the Hilbert space of a single fermionic mode and $\otimes_{g}$ denotes the $\mathbb{Z}_{2}$-graded tensor product \cite{Bultinck2017Fermionic}. We introduce a set of fermionic operators $\hat{\xi}^{a}$, $a=1,2,\ldots,2N$, satisfying
\begin{equation}
    \{\hat{\xi}^{a},\hat{\xi}^{b}\}
    =
    G^{ab}\hat{I},
\end{equation}
where $G^{ab}$ is a symmetric, nondegenerate matrix. In the Majorana basis,
$\mathbf{\hat{\xi}}\overset{q,p}{=}(\mathbf{\hat{q}},\mathbf{\hat{p}})$,
one has
\begin{equation}
    G\overset{q,p}{=}
    \left(
    \begin{array}{cc}
        I & 0\\
        0 & I
    \end{array}
    \right),
\end{equation}
whereas in the creation-annihilation basis,
$\mathbf{\hat{\xi}}\overset{c,c^{\dagger}}{=}(\mathbf{\hat{c}},\mathbf{\hat{c}}^{\dagger})$,
\begin{equation}
    G\overset{c,c^{\dagger}}{=}
    \left(
    \begin{array}{cc}
        0 & I\\
        I & 0
    \end{array}
    \right).
\end{equation}

Fermionic Gaussian unitaries are generated by quadratic fermionic operators and act linearly on the fermionic modes under conjugation. In the Majorana representation, this action is described by an orthogonal transformation. Denoting the group of fermionic Gaussian unitaries by
$\mathrm{U}_{fG}(\mathcal{H}_{f}^{\otimes_{g}N})$, the corresponding single-particle transformation is represented projectively as \cite{Hackl_2021, guaita2024representationtheorygaussianunitary}
\begin{equation}
\begin{aligned}
    &\hat{\rho}(e^{K})=\pm e^{\hat{K}},\\
    &\hat{K}
    =
    \frac{1}{2}
    g_{ac}K^{c}_{\;b}\,
    \hat{\xi}^{\dagger}_{a}\hat{\xi}_{b},
\end{aligned}
\end{equation}
where $g_{ab}$ is the inverse of $G^{ab}$, $g_{ac}G^{cb}=\delta_{a}^{b}$.
The relation between the single-particle transformation $R$ and its many-body realization is characterized by
\begin{equation}
    \hat{\rho}^{\dagger}(R)\hat{\xi}^{a}\hat{\rho}(R)
    =
    R^{a}{}_{b}\hat{\xi}^{b},
\end{equation}
which follows directly from the Baker--Campbell--Hausdorff formula. It is useful to define the map
\begin{equation}
    \rho^{-1}:
    \mathrm{U}_{fG}(\mathcal{H}_{f}^{\otimes_{g}N})
    \rightarrow
    \mathrm{SO}(2N),
\end{equation}
where $\rho^{-1}(\hat{U})$ is the single-particle transformation induced by conjugation with $\hat{U}$. This map is a group homomorphism:
\begin{equation}
    \rho^{-1}(\hat{U}_{1}\hat{U}_{2})
    =
    \rho^{-1}(\hat{U}_{1})\rho^{-1}(\hat{U}_{2}).
\end{equation}
Thus, composition of Gaussian many-body unitaries corresponds directly to matrix multiplication at the single-particle level.

For a particle-number-conserving fermionic Gaussian unitary, the transformation reduces to an $N\times N$ single-particle unitary matrix $U\in\mathrm{U}(N)$. Throughout this paper, we use $U$ to denote the single-particle transformation and $\hat{U}$ its many-body realization. Writing $U=e^{-ih}$, with $h$ Hermitian, the corresponding many-body unitary is generated by the quadratic Hamiltonian
\begin{equation}
\begin{aligned}
    &\hat{U}=\hat{\rho}(U)=e^{-i\hat{H}},\\
    &\hat{H}=h_{ab}\hat{c}^{\dagger}_{a}\hat{c}_{b}.
\end{aligned}
\end{equation}
Thus, a particle-number-conserving fermionic Gaussian unitary is determined by its single-particle unitary matrix $U$ up to an overall phase. Indeed, multiplying the many-body operator by an arbitrary phase does not change its conjugation action on the fermionic operators:
\begin{equation}
    (e^{i\phi}\hat{U})^{\dagger}
    \hat{\xi}^{a}
    (e^{i\phi}\hat{U})
    =
    \hat{U}^{\dagger}\hat{\xi}^{a}\hat{U}.
\end{equation}

\subsection{Orthogonal matrix representation}

A fermionic Gaussian unitary operator $\hat{U}$ induces a linear transformation on the Majorana operators $\mathbf{\hat{\xi}} \overset{q,p}{=} (\mathbf{\hat{q}}, \mathbf{\hat{p}})$:
\begin{equation}
    \hat{U}^{\dagger} \hat{\xi}^{a} \hat{U} = O^{a}_{\;b} \hat{\xi}^{b},
\end{equation}
where $O \in \mathrm{SO}(2N)$ is the corresponding single-particle orthogonal matrix. Explicitly,
\begin{equation}
\label{eq: orthorgonal matrix form}
   \left(
    \begin{array}{c}
     \mathbf{\hat{q}}^{\prime}   \\
     \mathbf{\hat{p}}^{\prime}
   \end{array} 
   \right)  =  O \left(
    \begin{array}{c}
     \mathbf{\hat{q}}   \\
     \mathbf{\hat{p}}  
   \end{array} 
   \right). 
\end{equation}
The relation between the single-particle transformation $O$ and the corresponding many-body Gaussian unitary can equivalently be expressed through the projective representation
$\hat{\rho}: \mathrm{SO}(2N) \rightarrow \mathrm{U}_{fG}(\mathcal{H}_{f}^{\otimes_{g} N})$:
\begin{equation}
\begin{aligned}
    &\hat{\rho}(e^{K}) = \pm e^{\hat{K}}, \\
    &\hat{K} = \frac{1}{2} g_{ac} K^{c}_{\; b}\,\hat{\xi}^{\dagger}_{a}\hat{\xi}_{b},
\end{aligned}
\end{equation}
where $g_{ab}$ is the inverse of $G^{ab}$, satisfying $g_{ac}G^{cb} = \delta_{a}^{b}$.

\subsection{Bogoliubov transformation representation}
Alternatively, a fermionic Gaussian unitary can be characterized by its action on the creation and annihilation operators,
$\mathbf{\hat{\xi}} \overset{c,c^{\dagger}}{=} (\mathbf{\hat{c}}, \mathbf{\hat{c}}^{\dagger})$:
\begin{equation}
\label{eq: unitary_transformation}
    \hat{U}^{\dagger} \hat{\xi}^{a} \hat{U}
    =
    U^{a}_{\;b} \hat{\xi}^{b},
\end{equation}
where $U\in\mathrm{U}(2N)$ is the corresponding Bogoliubov transformation. The Majorana and creation-annihilation representations are related by the change-of-basis matrix
\begin{equation}
\left(
    \begin{array}{c}
     q   \\
    p
   \end{array} 
   \right)  = T \left(
    \begin{array}{c}
     \hat{c}   \\
     \hat{c}^{\dagger}  
   \end{array} 
   \right), \quad
    T = \frac{1}{\sqrt{2}}\left(
    \begin{array}{cc}
     I  & I \\
     iI & -iI 
   \end{array} 
   \right).
\end{equation}
Accordingly, if the orthogonal matrix is written as
\[
O=
\left(
\begin{array}{cc}
a & b\\
c & d
\end{array}
\right),
\]
the corresponding Bogoliubov transformation is
\begin{equation}
\begin{aligned}
    U &= T^{\dagger}OT \\
   & = \frac{1}{2} \left(
    \begin{array}{cc}
     I  & -iI \\
     I & iI 
   \end{array} 
   \right)
   \left(
    \begin{array}{cc}
     a  & b \\
     c & d 
   \end{array} 
   \right)
   \left(
    \begin{array}{cc}
     I  & I \\
     iI & -iI 
   \end{array} 
   \right) \\
  &= \frac{1}{2} \left(
    \begin{array}{cc}
     (a+d) +i(b-c)  & (a-d) - i(b+c) \\
     (a-d)+i(b+c) &  (a+d) - i(b-c)
   \end{array} 
   \right) \\
   & =  \left(
    \begin{array}{cc}
     u  & v \\
     v^{*} & u^{*} 
   \end{array} 
   \right).
\end{aligned}
\end{equation}
Thus, the Bogoliubov transformation acts explicitly as
\begin{equation}
\label{eq: Bogoliubov transformation matrix form}
   \left(
    \begin{array}{c}
     \hat{c}^{\prime}   \\
     \hat{c}^{\dagger\prime}
   \end{array} 
   \right)  = \left(
    \begin{array}{cc}
     u  & v \\
     v^{*}  & u^{*} 
   \end{array} 
   \right) \left(
    \begin{array}{c}
     \hat{c}   \\
     \hat{c}^{\dagger}  
   \end{array} 
   \right) = U \left(
    \begin{array}{c}
     \hat{c}   \\
     \hat{c}^{\dagger}  
   \end{array} 
   \right).
\end{equation}

\subsection{Covariance matrix representation}

We introduce the antisymmetric matrix $\Gamma$ as an alternative representation of a fermionic Gaussian unitary, providing a convenient link between the Bogoliubov representation and the Fock-space representation. Starting from the Bogoliubov matrix $U$, we define \cite{hara_quantum_1979, mizusaki_grassmann_2013}
\begin{equation}
\label{eq: covariance matrix}
    \Gamma =  \left(
    \begin{array}{cc}
     B  & C \\
     -C^{T} & A 
   \end{array} 
   \right) =  \left(
    \begin{array}{cc}
     \Gamma_{cc}  & \Gamma_{cc^{\dagger}} \\
     \Gamma_{c^{\dagger}c} & \Gamma_{c^{\dagger}c^{\dagger}} 
   \end{array} 
   \right),
\end{equation}
where, assuming $u$ is invertible,
\begin{equation}
\label{eq: covariance entry}
\begin{aligned}
   &  A_{ij} = \frac{\langle \emptyset| \hat{U}^\dagger \hat{c}^{\dagger}_{i} \hat{c}^{\dagger}_{j} |\emptyset \rangle}{\langle \emptyset| \hat{U}^\dagger  |\emptyset \rangle}  = v^{*}u^{-1}, \\
  &   B_{ij} = \frac{\langle \emptyset| \hat{c}_{i} \hat{c}_{j}\hat{U}^\dagger |\emptyset \rangle}{\langle \emptyset| \hat{U}^\dagger  |\emptyset \rangle} =u^{-1}v, \\
  & C_{ij} = \frac{\langle \emptyset| \hat{c}_{i} \hat{U}^\dagger \hat{c}^{\dagger}_{j} |\emptyset \rangle}{\langle \emptyset| \hat{U}^\dagger  |\emptyset \rangle} =u^{-1}.
\end{aligned}
\end{equation}
Using $\hat{U}^\dagger$ instead of $\hat{U}$ makes the convention in
Eq.~\eqref{eq: unitary_transformation} consistent with the results of Ref.~\cite{mizusaki_grassmann_2013}. We denote this correspondence by $\Gamma=\operatorname{cov}(U)$.

The Bogoliubov matrix $U$ and the covariance matrix $\Gamma$ encode the same Gaussian transformation, but with different groupings of the input and output operators. Specifically, for
\begin{equation}
   \left(
    \begin{array}{c}
     \hat{c}^{\prime}   \\
     \hat{c}^{\dagger\prime}
   \end{array} 
   \right)  = \left(
    \begin{array}{cc}
     u  & v \\
     v^{*}  & u^{*} 
   \end{array} 
   \right) \left(
    \begin{array}{c}
     \hat{c}   \\
     \hat{c}^{\dagger}  
   \end{array} 
   \right), 
\end{equation}
the same transformation can equivalently be written as
\begin{equation}
\left(
    \begin{array}{c}
         \hat{c}  \\
         \hat{c}^{\dagger \prime}
    \end{array} \right)
    = 
    \left(
    \begin{array}{cc}
         B & C  \\
         -C^T & A
    \end{array} \right)
    \left(
    \begin{array}{c}
         -\hat{c}^{\dagger}  \\
         \hat{c}^{\prime}
    \end{array}
    \right) = \Gamma  \left(
    \begin{array}{c}
         -\hat{c}^{\dagger}  \\
         \hat{c}^{\prime}
    \end{array}
    \right).
\end{equation}
To verify this relation, the Bogoliubov transformation gives
\begin{equation}
\label{eq: bogoliubov trans}
    \begin{aligned}
        & \hat{c}^{\prime} = u\hat{c} +v\hat{c}^{\dagger}, \\
        & \hat{c}^{\dagger\prime} = v^{*}\hat{c} + u^{*}\hat{c}^{\dagger}.
    \end{aligned}
\end{equation}
Solving the first equation for $\hat{c}$ yields
\begin{equation}
\begin{aligned}
   & u\hat{c}  = \hat{c}^{\prime} - v\hat{c}^{\dagger}  \\
    \implies  & \hat{c} = u^{-1}\hat{c}^{\prime} - u^{-1}v \hat{c}^{\dagger}
    = C\hat{c}^{\prime} + B(-\hat{c}^{\dagger}).
\end{aligned}
\end{equation}
Substituting this expression into the second equation in Eq.~\eqref{eq: bogoliubov trans} gives
\begin{equation}
\begin{aligned}
  \hat{c}^{\dagger\prime}
  &= v^*(u^{-1}\hat{c}^{\prime} - u^{-1}v \hat{c}^{\dagger}) + u^* \hat{c}^{\dagger} \\
  &= v^*u^{-1} \hat{c}^{\prime}
  + (u^* - v^*u^{-1}v)\hat{c}^{\dagger}.
\end{aligned}
\end{equation}
Using the unitarity of $U$ together with the block-matrix inversion formula in Eq.~\eqref{eq: block inverse}, one finds
\begin{equation}
    u^* - v^*u^{-1}v = u^{-T}.
\end{equation}
Therefore,
\begin{equation}
    \hat{c}^{\dagger\prime}
    = A\hat{c}^{\prime}
    - C^T(-\hat{c}^{\dagger}),
\end{equation}
which proves the stated relation between $U$ and $\Gamma$.

As shown in the next subsection, the covariance matrix $\Gamma$ also provides a direct route to the Fock-space matrix elements of the fermionic Gaussian unitary operator. It therefore serves as a convenient intermediate representation between the Bogoliubov transformation and the many-body Fock-space representation.

\subsection{Fock Basis representation}

For a fermionic system with modes $S=\{1,2,\cdots,n\}$, the Hilbert space is spanned by the Fock basis. By the Pauli exclusion principle, each mode is either empty or occupied by a single fermion. A Fock state is therefore uniquely specified by a subset $M\subseteq S$ of occupied modes:
\begin{equation}
    |M\rangle, \qquad M\subseteq S, \qquad M\in 2^S.
\end{equation}
The power set $2^S$ contains $2^n$ subsets and hence labels the $2^n$ Fock states. In particular,
\begin{itemize}
    \item the vacuum state is
    $|\Omega\rangle=|\emptyset\rangle=|\{\}\rangle$;

    \item a one-particle state is
    $|\{i\}\rangle=\hat{c}^{\dagger}_{i}|\Omega\rangle$;

    \item a two-particle state is
    $|\{i,j\}\rangle=\hat{c}^{\dagger}_{i}\hat{c}^{\dagger}_{j}|\Omega\rangle$,
    with $i<j$;

    \item more generally, for
    $M=\{m_{1},m_{2},\cdots,m_{|M|}\}$ with
    $m_{1}<m_{2}<\cdots<m_{|M|}$,
    \begin{equation}
        |M\rangle = \prod_{i \in M}\hat{c}^{\dagger}_{i}|\Omega \rangle
        = \hat{c}^{\dagger}_{m_{1}}\hat{c}^{\dagger}_{m_{2}}\cdots
        \hat{c}^{\dagger}_{m_{|M|}}|\Omega \rangle.
    \end{equation}
    The corresponding bra is
    \begin{equation}
        \langle M|
        = \langle \Omega |
        \hat{c}_{m_{|M|}}\cdots
        \hat{c}_{m_{2}}\hat{c}_{m_{1}}.
    \end{equation}
\end{itemize}

A parity-preserving fermionic unitary operator can then be expanded in the Fock basis as
\begin{equation}
\label{eq:fG unitary}
    \hat{U} = \sum_{M,N} U_{MN} |M\rangle \langle N|,
\end{equation}
where $M,N\subseteq S$ label the occupied output and input modes, respectively. Owing to the fermion-parity superselection rule \cite{Wick_Intrinsic_1952, Vidal_Quantum_2021}, the matrix element $U_{MN}$ can be nonzero only when $|M|$ and $|N|$ have the same parity.

We now derive the Fock-basis matrix elements of a fermionic Gaussian unitary $\hat{U}$ from the covariance matrix $\Gamma$ using the Pfaffian formula \cite{hara_quantum_1979, mizusaki_grassmann_2013, rajabpour2025matrixelementsfermionicgaussian}:
\begin{equation}
\label{eq: matrix element of fG-unitary}
\begin{aligned}
     \frac{U_{NM}^{*}}{U^*_{\emptyset\emptyset}} &= \frac{\langle \emptyset| (\prod_{i \in M}\hat{c}_{i}) \hat{U}^\dagger (\prod_{j \in N}\hat{c}^{\dagger}_{j})|\emptyset\rangle}{\langle \emptyset|  \hat{U}^\dagger |\emptyset\rangle} \\
     &= (-1)^{\frac{|M|(|M|-1)}{2}}\mathrm{Pf} ( \Gamma[M_{c}\cup N_{c^{\dagger}},M_{c}\cup N_{c^{\dagger}}]) \\
     & = (-1)^{\frac{|M|(|M|-1)}{2}}\mathrm{Pf} \left[ \left(
    \begin{array}{cc}
     \Gamma_{cc}[M,M]  & \Gamma_{cc^{\dagger}}[M,N]\\
     \Gamma_{c^{\dagger}c} [N,M] & \Gamma_{c^{\dagger}c^{\dagger}} [N,N]
   \end{array} 
   \right) \right].
\end{aligned}
\end{equation} 
Again, we use $U_{NM}^{*}$ and $\hat{U}^\dagger$ instead of $U_{MN}$ and $\hat{U}$ to
make the convention in Eq.~\eqref{eq: unitary_transformation}
consistent with that of Ref.~\cite{mizusaki_grassmann_2013}.
Here, $\Gamma_{cc}[M,N]$ denotes the submatrix of $\Gamma_{cc}$ obtained by selecting the rows indexed by $M$ and the columns indexed by $N$. The factor $(-1)^{\frac{|M|(|M|-1)}{2}}$ arises from reversing the order of the $|M|$ fermionic operators and is the simplified form of the sign convention used in Ref.~\cite{rajabpour2025matrixelementsfermionicgaussian}.

The covariance matrix therefore determines all Fock-space matrix elements relative to the vacuum amplitude
\begin{equation}
    U^{*}_{\emptyset\emptyset}
    =
    \langle \emptyset|\hat{U}^\dagger|\emptyset\rangle
    =
    [\det u]^{1/2}\mathrm{e}^{i\theta},
\end{equation}
where $u$ is the corresponding block of the Bogoliubov transformation and $\mathrm{e}^{i\theta}$ is an overall phase \cite{hara_quantum_1979}. Once this overall phase convention is fixed, we may associate $\Gamma$ with the corresponding many-body fermionic Gaussian unitary through
\begin{equation}
\label{eq: covariance to many-body mapping}
    \hat{\sigma}(\Gamma) = \hat{U} = \sum_{M,N} U_{MN} |M\rangle \langle N|.
\end{equation}

Since the entries of $\Gamma$ in Eq.~\eqref{eq: covariance entry} are normalized two-point contractions, Eq.~\eqref{eq: matrix element of fG-unitary} expresses higher-order matrix elements entirely in terms of these pairwise contractions. It can therefore be viewed as a generalized Wick theorem, as originally formulated in Ref.~\cite{hara_quantum_1979}.

\subsection{Particle-number-conserving fermionic Gaussian unitary}

For a particle-number-conserving fermionic Gaussian unitary, the transformation is characterized by a single-particle unitary matrix $u\in\mathrm{U}(N)$. In the Bogoliubov representation, particle-number conservation implies $v=0$, so that
\begin{equation}
    U=
    \left(
    \begin{array}{cc}
        u & 0\\
        0 & u^{*}
    \end{array}
    \right).
\end{equation}
The corresponding many-body operator is obtained by second quantization,
\begin{equation}
\begin{aligned}
    &\hat{\rho}(e^{-ih})=e^{-i\hat{H}},\\
    &\hat{H}=h_{ab}\hat{c}^{\dagger}_{a}\hat{c}_{b},
\end{aligned}
\end{equation}
where $u=e^{-ih}$ \cite{Hackl_2021}.

Because fermion number is conserved, a Fock-space matrix element can be nonzero only when the input and output states contain the same number of particles,
\[
    |M|=|N|.
\]
For such matrix elements, the many-body representation is given directly by the determinant formula \cite{rajabpour2025matrixelementsfermionicgaussian}
\begin{equation}
    \hat{\rho}(u)_{MN}
    =
    \mathrm{det}\,u[M,N],
\end{equation}
where $u[M,N]$ denotes the submatrix of $u$ obtained by selecting the rows indexed by $M$ and the columns indexed by $N$. When all row or column indices are retained, we use the shorthand $u[:,N]$ and $u[M,:]$, respectively.

The determinant representation also makes the composition law transparent. For $|M|=|N|$, one has
\begin{equation}
\begin{aligned}
    \bigl[\hat{\rho}(u)\hat{\rho}(v)\bigr]_{MN}
    &=
    \sum_{P:\,|P|=|M|}
    \mathrm{det}\,u[M,P]\,
    \mathrm{det}\,v[P,N] \\
    &=
    \mathrm{det}\!\left(u[M,:]\,v[:,N]\right)\\
    &=
    \mathrm{det}\,(uv)[M,N]\\
    &=
    \hat{\rho}(uv)_{MN}.
\end{aligned}
\end{equation}
In the second equality, we used the Cauchy--Binet formula,
\begin{equation}
    \mathrm{det}(AB)
    =
    \sum_{\alpha\subseteq\{1,\ldots,n\},\,|\alpha|=m}
    \mathrm{det}(A[:,\alpha])\,
    \mathrm{det}(B[\alpha,:]).
\end{equation}
Thus, multiplication of single-particle unitary matrices is mapped directly to composition of the corresponding many-body Gaussian unitaries.

When the resulting many-body operators are interpreted as fermionic tensors, however, permutations and contractions of tensor legs generally generate additional fermionic sign factors arising from the $\mathbb{Z}_{2}$ grading. Their contraction rules therefore differ from the bosonic case and must incorporate the appropriate fermionic signs \cite{Barthel_2009, Mortier_2025}.

\section{Self-contraction rule with fermion number conservation}
\label{Append: Self-contraction rule with fermion number conservation}

In this Appendix, we derive the self-contraction rule for fermionic Gaussian unitary operators. As shown in Subsection~\ref{subsubsec: Contract without causal order}, contracting a Dirac mode corresponds to taking a partial supertrace. We show that, when an arbitrary set of Dirac modes is contracted, a fermionic Gaussian unitary remains Gaussian and unitary up to an overall normalization factor.

For clarity, we first restrict to the fermion-number-conserving case and defer the general number-nonconserving case to Appendix~\ref{Append: Self-contraction rule without fermion number conservation}. In the number-conserving setting, the self-contraction can be expressed directly in terms of the single-particle unitary matrix:
\begin{equation}
\label{eq: self contraction many modes}
\begin{aligned}
   & \mathrm{STr}_{B}[\hat{\rho}(u)] = \mathrm{det}[1-u_{BB}] \, \hat{\rho}(\mathrm{str}_{B} u),
\end{aligned}
\end{equation}
where the single-particle self-contraction is defined as
\begin{equation}
\label{Eq: Self-contraction of a unitary matrix}
\mathrm{str}_{B} u := u_{AA} + u_{AB}(I-u_{BB})^{-1}u_{BA}.
\end{equation}

\subsection{Contract one mode}

As a first step, we consider the self-contraction of a single mode of a fermionic Gaussian unitary operator. Without loss of generality, we take the contracted mode to be the last mode, labeled by $n$, since any other mode can be brought to this position by an appropriate permutation. Let
\begin{equation}
    \hat{V} = \mathrm{STr}_{n}\hat{U}.
\end{equation}
We evaluate the matrix elements of $\hat{V}$ in the Fock basis. For $M,N\subseteq S\setminus\{n\}$, there are two contributions to $V_{MN}$. The first comes from states in which mode $n$ is unoccupied,
\begin{equation}
    \mathrm{STr}_{n} |M\rangle\langle N|
    =
    |M\rangle\langle N|,
\end{equation}
while the second comes from states in which mode $n$ is occupied,
\begin{equation}
    \mathrm{STr}_{n}
    |M\cup\{n\}\rangle
    \langle N\cup\{n\}|
    =
    -|M\rangle\langle N|.
\end{equation}
Therefore,
\begin{equation}
\label{eq: new unitary}
    V_{MN}
    =
    U_{MN}
    -
    U_{M\cup\{n\},\,N\cup\{n\}}.
\end{equation}

Using the determinant representation of a particle-number-conserving fermionic Gaussian unitary, we have
\begin{equation}
  U_{M\cup\{n\},\,N\cup\{n\}}
  =
  \det \left(
    \begin{array}{cc}
     u[M,N]  & u[M,n] \\
     u[n,N] &  u[n,n]
   \end{array}
   \right),
\end{equation}
whereas
\begin{equation}
\begin{aligned}
   U_{MN}
   &= \det\,u[M,N] \\
   &= \det\left(
    \begin{array}{cc}
     u[M,N]  & 0 \\
     u[n,N] & 1
   \end{array}
   \right).
\end{aligned}
\end{equation}
Using the linearity of the determinant in its last column, Eq.~\eqref{eq: new unitary} becomes
\begin{equation}
\begin{aligned}
    V_{MN}
    &=
    \det\left(
    \begin{array}{cc}
     u[M,N]  & 0 \\
     u[n,N] & 1
    \end{array}
    \right)
    -
    \det\left(
    \begin{array}{cc}
     u[M,N]  & u[M,n] \\
     u[n,N] & u[n,n]
    \end{array}
    \right)
    \\
    &=
    \det\left(
    \begin{array}{cc}
     u[M,N]  & -u[M,n] \\
     u[n,N] & 1-u[n,n]
    \end{array}
    \right)
    \\
    &=
    \det(1-u_{nn}) \cdot \\
     & \det\!\left[
    u[M,N]
    +
    u[M,n](1-u_{nn})^{-1}u[n,N]
    \right].
\end{aligned}
\end{equation}
In the last step, we used the Schur-complement formula. This motivates the following self-contraction rule at the single-particle level:
\begin{equation}
\label{eq: contract one mode}
    \mathrm{str}_{n} u
    =
    u_{\bar{n}\bar{n}}
    +
    u_{\bar{n}n}
    (I_{nn}-u_{nn})^{-1}
    u_{n\bar{n}},
\end{equation}
where $\bar{n}=S\setminus\{n\}$ denotes the uncontracted modes. For any $M,N\subseteq\bar{n}$,
\begin{equation}
    (\mathrm{str}_{n}u)[M,N]
    =
    u[M,N]
    +
    u[M,n](1-u_{nn})^{-1}u[n,N].
\end{equation}
Hence,
\begin{equation}
    V_{MN}
    =
    \det(1-u_{nn})\,
    \det\!\left[(\mathrm{str}_{n}u)[M,N]\right].
\end{equation}
Using the determinant representation of the corresponding many-body Gaussian operator,
\begin{equation}
    \hat{\rho}(\mathrm{str}_{n} u)
    =
    \sum_{M,N}
    \det\!\left[(\mathrm{str}_{n} u)[M,N]\right]
    |M\rangle\langle N|,
\end{equation}
we therefore obtain
\begin{equation}
    \hat{V}
    =
    \mathrm{STr}_{n}\hat{U}
    =
    \det(1-u_{nn})\,
    \hat{\rho}(\mathrm{str}_{n}u).
\end{equation}

Thus, the partial supertrace over a single Dirac mode preserves the fermionic Gaussian structure, up to the overall factor $\det(1-u_{nn})$. It remains to show that $\mathrm{str}_{n}u$ is itself unitary; this is established in the next subsection.

\subsection{Contraction over multiple modes}

We now generalize the single-mode result to the contraction of an arbitrary set of modes. The key observation is that self-contractions can be performed sequentially. For two disjoint sets of modes $B$ and $C$, we show that
\begin{equation}
    \mathrm{str}_{B}\!\left(\mathrm{str}_{C}u\right)
    =
    \mathrm{str}_{B\cup C}u.
\end{equation}
Thus, contracting the modes one by one is equivalent to contracting them simultaneously.

Let the full set of modes be $S=A\cup B\cup C$, and write
\begin{equation}
u =
\left(
    \begin{array}{ccc}
     u_{AA}  & u_{AB} & u_{AC} \\
     u_{BA} & u_{BB} & u_{BC} \\
     u_{CA} & u_{CB} & u_{CC}
   \end{array}
\right).
\end{equation}
Contracting $B\cup C$ directly gives
\begin{equation}
\begin{aligned}
    &\mathrm{str}_{B\cup C} u
    =
    u_{AA} \\
    &+
    \left(
    \begin{array}{cc}
     u_{AB}  & u_{AC}
    \end{array}
    \right)
    \left(
    \begin{array}{cc}
     I-u_{BB}  & -u_{BC} \\
     -u_{CB} & I-u_{CC}
    \end{array}
    \right)^{-1}
    \left(
    \begin{array}{c}
     u_{BA} \\
     u_{CA}
    \end{array}
    \right).
\end{aligned}
\end{equation}
For compactness, define
\begin{equation}
    \left(
    \begin{array}{cc}
     I-u_{BB}  & -u_{BC} \\
     -u_{CB} & I-u_{CC}
    \end{array}
    \right)
    =
    \left(
    \begin{array}{cc}
     a & b \\
     c & d
    \end{array}
    \right),
\end{equation}
and let
\begin{equation}
    \Sigma_d=a-bd^{-1}c
\end{equation}
denote the Schur complement of $d$. We now perform the contractions sequentially, beginning with $C$. This gives
\begin{equation}
\begin{aligned}
u'
&=
\mathrm{str}_{C}u
\\
&=
\left(
\begin{array}{c|c}
u_{AA}+u_{AC}d^{-1}u_{CA}
&
u_{AB}-u_{AC}d^{-1}c
\\
\hline
u_{BA}-bd^{-1}u_{CA}
&
u_{BB}+bd^{-1}c
\end{array}
\right)
\\
&=
\left(
\begin{array}{c|c}
u'_{AA} & u'_{AB}\\
\hline
u'_{BA} & u'_{BB}
\end{array}
\right).
\end{aligned}
\end{equation}
Since
\begin{equation}
    I-u'_{BB}
    =
    a-bd^{-1}c
    =
    \Sigma_d,
\end{equation}
the subsequent contraction over $B$ yields
\begin{equation}
\begin{aligned}
\mathrm{str}_{B}\!\left(\mathrm{str}_{C}u\right)
&=
u'_{AA}
+
u'_{AB}\Sigma_d^{-1}u'_{BA}
\\
&=
u_{AA} \\
+
\left(
\begin{array}{cc}
u_{AB} & u_{AC}
\end{array}
\right) &
\left(
\begin{array}{cc}
\Sigma_d^{-1}
&
-\Sigma_d^{-1}bd^{-1}
\\
-d^{-1}c\Sigma_d^{-1}
&
d^{-1}+d^{-1}c\Sigma_d^{-1}bd^{-1}
\end{array}
\right)
\left(
\begin{array}{c}
u_{BA}\\
u_{CA}
\end{array}
\right)
\\
&=
\mathrm{str}_{B\cup C}u,
\end{aligned}
\end{equation}
where the last equality follows from the block-matrix inverse formula in Eq.~\eqref{eq: block inverse}. Hence,
\begin{equation}
    \mathrm{str}_{B}\mathrm{str}_{C}u
    =
    \mathrm{str}_{B\cup C}u.
\end{equation}
It remains to verify that the normalization factors combine in the same way. From the sequential contractions,
\begin{equation}
\begin{aligned}
&\det(I-u'_{BB})\det(I-u_{CC})
\\
&\quad=
\det\!\left[
I-u_{BB}
-u_{BC}(I-u_{CC})^{-1}u_{CB}
\right]
\det(I-u_{CC})
\\
&\quad=
\det\left(
\begin{array}{cc}
I-u_{BB} & -u_{BC}\\
-u_{CB} & I-u_{CC}
\end{array}
\right)
\\
&\quad=
\det\!\left[
I-
\left(
\begin{array}{cc}
u_{BB} & u_{BC}\\
u_{CB} & u_{CC}
\end{array}
\right)
\right]
\\
&\quad=
\det\!\left(I-u_{B\cup C,B\cup C}\right),
\end{aligned}
\end{equation}
where we used the determinant identity associated with the Schur complement. Therefore, both the single-particle contraction and its normalization factor are independent of whether the modes are contracted sequentially or simultaneously.

\subsection{Unitarity}

We show that the matrix $\mathrm{str}_{B}u$, defined in Eq.~\eqref{Eq: Self-contraction of a unitary matrix}, remains unitary. Let
\begin{equation}
u^{\prime}
=
\mathrm{str}_{B} u
:=
u_{AA}
+
u_{AB}(I-u_{BB})^{-1}u_{BA},
\end{equation}
where we assume that $I-u_{BB}$ is invertible. To prove that $u^{\prime}$ is unitary, we evaluate
\begin{equation}
\label{eq: check unitarity}
\begin{aligned}
u^{\prime\dagger}u^{\prime}
&=
\left[
u^{\dagger}_{AA}
+
u_{BA}^{\dagger}(I-u^{\dagger}_{BB})^{-1}u_{AB}^{\dagger}
\right]
\\
&\quad\times
\left[
u_{AA}
+
u_{AB}(I-u_{BB})^{-1}u_{BA}
\right].
\end{aligned}
\end{equation}

Since $u$ is unitary,
\begin{equation}
    u^{\dagger}u
    =
    \left(
    \begin{array}{cc}
     u_{AA}^{\dagger}  & u_{BA}^{\dagger} \\
     u_{AB}^{\dagger} & u_{BB}^{\dagger}
   \end{array}
   \right)
   \left(
    \begin{array}{cc}
     u_{AA}  & u_{AB} \\
     u_{BA} & u_{BB}
   \end{array}
   \right)
   =
   \left(
    \begin{array}{cc}
     I  & 0 \\
     0 & I
   \end{array}
   \right),
\end{equation}
which implies
\begin{equation}
\label{eq: block unitary identity}
\begin{aligned}
   &u^{\dagger}_{AA}u_{AA}
   +u^{\dagger}_{BA}u_{BA}=I,\\
   &u^{\dagger}_{AA}u_{AB}
   +u^{\dagger}_{BA}u_{BB}=0,\\
   &u^{\dagger}_{AB}u_{AA}
   +u^{\dagger}_{BB}u_{BA}=0,\\
   &u^{\dagger}_{AB}u_{AB}
   +u^{\dagger}_{BB}u_{BB}=I.
\end{aligned}
\end{equation}
Using these identities in Eq.~\eqref{eq: check unitarity}, we obtain
\begin{equation}
\begin{aligned}
u^{\prime\dagger}u^{\prime}
&=
u^{\dagger}_{AA}u_{AA}
+
u^{\dagger}_{BA}
\Big[
-u_{BB}(I-u_{BB})^{-1}
\\
&\qquad
-(I-u^{\dagger}_{BB})^{-1}u^{\dagger}_{BB}
\\
&\qquad
+(I-u^{\dagger}_{BB})^{-1}
(I-u^{\dagger}_{BB}u_{BB})
(I-u_{BB})^{-1}
\Big]
u_{BA}.
\end{aligned}
\end{equation}
The expression in brackets simplifies to the identity. Indeed,
\begin{equation}
I-u^{\dagger}_{BB}u_{BB}
=
(I-u^{\dagger}_{BB})
+
u^{\dagger}_{BB}(I-u_{BB}),
\end{equation}
and hence
\begin{equation}
\begin{aligned}
&(I-u^{\dagger}_{BB})^{-1}
(I-u^{\dagger}_{BB}u_{BB})
(I-u_{BB})^{-1}
\\
&\qquad
=
(I-u_{BB})^{-1}
+
(I-u^{\dagger}_{BB})^{-1}u^{\dagger}_{BB}.
\end{aligned}
\end{equation}
Therefore,
\begin{equation}
\begin{aligned}
u^{\prime\dagger}u^{\prime}
&=
u^{\dagger}_{AA}u_{AA}
+
u^{\dagger}_{BA}
\left[
(I-u_{BB})^{-1}
-u_{BB}(I-u_{BB})^{-1}
\right]
u_{BA}
\\
&=
u^{\dagger}_{AA}u_{AA}
+
u^{\dagger}_{BA}u_{BA}
\\
&=
I.
\end{aligned}
\end{equation}
Since $u^{\prime}$ is square, this proves that
\begin{equation}
    \mathrm{str}_{B}u
\end{equation}
is unitary. Thus, self-contraction over an arbitrary set of modes preserves the single-particle unitary structure, provided that $I-u_{BB}$ is invertible.

\section{Self-contraction rule without fermion number conservation}
\label{Append: Self-contraction rule without fermion number conservation}

In this subsection, we derive the self-contraction rule for general fermionic Gaussian unitary operators without assuming fermion-number conservation. Our goal is to show that the partial supertrace of a fermionic Gaussian unitary can be evaluated directly in its orthogonal-matrix representation. For the contraction of a single Dirac mode $n$,
\begin{equation}
      \mathrm{STr}_{n}[\hat{\rho}(O)] = \mathcal{N}\hat{\rho}(\mathrm{str}_{n}[O]),
\end{equation}
where $\mathcal{N}$ is an overall normalization factor. More generally, for a set of Dirac modes $B$, the corresponding single-particle contraction is
\begin{equation}
\begin{aligned}
\mathrm{str}_{B} [O]
\coloneqq
\mathrm{str}_{\{q_{B},p_{B} \}} [O]
=
O_{AA} + O_{AB}(I-O_{BB})^{-1}O_{BA}.
\end{aligned}
\end{equation}
Because each Dirac mode consists of two Majorana modes, contracting a Dirac mode $n$ requires contracting the corresponding pair $q_{n}$ and $p_{n}$ simultaneously in the orthogonal representation.

The derivation proceeds by relating the different representations of a fermionic Gaussian unitary introduced in Appendix~\ref{Append: Fermionic Gaussian unitary operations}. The relevant transformations and contraction rules are summarized by the commuting diagram
\begin{equation}
\label{eq: commuting diagram}
    \begin{tikzcd} [row sep=large, column sep=large]
O \arrow[r, shift left, " T^{\dagger}  \cdot T"]
  \arrow[d, "\mathrm{str}_{\{q_{B}, p_{B}\}}"]
  & U  \arrow[l, shift left,  "T \cdot T^{\dagger}"]
  \arrow[r,shift left, "\mathrm{cov}"]
  \arrow[d, "\mathrm{str}_{\{c_{B}, c^{\dagger}_{B}\}}"] 
  & \Gamma \arrow[l, shift left, "\mathrm{cov}^{-1}"]
  \arrow[r,shift left, "\hat{\sigma}"]
  \arrow[d, "\tilde{\mathrm{str}}_{\{c_{B}, c^{\dagger}_{B}\}}"]
  & \hat{U} \arrow[l, shift left, "\sigma^{-1}"]
  \arrow[d, "\mathrm{Str}_{B}"]
  \\
O^{\prime} 
\arrow[r, shift left, " T^{\dagger}  \cdot T"]
  & U^{\prime} \arrow[l, shift left,  "T \cdot T^{\dagger}"]
  \arrow[r,shift left, "\mathrm{cov}"]
  & \Gamma^{\prime} \arrow[l, shift left, "\mathrm{cov}^{-1}"]
  \arrow[r,shift left, "\hat{\sigma}"]
  & \hat{U}^{\prime} \arrow[l, shift left, "\sigma^{-1}"]
\end{tikzcd}
\end{equation}

We read Eq.~\eqref{eq: commuting diagram} from right to left. We first evaluate the partial supertrace in the Fock-space representation and determine its effect on the matrix elements of $\hat{U}$. We then translate this contraction into a corresponding rule for the covariance matrix $\Gamma$. From the contracted covariance matrix, we derive the induced contraction rule for the Bogoliubov transformation $U$, and finally transform back to the Majorana basis to obtain the contraction rule for the orthogonal matrix $O$.

\subsection{Self-contraction rule for covariance matrix}

We begin by contracting a single Dirac mode $n$ of a fermionic Gaussian unitary operator. Let
\begin{equation}
    \hat{V}=\mathrm{STr}_{n}[\hat{U}].
\end{equation}
As in Eq.~\eqref{eq: new unitary}, for
$M,N\subseteq S\setminus\{n\}$ the corresponding Fock-space matrix element is
\begin{equation}
    V_{MN}
    =
    U_{MN}
    -
    U_{M\cup\{n\},\,N\cup\{n\}}.
\end{equation}
To simplify the notation, define
\begin{equation}
    X_{MN}=M_{c}\cup N_{c^{\dagger}},
    \qquad
    B_{n}=\{n_{c},n_{c^{\dagger}}\},
\end{equation}
and
\begin{equation}
    s_M=(-1)^{\frac{|M|(|M|-1)}{2}}.
\end{equation}
Using the Pfaffian representation in Eq.~\eqref{eq: matrix element of fG-unitary}, the first contribution can be written as
\begin{equation}
\begin{aligned}
    \frac{U_{NM}^*}{s_M U_{\emptyset\emptyset}^*}
    &=
    \mathrm{Pf}\!\left(
    \Gamma[X_{MN},X_{MN}]
    \right)
    \\
    &=
    \mathrm{Pf}\!\left[
    \left(
    \begin{array}{cc}
        \Gamma[X_{MN},X_{MN}] & 0\\
        0 & \Omega_{1}
    \end{array}
    \right)
    \right],
\end{aligned}
\end{equation}
where
\begin{equation}
    \Omega_{1}
    =
    \left(
    \begin{array}{cc}
        0 & 1\\
        -1 & 0
    \end{array}
    \right)
\end{equation}
is the standard antisymmetric form associated with one Dirac mode.

Similarly, applying Eq.~\eqref{eq: matrix element of fG-unitary} to
$U_{M\cup\{n\},\,N\cup\{n\}}$ and reordering the indices so that the contracted pair $B_n$ appears last gives the same overall sign factor $s_M$. Combining the two contributions, we obtain
\begin{widetext}
\begin{equation}
\label{eq: matrix element of fg unitary without number conserving}
\begin{aligned}
V^*_{NM}
&=
s_M U^*_{\emptyset\emptyset}
\mathrm{Pf}\!\left[
\left(
\begin{array}{cc}
\Gamma[X_{MN},X_{MN}]
&
\Gamma[X_{MN},B_n]
\\
\Gamma[B_n,X_{MN}]
&
\Omega_{1}-\Gamma[B_n,B_n]
\end{array}
\right)
\right]
\\
&=
s_M U^*_{\emptyset\emptyset}\,
\mathrm{Pf}\!\left(
\Omega_{1}-\Gamma[B_n,B_n]
\right)
\\
&\qquad\times
\mathrm{Pf}\!\left[
\Gamma[X_{MN},X_{MN}]
+
\Gamma[X_{MN},B_n]
\left(
\Omega_{1}-\Gamma[B_n,B_n]
\right)^{-1}
\Gamma[B_n,X_{MN}]
\right],
\end{aligned}
\end{equation}
\end{widetext}
where the second equality follows from the Pfaffian analogue of the Schur-complement identity.

Compared with the Pfaffian representation in Eq.~\eqref{eq: matrix element of fG-unitary}, Equation~\eqref{eq: matrix element of fg unitary without number conserving} naturally motivates a self-contraction rule for the covariance matrix. Consider a bipartition into uncontracted modes $A$ and contracted modes $B$, and reorder the creation and annihilation operators so that those belonging to each subsystem are grouped together:
\begin{equation}
    \Gamma
    =
    \left(
    \begin{array}{cc}
        \Gamma_{AA} & \Gamma_{AB}\\
        \Gamma_{BA} & \Gamma_{BB}
    \end{array}
    \right).
\end{equation}
We define
\begin{equation}
\label{eq: self-contraction rule for covariance matrix}
    \widetilde{\mathrm{str}}_{B}\Gamma
    :=
    \Gamma_{AA}
    +
    \Gamma_{AB}
    (\Omega_{B}-\Gamma_{BB})^{-1}
    \Gamma_{BA},
\end{equation}
where $\Omega_{B}$ is the standard antisymmetric form on the creation-annihilation modes associated with subsystem $B$.

The contracted matrix remains antisymmetric. Indeed, using
\begin{equation}
    \Gamma_{AA}^{T}=-\Gamma_{AA},
    \qquad
    \Gamma_{BB}^{T}=-\Gamma_{BB},
    \qquad
    \Gamma_{BA}=-\Gamma_{AB}^{T},
\end{equation}
together with $\Omega_{B}^{T}=-\Omega_{B}$, one immediately finds
\begin{equation}
    \left(
    \widetilde{\mathrm{str}}_{B}\Gamma
    \right)^{T}
    =
    -\widetilde{\mathrm{str}}_{B}\Gamma.
\end{equation}

The second Pfaffian in Eq.~\eqref{eq: matrix element of fg unitary without number conserving} is precisely the Pfaffian entering the Fock-space matrix elements associated with
$\widetilde{\mathrm{str}}_{n}\Gamma$. Therefore, all matrix elements of the partially contracted operator agree with those generated by the contracted covariance matrix up to a common normalization factor:
\begin{equation}
    \mathrm{STr}_{n}[\hat{U}]
    =
    \mathcal{N}\,
    \hat{\sigma}
    \left(
    \widetilde{\mathrm{str}}_{n}\Gamma
    \right),
\end{equation}
where $\hat{\sigma}$ is the map from a covariance matrix to the corresponding fermionic Gaussian operator introduced in Eq.~\eqref{eq: covariance to many-body mapping}, and $\mathcal{N}$ is independent of the Fock-space indices $M$ and $N$.

The contraction of multiple Dirac modes follows by applying the same procedure sequentially, in direct analogy with Appendix~\ref{Append: Self-contraction rule with fermion number conservation}. Consequently,
\begin{equation}
    \mathrm{STr}_{B}[\hat{U}]
    =
    \mathcal{N}\,
    \hat{\sigma}
    \left(
    \widetilde{\mathrm{str}}_{B}\Gamma
    \right).
\end{equation}
Thus, partial supertrace preserves the fermionic Gaussian structure: at the covariance-matrix level, it is represented by the fractional-linear transformation in Eq.~\eqref{eq: self-contraction rule for covariance matrix}, up to an overall normalization factor.

\subsection{Self-contraction rule for the Bogoliubov transformation}

In this subsection, we derive the self-contraction rule for the Bogoliubov transformation $U$ starting from the covariance-matrix rule in Eq.~\eqref{eq: self-contraction rule for covariance matrix}. The key observation is that the covariance matrix and the Bogoliubov matrix describe the same Gaussian transformation with different groupings of input and output operators.

We first recall the feedback interpretation of the covariance-matrix contraction. For a bipartition into subsystems $A$ and $B$, the covariance matrix acts as
\begin{equation}
\left(
    \begin{array}{c}
         \hat{c}_{A}  \\
         \hat{c}^{\dagger \prime}_{A} \\
         \hline
         \hat{c}_{B}  \\
         \hat{c}^{\dagger \prime}_{B}
    \end{array} \right)
     = 
     \left(
    \begin{array}{c|c}
         \Gamma_{AA} &  \Gamma_{AB} \\
         \hline
         \Gamma_{BA} &  \Gamma_{BB}
    \end{array}
    \right)
    \left(
    \begin{array}{c}
         -\hat{c}^{\dagger}_{A}  \\
         \hat{c}^{\prime}_{A} \\ 
         \hline
         -\hat{c}^{\dagger}_{B}  \\
         \hat{c}^{\prime}_{B}  
    \end{array}
    \right).
\end{equation}
Equivalently,
\begin{equation}
\label{eq: covariance tranformation into two eqs}
\begin{aligned}
 \left(
    \begin{array}{c}
         \hat{c}_{A}  \\
         \hat{c}^{\dagger \prime}_{A}
    \end{array} \right)
&=
\Gamma_{AA}
\left(
    \begin{array}{c}
         -\hat{c}^{\dagger}_{A}  \\
         \hat{c}^{\prime}_{A}
    \end{array}
\right)
+
\Gamma_{AB}
\left(
    \begin{array}{c}
         -\hat{c}^{\dagger}_{B}  \\
         \hat{c}^{\prime}_{B}
    \end{array}
\right),
\\
 \left(
    \begin{array}{c}
         \hat{c}_{B}  \\
         \hat{c}^{\dagger \prime}_{B}
    \end{array} \right)
&=
\Gamma_{BA}
\left(
    \begin{array}{c}
         -\hat{c}^{\dagger}_{A}  \\
         \hat{c}^{\prime}_{A}
    \end{array}
\right)
+
\Gamma_{BB}
\left(
    \begin{array}{c}
         -\hat{c}^{\dagger}_{B}  \\
         \hat{c}^{\prime}_{B}
    \end{array}
\right).
\end{aligned}
\end{equation}

Self-contraction over subsystem $B$ identifies the outgoing and incoming operators,
\begin{equation}
    \hat{c}^{\prime}_{B}=\hat{c}_{B},
    \qquad
    \hat{c}^{\dagger\prime}_{B}=\hat{c}^{\dagger}_{B}.
\end{equation}
In the covariance-matrix convention, this condition is written as
\begin{equation}
    \left(
    \begin{array}{c}
         \hat{c}_{B}  \\
         \hat{c}^{\dagger \prime}_{B}
    \end{array} \right) 
    =
    \Omega
    \left(
    \begin{array}{c}
         -\hat{c}^{\dagger}_{B}  \\
         \hat{c}^{\prime}_{B}
    \end{array} \right),
\qquad
\Omega=
\left(
\begin{array}{cc}
0&I\\
-I&0
\end{array}
\right).
\end{equation}
Substituting this condition into the second equation of Eq.~\eqref{eq: covariance tranformation into two eqs} gives
\begin{equation}
\left(
    \begin{array}{c}
         -\hat{c}^{\dagger}_{B}  \\
         \hat{c}^{\prime}_{B}
    \end{array} \right)
=
(\Omega-\Gamma_{BB})^{-1}
\Gamma_{BA}
\left(
    \begin{array}{c}
         -\hat{c}^{\dagger}_{A}  \\
         \hat{c}^{\prime}_{A}
    \end{array}
\right).
\end{equation}
Substitution into the first equation of Eq.~\eqref{eq: covariance tranformation into two eqs} then yields
\begin{equation}
     \left(
    \begin{array}{c}
         \hat{c}_{A}  \\
         \hat{c}^{\dagger \prime}_{A}
    \end{array} \right)
=
\left[
\Gamma_{AA}
+
\Gamma_{AB}
(\Omega-\Gamma_{BB})^{-1}
\Gamma_{BA}
\right]
\left(
    \begin{array}{c}
         -\hat{c}^{\dagger}_{A}  \\
         \hat{c}^{\prime}_{A}
    \end{array}
\right).
\end{equation}
Thus, the covariance-matrix contraction
\begin{equation}
    \widetilde{\mathrm{str}}_{B}\Gamma
    =
    \Gamma_{AA}
    +
    \Gamma_{AB}
    (\Omega-\Gamma_{BB})^{-1}
    \Gamma_{BA}
\end{equation}
is precisely the effective transformation obtained after closing subsystem $B$ into a feedback loop.

We now apply the same reasoning directly to the Bogoliubov transformation. For the bipartite system,
\begin{equation}
\label{eq: covariance tranformation matrix form}
\left(
    \begin{array}{c}
         \hat{c}^{\prime}_{A}  \\
         \hat{c}^{\dagger \prime}_{A} \\
         \hline
         \hat{c}^{\prime}_{B}  \\
         \hat{c}^{\dagger \prime}_{B}
    \end{array} \right)
     = 
     \left(
    \begin{array}{c|c}
         U_{AA} &  U_{AB} \\
         \hline
         U_{BA} &  U_{BB}
    \end{array}
    \right)
    \left(
    \begin{array}{c}
         \hat{c}_{A}  \\
         \hat{c}^{\dagger}_{A} \\
         \hline
         \hat{c}_{B}  \\
         \hat{c}^{\dagger}_{B}
    \end{array}
    \right).
\end{equation}
This gives
\begin{equation}
\label{eq: Bogoliubov tranformation into two eqs}
\begin{aligned}
\left(
    \begin{array}{c}
         \hat{c}^{\prime}_{A}  \\
         \hat{c}^{\dagger \prime}_{A}
    \end{array} \right)
&=
U_{AA}
\left(
    \begin{array}{c}
         \hat{c}_{A}  \\
         \hat{c}^{\dagger}_{A}
    \end{array}
\right)
+
U_{AB}
\left(
    \begin{array}{c}
         \hat{c}_{B}  \\
         \hat{c}^{\dagger}_{B}
    \end{array}
\right),
\\
\left(
    \begin{array}{c}
         \hat{c}^{\prime}_{B}  \\
         \hat{c}^{\dagger \prime}_{B}
    \end{array} \right)
&=
U_{BA}
\left(
    \begin{array}{c}
         \hat{c}_{A}  \\
         \hat{c}^{\dagger}_{A}
    \end{array}
\right)
+
U_{BB}
\left(
    \begin{array}{c}
         \hat{c}_{B}  \\
         \hat{c}^{\dagger}_{B}
    \end{array}
\right).
\end{aligned}
\end{equation}
The same feedback condition,
\begin{equation}
    \left(
    \begin{array}{c}
         \hat{c}^{\prime}_{B}  \\
         \hat{c}^{\dagger \prime}_{B}
    \end{array} \right)
=
\left(
    \begin{array}{c}
         \hat{c}_{B}  \\
         \hat{c}^{\dagger}_{B}
    \end{array} \right),
\end{equation}
implies the following from the second equation of Eq.~\eqref{eq: Bogoliubov tranformation into two eqs}
\begin{equation}
    \left(
    \begin{array}{c}
         \hat{c}_{B}  \\
         \hat{c}^{\dagger}_{B}
    \end{array} \right)
=
(I-U_{BB})^{-1}U_{BA}
\left(
    \begin{array}{c}
         \hat{c}_{A}  \\
         \hat{c}^{\dagger}_{A}
    \end{array}
\right).
\end{equation}
Substituting this result into the first equation of Eq.~\eqref{eq: Bogoliubov tranformation into two eqs}, we obtain
\begin{equation}
    \left(
    \begin{array}{c}
         \hat{c}^{\prime}_{A}  \\
         \hat{c}^{\dagger \prime}_{A}
    \end{array} \right)
=
\left[
U_{AA}
+
U_{AB}(I-U_{BB})^{-1}U_{BA}
\right]
\left(
    \begin{array}{c}
         \hat{c}_{A}  \\
         \hat{c}^{\dagger}_{A}
    \end{array}
\right).
\end{equation}
This motivates the self-contraction rule
\begin{equation}
    \mathrm{str}_{B}U
    :=
    U_{AA}
    +
    U_{AB}(I-U_{BB})^{-1}U_{BA}.
\end{equation}
Equivalently, since both the creation and annihilation operators associated with the Dirac modes in $B$ are contracted, we may write
\begin{equation}
    \mathrm{str}_{B}U
    =
    \mathrm{str}_{\{c_{B},c_{B}^{\dagger}\}}U.
\end{equation}

Thus, both $\widetilde{\mathrm{str}}_{B}\Gamma$ and $\mathrm{str}_{B}U$ describe the same feedback contraction in their respective representations. Since the covariance matrix and the Bogoliubov matrix encode the same Gaussian transformation, their contraction rules are compatible:
\begin{equation}
    \mathrm{cov}\!\left(\mathrm{str}_{B}U\right)
    =
    \widetilde{\mathrm{str}}_{B}\!\left(\mathrm{cov}(U)\right).
\end{equation}
This identity provides the desired self-contraction rule for the Bogoliubov representation.

\subsection{Self-contraction rule for the orthogonal matrix}

In the previous subsection, we established the self-contraction rule for a Bogoliubov transformation $U$, where both the creation and annihilation operators associated with the Dirac modes in subsystem $B$ are contracted:
\begin{equation}
    \mathrm{str}_{B}U
    =
    \mathrm{str}_{\{c_{B},c_{B}^{\dagger}\}}U.
\end{equation}
We now show that the same contraction can be performed directly in the Majorana basis $(\mathbf{\hat{q}},\mathbf{\hat{p}})$. In this basis, the Gaussian transformation is represented by an orthogonal matrix $O$,
\begin{equation}
   \left(
    \begin{array}{c}
     \mathbf{\hat{q}}^{\prime}   \\
     \mathbf{\hat{p}}^{\prime}
   \end{array} 
   \right)
   =
   O
   \left(
    \begin{array}{c}
     \mathbf{\hat{q}}   \\
     \mathbf{\hat{p}}
   \end{array} 
   \right).
\end{equation}
Contracting a Dirac mode in subsystem $B$ therefore corresponds to contracting its two associated Majorana modes, $q_{B}$ and $p_{B}$.

Before performing the contraction, we reorder the modes so that the degrees of freedom belonging to subsystems $A$ and $B$ are grouped together:
\begin{equation}
\begin{aligned}
    &(\mathbf{q},\mathbf{p})
    \rightarrow
    (\mathbf{q}_{A},\mathbf{p}_{A},\mathbf{q}_{B},\mathbf{p}_{B}),\\
    &(\mathbf{c},\mathbf{c}^{\dagger})
    \rightarrow
    (\mathbf{c}_{A},\mathbf{c}^{\dagger}_{A},
    \mathbf{c}_{B},\mathbf{c}^{\dagger}_{B}).
\end{aligned}
\end{equation}
With this ordering, the basis transformation between the Majorana and creation-annihilation representations becomes block diagonal,
\begin{equation}
    T =
    \left(
    \begin{array}{cc}
     T_{A}  & 0 \\
     0 & T_{B}
   \end{array} 
   \right).
\end{equation}
The two single-particle representations are related by
\begin{equation}
    O=TUT^{\dagger},
    \qquad
    U=T^{\dagger}OT.
\end{equation}
Accordingly,
\begin{equation}
\begin{aligned}
    U
    &= T^{\dagger}OT \\
    &=
    \left(
    \begin{array}{cc}
     T^{\dagger}_{A}  & 0 \\
     0 & T^{\dagger}_{B}
   \end{array} 
   \right)
   \left(
    \begin{array}{cc}
     O_{AA}  & O_{AB} \\
     O_{BA} & O_{BB}
   \end{array} 
   \right)
   \left(
    \begin{array}{cc}
     T_{A}  & 0 \\
     0 & T_{B}
   \end{array} 
   \right)\\
    &=
    \left(
    \begin{array}{cc}
    T^{\dagger}_{A}O_{AA}T_{A}
    &
    T^{\dagger}_{A}O_{AB}T_{B}
    \\
    T^{\dagger}_{B}O_{BA}T_{A}
    &
    T^{\dagger}_{B}O_{BB}T_{B}
   \end{array} 
   \right).
\end{aligned}
\end{equation}

Applying the Bogoliubov self-contraction rule then gives
\begin{equation}
\label{eq: from u to o}
\begin{aligned}
  & \quad \mathrm{str}_{B}U \\
  &=
  \mathrm{str}_{\{c_{B},c_{B}^{\dagger}\}}
  \left[T^{\dagger}OT\right]
  \\
  &=
  T^{\dagger}_{A}O_{AA}T_{A}
  +
  T^{\dagger}_{A}O_{AB}T_{B}
  \left(
  I-T^{\dagger}_{B}O_{BB}T_{B}
  \right)^{-1}
  T^{\dagger}_{B}O_{BA}T_{A}
  \\
  &=
  T^{\dagger}_{A}
  \left[
  O_{AA}
  +
  O_{AB}(I-O_{BB})^{-1}O_{BA}
  \right]
  T_{A}
  \\
  &=
  T_{A}^{\dagger}
  \left(
  \mathrm{str}_{\{q_{B},p_{B}\}}[O]
  \right)
  T_{A}.
\end{aligned}
\end{equation}
In the third equality, we used the unitarity of $T_{B}$ to obtain
\begin{equation}
    \left(
    I-T_{B}^{\dagger}O_{BB}T_{B}
    \right)^{-1}
    =
    T_{B}^{\dagger}
    (I-O_{BB})^{-1}
    T_{B}.
\end{equation}

This motivates the self-contraction rule for an orthogonal matrix:
\begin{equation}
\begin{aligned}
\mathrm{str}_{B}[O]
\coloneqq
\mathrm{str}_{\{q_{B},p_{B}\}}[O]
=
O_{AA}
+
O_{AB}(I-O_{BB})^{-1}O_{BA}.
\end{aligned}
\end{equation}
Equation~\eqref{eq: from u to o} therefore implies
\begin{equation}
   T_{A}^{\dagger}
   \left(
   \mathrm{str}_{\{q_{B},p_{B}\}}[O]
   \right)
   T_{A}
   =
   \mathrm{str}_{\{c_{B},c_{B}^{\dagger}\}}
   \left[T^{\dagger}OT\right].
\end{equation}
Thus, self-contraction is compatible with the change of basis between the Majorana and creation-annihilation representations. Equivalently,
\begin{equation}
    \mathrm{str}_{B}U
    =
    T_{A}^{\dagger}
    \left(\mathrm{str}_{B}O\right)
    T_{A},
\end{equation}
showing that the contraction may be evaluated directly in either representation.

\section{Compression of the uniform evolution tensor}
\label{app:mpo-compression}

We compress the local tensor of a translation-invariant time-evolution operator by adapting the transfer-operator method of Ref.~\cite{Pirvu_2010}.

Let $W^{pq}$ denote the local MPO matrices for a single time step and $A^{pq}$ the compressed matrices accumulated from the preceding steps. Adding one layer gives
\begin{equation}
 C^{pq}=\sum_s W^{ps}\otimes A^{sq},
\end{equation}
where $p$ and $q$ label the physical output and input indices, respectively. Stacking the two layers therefore increases the virtual dimension to
\begin{equation}
    D=D_WD_A,
\end{equation}
although many of the resulting virtual directions may carry little weight in the thermodynamic environment.

Treating $(p,q)$ as a single operator-space index, we define the transfer map and its adjoint as
\begin{equation}
 \mathcal{E}(X)=\sum_{p,q}C^{pq}X(C^{pq})^\dagger,
 \qquad
 \mathcal{E}^\dagger(X)=\sum_{p,q}(C^{pq})^\dagger X C^{pq}.
\end{equation}
Their positive dominant right and left eigenmatrices satisfy
\begin{equation}
 \mathcal{E}(R)=\lambda R,
 \qquad
 \mathcal{E}^\dagger(L)=\lambda L .
\end{equation}
These matrices characterize the virtual environments seen from the two sides of a bond in the infinite uniform MPO.

Since $R\geq0$, its square root is well defined. Restricting to the numerical support of $R$, we introduce the gauge transformation
\begin{equation}
 B^{pq}=R^{-1/2}C^{pq}R^{1/2}.
\end{equation}
This transformation is only a change of basis on the virtual indices before truncation: neighboring factors $R^{1/2}R^{-1/2}$ cancel in the MPO contraction. In this gauge,
\begin{equation}
 \sum_{p,q}B^{pq}(B^{pq})^\dagger=\lambda I,
\end{equation}
so the right environment is proportional to the identity. The remaining nontrivial virtual environment is encoded by
\begin{equation}
 \rho=R^{1/2}LR^{1/2}
     =V\,\operatorname{diag}(\omega_1,\omega_2,\ldots)\,V^\dagger,
\end{equation}
where
\begin{equation}
    \omega_1\geq\omega_2\geq\cdots\geq0.
\end{equation}
The eigenvectors of $\rho$ identify orthogonal directions selected by the infinite-chain environment, while the eigenvalues $\omega_a$ quantify their corresponding weights. By retaining the $r$ largest weights, we obtain the rank-$r$ subspace
\begin{equation}
 P=V_{[:,1:r]} .
\end{equation}
The compressed local tensor is then
\begin{equation}
 \widetilde C^{pq}=P^\dagger B^{pq}P.
\end{equation}
The discarded weight
\begin{equation}
    \delta_r=
    \frac{\sum_{a>r}\omega_a}{\sum_a\omega_a}
\end{equation}
measures the fraction of the dominant-environment weight removed by the rank-$r$ projection. After compression, the bond dimension of the local tensor is reduced to $r$.

For fermionic tensors, the fermionic signs are incorporated into the local tensor before contraction by fixing an ordering of the physical and virtual modes. Exchanging two groups of modes with occupation parities $p_i,p_j\in\{0,1\}$ contributes a factor $(-1)^{p_i p_j}$. Multiplying the relevant tensor elements by the corresponding factors accounts for the fermionic reordering signs, after which the remaining contractions can be carried out using ordinary tensor operations. For periodic boundary conditions, an additional parity-dependent factor is included when closing the virtual ring to account for fermionic exchanges across the boundary.

To preserve fermionic parity during compression, we impose
the selection rule
\[
C^{pq}_{\alpha\beta}\neq 0
\quad\Rightarrow\quad
\pi(\alpha)+\pi(\beta)
= \pi(p)+\pi(q)\pmod 2,
\]
where $\pi$ denotes the parity of each tensor index.  We solve the transfer fixed-point equations
within the parity-block-diagonal subspace and diagonalize
$\rho$ separately in the even and odd sectors. The $r$
largest weights across both sectors are retained, together
with their parity labels. For periodic boundary conditions,
the virtual trace includes the parity-dependent factor
\[
(K_\nu)_{aa}=(-1)^{\eta_a(\nu+1)},
\qquad \nu=N\bmod 2.
\]

An important caveat is that the compression criterion is defined with respect to the thermodynamic-limit environment. It therefore does not, by itself, provide a bound on finite-ring errors, guarantee preservation of global unitarity, or ensure accurate reproduction of subleading transfer-matrix eigenvalues and their associated correlation lengths. These quantities must be assessed separately when they are relevant.

\bibliography{bib}

@article{bravyi2004lagrangian,
  title={Lagrangian representation for fermionic linear optics},
  author={Bravyi, Sergey},
  journal={arXiv preprint quant-ph/0404180},
  year={2004}
}

@article{Hackl_2021,
   title={Bosonic and fermionic Gaussian states from Kähler structures},
   volume={4},
   ISSN={2666-9366},
   url={http://dx.doi.org/10.21468/SciPostPhysCore.4.3.025},
   DOI={10.21468/scipostphyscore.4.3.025},
   number={3},
   journal={SciPost Physics Core},
   publisher={Stichting SciPost},
   author={Hackl, Lucas and Bianchi, Eugenio},
   year={2021},
   month=Sept }

@article{surace2022fermionic,
  title={Fermionic Gaussian states: an introduction to numerical approaches},
  author={Surace, Jacopo and Tagliacozzo, Luca},
  journal={SciPost Physics Lecture Notes},
  pages={054},
  year={2022}
}

@misc{ren2025graphicalcalculusfermionictensors,
      title={Graphical Calculus for Fermionic Tensors}, 
      author={Yuanjie Ren and Kaifeng Bu and Andreas Bauer},
      year={2025},
      eprint={2508.03976},
      archivePrefix={arXiv},
      primaryClass={quant-ph},
      url={https://arxiv.org/abs/2508.03976}, 
}

@article{mizusaki_grassmann_2013,
	title = {Grassmann integral and {Balian}–{Brézin} decomposition in {Hartree}–{Fock}–{Bogoliubov} matrix elements},
	volume = {725},
	issn = {0370-2693},
	url = {https://www.sciencedirect.com/science/article/pii/S0370269313005558},
	doi = {https://doi.org/10.1016/j.physletb.2013.07.005},
	number = {1},
	journal = {Physics Letters B},
	author = {Mizusaki, Takahiro and Oi, Makito and Chen, Fang-Qi and Sun, Yang},
	year = {2013},
	pages = {175--179},
}

@misc{rajabpour2025matrixelementsfermionicgaussian,
      title={Matrix Elements of Fermionic Gaussian Operators in Arbitrary Pauli Bases: A Pfaffian Formula}, 
      author={M. A. Rajabpour and MirAdel Seifi MirJafarlou and Reyhaneh Khasseh},
      year={2025},
      eprint={2506.02809},
      archivePrefix={arXiv},
      primaryClass={quant-ph},
      url={https://arxiv.org/abs/2506.02809}, 
}

@article{hara_quantum_1979,
	title = {On the quantum number projection: ({I}). {General} theory},
	volume = {332},
	issn = {0375-9474},
	shorttitle = {On the quantum number projection},
	url = {https://www.sciencedirect.com/science/article/pii/0375947479900940},
	doi = {10.1016/0375-9474(79)90094-0},
	number = {1},
	urldate = {2026-03-13},
	journal = {Nuclear Physics A},
	author = {Hara, K. and Iwasaki, S.},
	month = dec,
	year = {1979},
	pages = {61--68},
}

@article{Vidal_Quantum_2021,
  title = {Quantum operations in an information theory for fermions},
  author = {Vidal, Nicetu Tibau and Bera, Mohit Lal and Riera, Arnau and Lewenstein, Maciej and Bera, Manabendra Nath},
  journal = {Phys. Rev. A},
  volume = {104},
  issue = {3},
  pages = {032411},
  numpages = {23},
  year = {2021},
  month = {Sep},
  publisher = {American Physical Society},
  doi = {10.1103/PhysRevA.104.032411},
  url = {https://link.aps.org/doi/10.1103/PhysRevA.104.032411}
}

@article{Wick_Intrinsic_1952,
  title = {The Intrinsic Parity of Elementary Particles},
  author = {Wick, G. C. and Wightman, A. S. and Wigner, E. P.},
  journal = {Phys. Rev.},
  volume = {88},
  issue = {1},
  pages = {101--105},
  numpages = {0},
  year = {1952},
  month = {Oct},
  publisher = {American Physical Society},
  doi = {10.1103/PhysRev.88.101},
  url = {https://link.aps.org/doi/10.1103/PhysRev.88.101}
}

@article{redheffer1959difference,
  title={Difference and functional equations in transmission line theory},
  author={Redheffer, Raymond},
  journal={Proceedings of the IRE},
  year={1959}
}

@book{zhou1996robust,
  title={Robust and optimal control},
  author={Zhou, Kemin and Doyle, John Comstock and Glover, Keith and others},
  volume={40},
  year={1996},
  publisher={Prentice hall New Jersey}
}

@article{Qi2006Topological,
  title = {Topological quantization of the spin Hall effect in two-dimensional paramagnetic semiconductors},
  author = {Qi, Xiao-Liang and Wu, Yong-Shi and Zhang, Shou-Cheng},
  journal = {Phys. Rev. B},
  volume = {74},
  issue = {8},
  pages = {085308},
  numpages = {7},
  year = {2006},
  month = {Aug},
  publisher = {American Physical Society},
  doi = {10.1103/PhysRevB.74.085308},
  url = {https://link.aps.org/doi/10.1103/PhysRevB.74.085308}
}

@article{Dubail2015Tensor,
  title = {Tensor network trial states for chiral topological phases in two dimensions and a no-go theorem in any dimension},
  author = {Dubail, J. and Read, N.},
  journal = {Phys. Rev. B},
  volume = {92},
  issue = {20},
  pages = {205307},
  numpages = {15},
  year = {2015},
  month = {Nov},
  publisher = {American Physical Society},
  doi = {10.1103/PhysRevB.92.205307},
  url = {https://link.aps.org/doi/10.1103/PhysRevB.92.205307}
}

@article{Wahl2013Projected,
  title = {Projected Entangled-Pair States Can Describe Chiral Topological States},
  author = {Wahl, T. B. and Tu, H.-H. and Schuch, N. and Cirac, J. I.},
  journal = {Phys. Rev. Lett.},
  volume = {111},
  issue = {23},
  pages = {236805},
  numpages = {5},
  year = {2013},
  month = {Dec},
  publisher = {American Physical Society},
  doi = {10.1103/PhysRevLett.111.236805},
  url = {https://link.aps.org/doi/10.1103/PhysRevLett.111.236805}
}

@article{Vidal2008Class,
  title = {Class of Quantum Many-Body States That Can Be Efficiently Simulated},
  author = {Vidal, G.},
  journal = {Phys. Rev. Lett.},
  volume = {101},
  issue = {11},
  pages = {110501},
  numpages = {4},
  year = {2008},
  month = {Sep},
  publisher = {American Physical Society},
  doi = {10.1103/PhysRevLett.101.110501},
  url = {https://link.aps.org/doi/10.1103/PhysRevLett.101.110501}
}

@article{Vidal2007Entanglement,
  title = {Entanglement Renormalization},
  author = {Vidal, G.},
  journal = {Phys. Rev. Lett.},
  volume = {99},
  issue = {22},
  pages = {220405},
  numpages = {4},
  year = {2007},
  month = {Nov},
  publisher = {American Physical Society},
  doi = {10.1103/PhysRevLett.99.220405},
  url = {https://link.aps.org/doi/10.1103/PhysRevLett.99.220405}
}

@article{Eisert2010Colloquium,
  title = {Colloquium: Area laws for the entanglement entropy},
  author = {Eisert, J. and Cramer, M. and Plenio, M. B.},
  journal = {Rev. Mod. Phys.},
  volume = {82},
  issue = {1},
  pages = {277--306},
  numpages = {0},
  year = {2010},
  month = {Feb},
  publisher = {American Physical Society},
  doi = {10.1103/RevModPhys.82.277},
  url = {https://link.aps.org/doi/10.1103/RevModPhys.82.277}
}

@article{Hastings_2007,
   title={An area law for one-dimensional quantum systems},
   volume={2007},
   ISSN={1742-5468},
   url={http://dx.doi.org/10.1088/1742-5468/2007/08/P08024},
   DOI={10.1088/1742-5468/2007/08/p08024},
   number={08},
   journal={Journal of Statistical Mechanics: Theory and Experiment},
   publisher={IOP Publishing},
   author={Hastings, M B},
   year={2007},
   month=Aug, pages={P08024–P08024} }

@misc{verstraete2004renormalizationalgorithmsquantummanybody,
      title={Renormalization algorithms for Quantum-Many Body Systems in two and higher dimensions}, 
      author={F. Verstraete and J. I. Cirac},
      year={2004},
      eprint={cond-mat/0407066},
      archivePrefix={arXiv},
      primaryClass={cond-mat.str-el},
      url={https://arxiv.org/abs/cond-mat/0407066}, 
}

@article{Verstraete_2008,
   title={Matrix product states, projected entangled pair states, and variational renormalization group methods for quantum spin systems},
   volume={57},
   ISSN={1460-6976},
   url={http://dx.doi.org/10.1080/14789940801912366},
   DOI={10.1080/14789940801912366},
   number={2},
   journal={Advances in Physics},
   publisher={Informa UK Limited},
   author={Verstraete, F. and Murg, V. and Cirac, J.I.},
   year={2008},
   month=Mar, pages={143–224} }

@article{Cirac2021Matrix,
  title = {Matrix product states and projected entangled pair states: Concepts, symmetries, theorems},
  author = {Cirac, J. Ignacio and P\'erez-Garc\'{\i}a, David and Schuch, Norbert and Verstraete, Frank},
  journal = {Rev. Mod. Phys.},
  volume = {93},
  issue = {4},
  pages = {045003},
  numpages = {65},
  year = {2021},
  month = {Dec},
  publisher = {American Physical Society},
  doi = {10.1103/RevModPhys.93.045003},
  url = {https://link.aps.org/doi/10.1103/RevModPhys.93.045003}
}

@article{perezgarcia2007matrixproductstaterepresentations,
author = {Perez-Garcia, D. and Verstraete, F. and Wolf, M. M. and Cirac, J. I.},
title = {Matrix product state representations},
year = {2007},
issue_date = {July 2007},
publisher = {Rinton Press, Incorporated},
address = {Paramus, NJ},
volume = {7},
number = {5},
issn = {1533-7146},
journal = {Quantum Info. Comput.},
month = jul,
pages = {401–430},
numpages = {30}
}

@article{Schollw_ck_2011,
   title={The density-matrix renormalization group in the age of matrix product states},
   volume={326},
   ISSN={0003-4916},
   url={http://dx.doi.org/10.1016/j.aop.2010.09.012},
   DOI={10.1016/j.aop.2010.09.012},
   number={1},
   journal={Annals of Physics},
   publisher={Elsevier BV},
   author={Schollwöck, Ulrich},
   year={2011},
   month=Jan, pages={96–192} }

@article{Haghshenas2021Optimization,
  title = {Optimization schemes for unitary tensor-network circuit},
  author = {Haghshenas, Reza},
  journal = {Phys. Rev. Res.},
  volume = {3},
  issue = {2},
  pages = {023148},
  numpages = {10},
  year = {2021},
  month = {May},
  publisher = {American Physical Society},
  doi = {10.1103/PhysRevResearch.3.023148},
  url = {https://link.aps.org/doi/10.1103/PhysRevResearch.3.023148}
}

@article{Haghshenas2022Variational,
  title = {Variational Power of Quantum Circuit Tensor Networks},
  author = {Haghshenas, Reza and Gray, Johnnie and Potter, Andrew C. and Chan, Garnet Kin-Lic},
  journal = {Phys. Rev. X},
  volume = {12},
  issue = {1},
  pages = {011047},
  numpages = {14},
  year = {2022},
  month = {Mar},
  publisher = {American Physical Society},
  doi = {10.1103/PhysRevX.12.011047},
  url = {https://link.aps.org/doi/10.1103/PhysRevX.12.011047}
}

@article{Chen2011Two-dimensional,
  title = {Two-dimensional symmetry-protected topological orders and their protected gapless edge excitations},
  author = {Chen, Xie and Liu, Zheng-Xin and Wen, Xiao-Gang},
  journal = {Phys. Rev. B},
  volume = {84},
  issue = {23},
  pages = {235141},
  numpages = {13},
  year = {2011},
  month = {Dec},
  publisher = {American Physical Society},
  doi = {10.1103/PhysRevB.84.235141},
  url = {https://link.aps.org/doi/10.1103/PhysRevB.84.235141}
}

@article{Ignacio_Cirac_2017,
    doi = {10.1088/1742-5468/aa7e55},
    url = {https://doi.org/10.1088/1742-5468/aa7e55},
    year = {2017},
    month = {aug},
    publisher = {IOP Publishing and SISSA},
    volume = {2017},
    number = {8},
    pages = {083105},
    author = {Ignacio Cirac, J and Perez-Garcia, David and Schuch, Norbert and Verstraete, Frank},
    title = {Matrix product unitaries: structure, symmetries, and topological invariants},
    journal = {Journal of Statistical Mechanics: Theory and Experiment},
}

@article{Burak2018Matrix,
  title = {Matrix product representation of locality preserving unitaries},
  author = {\ifmmode \mbox{\c{S}}\else \c{S}\fi{}ahino\ifmmode \breve{g}\else \u{g}\fi{}lu, M. Burak and Shukla, Sujeet K. and Bi, Feng and Chen, Xie},
  journal = {Phys. Rev. B},
  volume = {98},
  issue = {24},
  pages = {245122},
  numpages = {14},
  year = {2018},
  month = {Dec},
  publisher = {American Physical Society},
  doi = {10.1103/PhysRevB.98.245122},
  url = {https://link.aps.org/doi/10.1103/PhysRevB.98.245122}
}

@misc{xie2025unitarynetworktensornetwork,
      title={Unitary network: Tensor network unitaries with local unitarity}, 
      author={Wenqing Xie and Seishiro Ono and Hoi Chun Po},
      year={2025},
      eprint={2508.16890},
      archivePrefix={arXiv},
      primaryClass={quant-ph},
      url={https://arxiv.org/abs/2508.16890}, 
}

@article{Calabrese_2005,
doi = {10.1088/1742-5468/2005/04/P04010},
url = {https://doi.org/10.1088/1742-5468/2005/04/P04010},
year = {2005},
month = {apr},
publisher = {},
volume = {2005},
number = {04},
pages = {P04010},
author = {Calabrese, Pasquale and Cardy, John},
title = {Evolution of entanglement entropy in one-dimensional systems},
journal = {Journal of Statistical Mechanics: Theory and Experiment},
}

@article{Fagotti2008Evolution,
  title = {Evolution of entanglement entropy following a quantum quench: Analytic results for the $XY$ chain in a transverse magnetic field},
  author = {Fagotti, Maurizio and Calabrese, Pasquale},
  journal = {Phys. Rev. A},
  volume = {78},
  issue = {1},
  pages = {010306(R)},
  numpages = {4},
  year = {2008},
  month = {Jul},
  publisher = {American Physical Society},
  doi = {10.1103/PhysRevA.78.010306},
  url = {https://link.aps.org/doi/10.1103/PhysRevA.78.010306}
}

@article{Alba_2017,
   title={Entanglement and thermodynamics after a quantum quench in integrable systems},
   volume={114},
   ISSN={1091-6490},
   url={http://dx.doi.org/10.1073/pnas.1703516114},
   DOI={10.1073/pnas.1703516114},
   number={30},
   journal={Proceedings of the National Academy of Sciences},
   publisher={Proceedings of the National Academy of Sciences},
   author={Alba, Vincenzo and Calabrese, Pasquale},
   year={2017},
   month=July, pages={7947–7951} }

@article{Bultinck2017Fermionic,
  title = {Fermionic matrix product states and one-dimensional topological phases},
  author = {Bultinck, Nick and Williamson, Dominic J. and Haegeman, Jutho and Verstraete, Frank},
  journal = {Phys. Rev. B},
  volume = {95},
  issue = {7},
  pages = {075108},
  numpages = {20},
  year = {2017},
  month = {Feb},
  publisher = {American Physical Society},
  doi = {10.1103/PhysRevB.95.075108},
  url = {https://link.aps.org/doi/10.1103/PhysRevB.95.075108}
}

@article{Kraus2010Fermionic,
  title = {Fermionic projected entangled pair states},
  author = {Kraus, Christina V. and Schuch, Norbert and Verstraete, Frank and Cirac, J. Ignacio},
  journal = {Phys. Rev. A},
  volume = {81},
  issue = {5},
  pages = {052338},
  numpages = {6},
  year = {2010},
  month = {May},
  publisher = {American Physical Society},
  doi = {10.1103/PhysRevA.81.052338},
  url = {https://link.aps.org/doi/10.1103/PhysRevA.81.052338}
}

@article{Chiral2024Niu,
  title = {Chiral spin liquids with projected Gaussian fermionic entangled pair states},
  author = {Niu, Sen and Li, Jheng-Wei and Chen, Ji-Yao and Poilblanc, Didier},
  journal = {Phys. Rev. B},
  volume = {109},
  issue = {8},
  pages = {L081107},
  numpages = {6},
  year = {2024},
  month = {Feb},
  publisher = {American Physical Society},
  doi = {10.1103/PhysRevB.109.L081107},
  url = {https://link.aps.org/doi/10.1103/PhysRevB.109.L081107}
}

@article{Tensor2022Mortier,
  title = {Tensor Networks Can Resolve Fermi Surfaces},
  author = {Mortier, Quinten and Schuch, Norbert and Verstraete, Frank and Haegeman, Jutho},
  journal = {Phys. Rev. Lett.},
  volume = {129},
  issue = {20},
  pages = {206401},
  numpages = {6},
  year = {2022},
  month = {Nov},
  publisher = {American Physical Society},
  doi = {10.1103/PhysRevLett.129.206401},
  url = {https://link.aps.org/doi/10.1103/PhysRevLett.129.206401}
}

@article{Fermionic2009Corboz,
  title = {Fermionic multiscale entanglement renormalization ansatz},
  author = {Corboz, Philippe and Vidal, Guifr\'e},
  journal = {Phys. Rev. B},
  volume = {80},
  issue = {16},
  pages = {165129},
  numpages = {12},
  year = {2009},
  month = {Oct},
  publisher = {American Physical Society},
  doi = {10.1103/PhysRevB.80.165129},
  url = {https://link.aps.org/doi/10.1103/PhysRevB.80.165129}
}

@misc{gu2010grassmanntensornetworkstates,
      title={Grassmann tensor network states and its renormalization for strongly correlated fermionic and bosonic states}, 
      author={Zheng-Cheng Gu and Frank Verstraete and Xiao-Gang Wen},
      year={2010},
      eprint={1004.2563},
      archivePrefix={arXiv},
      primaryClass={cond-mat.str-el},
      url={https://arxiv.org/abs/1004.2563}, 
}

@article{Simulation2010Corboz,
  title = {Simulation of strongly correlated fermions in two spatial dimensions with fermionic projected entangled-pair states},
  author = {Corboz, Philippe and Or\'us, Rom\'an and Bauer, Bela and Vidal, Guifr\'e},
  journal = {Phys. Rev. B},
  volume = {81},
  issue = {16},
  pages = {165104},
  numpages = {22},
  year = {2010},
  month = {Apr},
  publisher = {American Physical Society},
  doi = {10.1103/PhysRevB.81.165104},
  url = {https://link.aps.org/doi/10.1103/PhysRevB.81.165104}
}

@article{interacting2010Corboz,
  title = {Simulation of interacting fermions with entanglement renormalization},
  author = {Corboz, Philippe and Evenbly, Glen and Verstraete, Frank and Vidal, Guifr\'e},
  journal = {Phys. Rev. A},
  volume = {81},
  issue = {1},
  pages = {010303(R)},
  numpages = {4},
  year = {2010},
  month = {Jan},
  publisher = {American Physical Society},
  doi = {10.1103/PhysRevA.81.010303},
  url = {https://link.aps.org/doi/10.1103/PhysRevA.81.010303}
}

@misc{wu2025alternatinggaussianfermionicisometric,
      title={Alternating and Gaussian fermionic Isometric Tensor Network States}, 
      author={Yantao Wu and Zhehao Dai and Sajant Anand and Sheng-Hsuan Lin and Qi Yang and Lei Wang and Frank Pollmann and Michael P. Zaletel},
      year={2025},
      eprint={2502.10695},
      archivePrefix={arXiv},
      primaryClass={quant-ph},
      url={https://arxiv.org/abs/2502.10695}, 
}

@article{Disentangling2025Wu,
  title = {Disentangling interacting systems with fermionic Gaussian circuits: Application to quantum impurity models},
  author = {Wu, Ang-Kun and Kloss, Benedikt and Krinitsin, Wladislaw and Fishman, Matthew T. and Pixley, J. H. and Stoudenmire, E. M.},
  journal = {Phys. Rev. B},
  volume = {111},
  issue = {3},
  pages = {035119},
  numpages = {18},
  year = {2025},
  month = {Jan},
  publisher = {American Physical Society},
  doi = {10.1103/PhysRevB.111.035119},
  url = {https://link.aps.org/doi/10.1103/PhysRevB.111.035119}
}

@article{ferris2012perfect,
	author = {Ferris, Andrew J. and Vidal, Guifre},
	doi = {10.1103/PhysRevB.85.165146},
	issue = {16},
	journal = {Phys. Rev. B},
	month = {Apr},
	numpages = {10},
	pages = {165146},
	publisher = {American Physical Society},
	title = {Perfect sampling with unitary tensor networks},
	url = {https://link.aps.org/doi/10.1103/PhysRevB.85.165146},
	volume = {85},
	year = {2012}}

@article{Barthel_2009,
   title={Contraction of fermionic operator circuits and the simulation of strongly correlated fermions},
   volume={80},
   ISSN={1094-1622},
   url={http://dx.doi.org/10.1103/PhysRevA.80.042333},
   DOI={10.1103/physreva.80.042333},
   number={4},
   journal={Physical Review A},
   publisher={American Physical Society (APS)},
   author={Barthel, Thomas and Pineda, Carlos and Eisert, Jens},
   year={2009},
   month=Oct }

@article{Mortier_2025,
   title={Fermionic tensor network methods},
   volume={18},
   ISSN={2542-4653},
   url={http://dx.doi.org/10.21468/SciPostPhys.18.1.012},
   DOI={10.21468/scipostphys.18.1.012},
   number={1},
   journal={SciPost Physics},
   publisher={Stichting SciPost},
   author={Mortier, Quinten and Devos, Lukas and Burgelman, Lander and Vanhecke, Bram and Bultinck, Nick and Verstraete, Frank and Haegeman, Jutho and Vanderstraeten, Laurens},
   year={2025},
   month=Jan }

@article{Wolf2006Violation,
  title = {Violation of the Entropic Area Law for Fermions},
  author = {Wolf, Michael M.},
  journal = {Phys. Rev. Lett.},
  volume = {96},
  issue = {1},
  pages = {010404},
  numpages = {4},
  year = {2006},
  month = {Jan},
  publisher = {American Physical Society},
  doi = {10.1103/PhysRevLett.96.010404},
  url = {https://link.aps.org/doi/10.1103/PhysRevLett.96.010404}
}

@article{Verstraete_2006,
   title={Criticality, the Area Law, and the Computational Power of Projected Entangled Pair States},
   volume={96},
   ISSN={1079-7114},
   url={http://dx.doi.org/10.1103/PhysRevLett.96.220601},
   DOI={10.1103/physrevlett.96.220601},
   number={22},
   journal={Physical Review Letters},
   publisher={American Physical Society (APS)},
   author={Verstraete, F. and Wolf, M. M. and Perez-Garcia, D. and Cirac, J. I.},
   year={2006},
   month=June }

@misc{wong2022zipperentanglementrenormalizationfree,
      title={Zipper Entanglement Renormalization for Free Fermions}, 
      author={Sing Lam Wong and Ka Chun Pang and Hoi Chun Po},
      year={2022},
      eprint={2206.11761},
      archivePrefix={arXiv},
      primaryClass={quant-ph},
      url={https://arxiv.org/abs/2206.11761}, 
}

@article{Pasquale_Calabrese_2004,
   title={Entanglement entropy and quantum field theory},
   volume={2004},
   ISSN={1742-5468},
   url={http://dx.doi.org/10.1088/1742-5468/2004/06/P06002},
   DOI={10.1088/1742-5468/2004/06/p06002},
   number={06},
   journal={Journal of Statistical Mechanics: Theory and Experiment},
   publisher={IOP Publishing},
   author={Pasquale Calabrese and John Cardy},
   year={2004},
   month=June, pages={P06002} }

@article{A_J_Daley_2004,
   title={Time-dependent density-matrix renormalization-group using adaptive effective Hilbert
spaces},
   volume={2004},
   ISSN={1742-5468},
   url={http://dx.doi.org/10.1088/1742-5468/2004/04/P04005},
   DOI={10.1088/1742-5468/2004/04/p04005},
   number={04},
   journal={Journal of Statistical Mechanics: Theory and Experiment},
   publisher={IOP Publishing},
   author={A J Daley and C Kollath and U Schollwöck and G Vidal},
   year={2004},
   month=Apr, pages={P04005} }

@article{Vidal_2004,
   title={Efficient Simulation of One-Dimensional Quantum Many-Body Systems},
   volume={93},
   ISSN={1079-7114},
   url={http://dx.doi.org/10.1103/PhysRevLett.93.040502},
   DOI={10.1103/physrevlett.93.040502},
   number={4},
   journal={Physical Review Letters},
   publisher={American Physical Society (APS)},
   author={Vidal, Guifré},
   year={2004},
   month=July }

@misc{guaita2024representationtheorygaussianunitary,
      title={Representation theory of Gaussian unitary transformations for bosonic and fermionic systems}, 
      author={Tommaso Guaita and Lucas Hackl and Thomas Quella},
      year={2024},
      eprint={2409.11628},
      archivePrefix={arXiv},
      primaryClass={quant-ph},
      url={https://arxiv.org/abs/2409.11628}, 
}

@article{Pirvu_2010,
   title={Matrix product operator representations},
   volume={12},
   ISSN={1367-2630},
   url={http://dx.doi.org/10.1088/1367-2630/12/2/025012},
   DOI={10.1088/1367-2630/12/2/025012},
   number={2},
   journal={New Journal of Physics},
   publisher={IOP Publishing},
   author={Pirvu, B and Murg, V and Cirac, J I and Verstraete, F},
   year={2010},
   month=Feb, pages={025012} }

@article{Lieb1989Two,
  title = {Two theorems on the Hubbard model},
  author = {Lieb, Elliott H.},
  journal = {Phys. Rev. Lett.},
  volume = {62},
  issue = {10},
  pages = {1201--1204},
  numpages = {0},
  year = {1989},
  month = {Mar},
  publisher = {American Physical Society},
  doi = {10.1103/PhysRevLett.62.1201},
  url = {https://link.aps.org/doi/10.1103/PhysRevLett.62.1201}
}

@article{strang1968construction,
  title={On the construction and comparison of difference schemes},
  author={Strang, Gilbert},
  journal={SIAM journal on numerical analysis},
  volume={5},
  number={3},
  pages={506--517},
  year={1968},
  publisher={SIAM}
}

@article{suzuki1976generalized,
  title={Generalized Trotter's formula and systematic approximants of exponential operators and inner derivations with applications to many-body problems},
  author={Suzuki, Masuo},
  journal={Communications in Mathematical Physics},
  volume={51},
  number={2},
  pages={183--190},
  year={1976},
  publisher={Springer}
}

@article{Julku_2016,
   title={Geometric Origin of Superfluidity in the Lieb-Lattice Flat Band},
   volume={117},
   ISSN={1079-7114},
   url={http://dx.doi.org/10.1103/PhysRevLett.117.045303},
   DOI={10.1103/physrevlett.117.045303},
   number={4},
   journal={Physical Review Letters},
   publisher={American Physical Society (APS)},
   author={Julku, Aleksi and Peotta, Sebastiano and Vanhala, Tuomas I. and Kim, Dong-Hee and Törmä, Päivi},
   year={2016},
   month=July }

@article{Rhim_2019,
   title={Classification of flat bands according to the band-crossing singularity of Bloch wave functions},
   volume={99},
   ISSN={2469-9969},
   url={http://dx.doi.org/10.1103/PhysRevB.99.045107},
   DOI={10.1103/physrevb.99.045107},
   number={4},
   journal={Physical Review B},
   publisher={American Physical Society (APS)},
   author={Rhim, Jun-Won and Yang, Bohm-Jung},
   year={2019},
   month=Jan }

@article{Rhim01012021,
    author = {Rhim, Jun-Won and Yang, Bohm-Jung},
    title = {Singular flat bands},
    journal = {Advances in Physics: X},
    volume = {6},
    number = {1},
    pages = {1901606},
    year = {2021},
    publisher = {Taylor \& Francis},
    doi = {10.1080/23746149.2021.1901606},
    URL = { https://doi.org/10.1080/23746149.2021.1901606
    },
    eprint = { https://doi.org/10.1080/23746149.2021.1901606
    }
}

@article{Dubail_2017,
doi = {10.1088/1751-8121/aa6f38},
url = {https://doi.org/10.1088/1751-8121/aa6f38},
year = {2017},
month = {may},
publisher = {IOP Publishing},
volume = {50},
number = {23},
pages = {234001},
author = {Dubail, J},
title = {Entanglement scaling of operators: a conformal field theory approach, with a glimpse of simulability of long-time dynamics in 1+1d},
journal = {Journal of Physics A: Mathematical and Theoretical},
}

@article{Pal_2018,
   title={Entangling power of time-evolution operators in integrable and nonintegrable many-body systems},
   volume={98},
   ISSN={2469-9969},
   url={http://dx.doi.org/10.1103/PhysRevB.98.174304},
   DOI={10.1103/physrevb.98.174304},
   number={17},
   journal={Physical Review B},
   publisher={American Physical Society (APS)},
   author={Pal, Rajarshi and Lakshminarayan, Arul},
   year={2018},
   month=Nov }

@article{Wolf_2006,
  title = {Violation of the Entropic Area Law for Fermions},
  author = {Wolf, Michael M.},
  journal = {Phys. Rev. Lett.},
  volume = {96},
  issue = {1},
  pages = {010404},
  numpages = {4},
  year = {2006},
  month = {Jan},
  publisher = {American Physical Society},
  doi = {10.1103/PhysRevLett.96.010404},
  url = {https://link.aps.org/doi/10.1103/PhysRevLett.96.010404}
}

@article{Swingle_2010,
  title = {Entanglement Entropy and the Fermi Surface},
  author = {Swingle, Brian},
  journal = {Phys. Rev. Lett.},
  volume = {105},
  issue = {5},
  pages = {050502},
  numpages = {4},
  year = {2010},
  month = {Jul},
  publisher = {American Physical Society},
  doi = {10.1103/PhysRevLett.105.050502},
  url = {https://link.aps.org/doi/10.1103/PhysRevLett.105.050502}
}

@article{Gioev_2006,
  title = {Entanglement Entropy of Fermions in Any Dimension and the Widom Conjecture},
  author = {Gioev, Dimitri and Klich, Israel},
  journal = {Phys. Rev. Lett.},
  volume = {96},
  issue = {10},
  pages = {100503},
  numpages = {4},
  year = {2006},
  month = {Mar},
  publisher = {American Physical Society},
  doi = {10.1103/PhysRevLett.96.100503},
  url = {https://link.aps.org/doi/10.1103/PhysRevLett.96.100503}
}

@misc{chan2026cherninsulatorsfixedpoints,
      title={Chern Insulators as Fixed Points of Entanglement Renormalization}, 
      author={Hing Nin Chan and Wenqing Xie and Sing Lam Wong and Hoi Chun Po},
      year={2026},
      eprint={2611.37788},
      archivePrefix={arXiv},
      primaryClass={cond-mat.str-el},
      url={https://arxiv.org/abs/2611.37788}, 
}

\end{document}